\documentclass[11pt]{article}
\pdfoutput=1
\usepackage{jcappub,natbib}
\input{colordvi.tex}

\def\ga{\mathrel{\raise.3ex\hbox{$>$\kern-.75em\lower1ex\hbox{$\sim$}}}}
\def\la{\mathrel{\raise.3ex\hbox{$<$\kern-.75em\lower1ex\hbox{$\sim$}}}}

\title{Rolling axions during inflation: perturbativity and signatures}

\author[a]{Marco Peloso,}
\author[b]{Lorenzo Sorbo,}
\author[a]{Caner Unal}

\affiliation[a]{School of Physics and Astronomy, and Minnesota Institute for Astrophysics, University of Minnesota, Minneapolis, 55455 (USA)}
\affiliation[b]{Amherst Center for Fundamental Interactions, Department of Physics, University of Massachusetts, Amherst, MA 01003 (USA)}

\abstract{The motion of a pseudo-scalar field $X$ during inflation naturally induces a significant amplification of the gauge fields to which it is coupled. 
 The amplified gauge fields can source characteristic scalar and tensor primordial perturbations. Several phenomenological implications have been discussed in the cases in which (i) $X$ is the inflation, and (ii) $X$ is a field different from the inflation, that experiences a  temporary speed up during inflation. In this second case, visible sourced gravitational waves (GW) can be produced at the CMB scales without affecting the scalar perturbations, even if the scale of inflation is several orders of magnitude below what is required to produce a visible vacuum GW signal. 
Perturbativity considerations can be used to limit the regime in which these results are under perturbative control.  We revised limits recently claimed for the case (i), and we extend these considerations to the case (ii). We show that, in both cases, these limits are satisfied by the applications that generate signals at CMB scales. Applications that generate gravitational waves and primordial black holes  at much smaller scales are at the limit of the validity of this perturbativity analysis, so we expect those results to be valid up to possibly order one corrections. 
}

\begin{document}

\begin{flushright}  ACFI-T16-16, UMN--TH--3531/16  \end{flushright}

\maketitle
\flushbottom

\section{Introduction}%
\label{sec:intro}

There is a one-to-one relationship between the energy scale of inflation $V_{\rm infl}^{1/4}$ and the amplitude ${\cal P}_{\rm GW}$ of primordial gravitational waves (GW) produced by the amplification of vacuum fluctuations on a quasi de-Sitter geometry,
\begin{align}\label{standard_pt}
{\cal P}_{\rm GW}=\frac{2}{\pi^2}\,\frac{V_{\rm infl}}{3\,M_P^4}\,. 
\end{align}
This result makes  the measurement of ${\cal P}_{\rm GW}$ one of the main science objectives of the current and upcoming CMB experiments \cite{Kamionkowski:2015yta}. It is therefore crucial to determine how robust the relation~(\ref{standard_pt}) is. In particular, this result  is generally invalid in presence of additional sources of GW that increase the left hand side of eq.~(\ref{standard_pt}) for a given value of $V_{\rm infl}$~\cite{Cook:2011hg,Senatore:2011sp}. 

However, introducing additional sources of primordial gravitational waves which are intense enough  is not easy, once we require that those sources do not spoil any  of the other successful predictions of inflation \cite{Barnaby:2012xt}. For instance, any GW source will also source  scalar perturbations $\zeta$ with a coupling that is at least of gravitational strength (or stronger, if the source is directly coupled to the field responsible for the observed density perturbations). This can result in a decrease of the observed tensor-to-scalar ratio 
\begin{equation}
r \equiv \frac{P_{\rm GW,vacuum}+P_{\rm GW,sourced}}{P_{\rm \zeta,vacuum}+P_{\rm \zeta,sourced}} \;, 
\end{equation}
even if $P_{\rm GW,sourced} \gg P_{\rm GW,vacuum}$  \cite{Barnaby:2012xt}. Moreover, the strong production required 
to obtain a large GW signal, typically leads to large non-Gaussianity, as first shown in \cite{Barnaby:2010vf} and \cite{Barnaby:2012xt}, and then  in \cite{Mirbabayi:2014jqa}. 

An effective and well studied mechanism that can result in a large sourced GW signal~\footnote{Other inflationary mechanisms for the generation of GW  which are alternative to the standard vacuum production include the use of  spectator fields  \cite{spectator}, of fields with fast-varying masses~\cite{Cook:2011hg,Senatore:2011sp,Carney:2012pk}, the modification of the dispersion relation of the tensor modes in the effective-field-theory approach \cite{Cannone:2014uqa,Cannone:2015rra}, breaking of space diffeomorphisms~\cite{Bartolo:2015qvr}, varying sound speed of the tensor \cite{Cai:2015dta,Cai:2016ldn},  the strong tachyonic growth of chiral tensor modes in the inflationary models with non-abelian gauge fields \cite{non-abelian}, tensor fossils \cite{Dimastrogiovanni:2014ina}, and  preheating \cite{preheating} (in this last case the produced GW signal is typically at scales much smaller than the CMB ones). It has also been investigated \cite{Antoniadis:2014xva} whether the presence of many light degrees of freedom 
could modify (\ref{standard_pt}). As shown in \cite{Kleban:2015daa}, this does not appear to be the case (see however \cite{Antoniadis:2015txa}).  See  \cite{Guzzetti:2016mkm} for a recent review on GW and inflation.} §  is the one where the rolling inflaton (or a rolling scalar spectator) $X$ source though an axion-like coupling vector modes which, in their turn, act as sources on gravitational waves. Two particular situations have been considered in the literature: either the field $X$ rolls at an approximately constant velocity~\cite{Sorbo:2011rz,Barnaby:2012xt}, or it experiences a transient of relatively fast~\footnote{More specifically, we assume that $\dot{X}$ is significantly different from zero only for a limited time. Also during this time we assume that $X$ is in a regime of slow-roll, namely that  $\vert \dot{X} \vert \ll H M_p$.}  roll~\cite{Namba:2015gja}.  

In the first case the additional contribution to the tensor spectrum is approximately scale invariant and slightly blue, since the amount of sourced GW is controlled by $\dot{X}$, which typically increases during inflation. In the second case the tensor spectrum will show a spike at the scales that left the horizon at the time when $\dot{X}$ was maximal. This second case is especially interesting if those scales correspond to those of the recombination bump in the CMB B-mode power spectrum. In this case, it is possible to generate a visible GW signal at arbitrary small scale of inflation $V_{\rm infl}$, without violating bounds from the power spectrum and bispectrum of the scalar perturbations \cite{Namba:2015gja}. We rediscuss this result in the present work,~\footnote{In doing so, we provide some detailed expression not given in \cite{Namba:2015gja}. In particular, we give analytic expressions for the peak of the sourced GW signal, and for the needed energy density in the gauge field.} in light of the extensive discussion in the literature on limits on these mechanisms~\cite{Ferreira:2014zia,Ferreira:2015omg} and, particularly, of a recent claim  that the sourced GW signal cannot be parametrically much stronger than the vacuum one \cite{Ozsoy:2014sba}.  
 
Since these mechanisms require that the sourced gravitational waves have a relatively large amplitude, in general the excited sector that sources the tensors must contain a sizable energy density. As a consequence one can wonder whether such large energy densities can bring us out of the perturbative regime in which such effects are analyzed, and put into question the validity of such calculations. Some of the perturbativity requirements are rather straightforward. For instance, one should obviously demand that all the energy in produced modes is smaller than the kinetic energy of the field  $X$, that is the source of the vector modes. 
 If this condition is not met, a more complete analysis that includes backreaction on the inflating background is needed, and it has been taken into account in several applications of this mechanism.  A less straightforward requirement is that the three point correlators be subdominant to the two point correlators. This question was tackled in Appendix F of~\cite{Namba:2015gja}.   More recently, reference~\cite{Ferreira:2015omg} has considered two additional requirements from perturbativity for the models of~\cite{Sorbo:2011rz,Barnaby:2012xt}, where $\dot{X}$ is approximately constant: first, that higher order effects do not spoil the leading order estimate for the amplitude of the gauge field amplified by the rolling of the $X$ field; second, that the fluctuations of the $X$ field  do not induce a variance $\sqrt{\langle X^2 \rangle}$ that is greater than the periodicity of the potential for $X$, and hence of the classical zero mode  of $X$.  The authors of~\cite{Ferreira:2015omg} find that in some regime some of these perturbativity conditions can be violated -- even if not by a parametrically large amount. 

In view of the fact that these models do exhibit interesting phenomenology (such as potential implications for inflationary magnetogenesis~\cite{Anber:2006xt,Caprini:2014mja,Bamba:2014vda}, CMB non-Gaussianity  \cite{Barnaby:2010vf,Barnaby:2011vw}, growth of the scalar power spectrum at CMB scales   \cite{Meerburg:2012id}, gravitational waves~\footnote{Further discussion of these, or similar mechanisms  can be found in~\cite{gauge-GW-discussion}.}  that might be detectable by gravitational interferometers~\cite{Cook:2011hg,Barnaby:2011qe,Domcke:2016bkh}, parity violation in the CMB~\cite{Sorbo:2011rz} and in interferometers~\cite{Crowder:2012ik}, primordial black holes~\cite{Linde:2012bt,Erfani:2015rqv,Cheng:2015oqa,McDonough:2016xvu}, blue tensor spectra~\cite{Mukohyama:2014gba} and large and parity violating tensor bispectra~\cite{odd-TTT})  in different portions of their parameter space,  we agree on the importance of studying the limits pointed out in~\cite{Ferreira:2015omg}. In the present paper, we show that  the models~\cite{Sorbo:2011rz,Barnaby:2012xt} are generally in better shape than what found in~\cite{Ferreira:2015omg}. In Appendix \ref{app:comparison}, we discuss the origin of the difference between our conclusions and those of~\cite{Ferreira:2015omg}.  

This work shows that the application of the mechanisms of~\cite{Sorbo:2011rz,Barnaby:2012xt} that produce signatures at CMB scales are well consistent with the limits from perturbativity. On the contrary, one of these limits is marginally violated by applications that produce signals at much smaller scales, as for instance GW at interferometers and primordial black-holes. The violation we find is much smaller than that obtained in  \cite{Ferreira:2015omg}: the gauge field production is controlled by a parameter $\xi$, that needs to be $\simeq 5$ for those signatures to be relevant, while the perturbativity criteria give $\xi \la 4.8$.  We therefore expect that these results remain valid, with possibly ${\rm O } \left( 1 \right)$ corrections.~\footnote{The criteria we formulate cannot be applied in the regime of very large $\xi$, as the one of~\cite{Anber:2009ua}, where the backreaction of the produced gauge field is strongly affecting the background dynamics, and the sourced scalar perturbations are much greater than the vacuum one. Ref.~\cite{Anber:2009ua} discussed how to deal with the scalar perturbations in that case.} 

 While the study~\cite{Ferreira:2015omg} focused on the case where $\dot{X}$ is approximately constant, here we  also consider the case~\cite{Namba:2015gja} where $X$ experiences a transient roll. This has been used in \cite{Namba:2015gja} to generate a visible sourced GW signal at CMB scales. We find that in this mechanism the limits for perturbativity can be satisfied in a large range of parameter space. 

Our work is structured as follows. In section~\ref{sec:gauge} we review in detail the amplification of gauge fields both in the case where $\dot{X}$ is constant and in the case where it rolls only for a few efoldings. In section~\ref{sec:phenomenology} we review some of the signatures from these mechanisms, and the phenomenologically interesting regions of parameter space of the model (these are the regions where the perturbative analysis is required to be consistent). In Section~\ref{sec:back} we compute the limits that ensure small backreaction of the produced gauge fields on the background dynamics. Our main results on perturbativity are presented in Section~\ref{sec:pertlimits}. The limits from backreaction and perturbativity are then studied in Section~\ref{sec:significance}. Section~\ref{sec:conclusions} contains our conclusions. The technical aspects of our calculations are relegated to several appendices.

\section{Gauge field amplification}
\label{sec:gauge}

The portion of the Lagrangian containing gauge field is
\begin{equation}
{\cal L}_A = - \frac{1}{4} F^2 - \frac{\alpha}{4 \, f} \, X \, F \, {\tilde F} \;, 
\label{lagrangian}
\end{equation}
where $X$ denotes the pseudo-scalar field being directly coupled to the vector field. In this work, we analyze 2 cases: 
\begin{itemize}
\item $X = \phi$ is the inflaton field, and $\dot{\phi}$ evolves adiabatically; 
\item $X = \sigma$ is a field with an energy density much smaller than that of the  inflaton,  which experiences a momentary speed-up for a few e-folds during inflation. 
\end{itemize}

We work in the $A_0 = \vec\nabla\cdot\vec{A}= 0$  gauge, and we decompose the  gauge field  in  two components of definite helicity 
\begin{equation}
\vec{A} =  \int \frac{d^3 k}{\left( 2 \pi \right)^{3/2}} \, {\rm e}^{i \vec{k} \cdot \vec{x}} \, \sum_{\lambda=+,-} \vec{\epsilon}_\lambda  \left( \vec{k} \right) \, {\hat A}_\lambda \left( t ,\, \vec{k} \right) \;, 
\label{A-deco} 
\end{equation}
where
\begin{equation}
{\hat A}_\lambda \left( t ,\, \vec{k} \right) = A_\lambda \left( t ,\, k \right) \, {\hat a}_\lambda \left( \vec{k} \right) +  A_\lambda^* \left( t ,\, k \right) \, {\hat a}_\lambda \left( - \vec{k} \right)^\dagger  \;. 
\end{equation}

The equation of motion for the mode function of the gauge field is 
\begin{equation}
\left( \partial_\tau^2 + k^2 \pm \frac{2\, k\, \xi}{\tau} \right) A_\pm = 0 \;\;\;,\;\;\; \xi \equiv \frac{\alpha \,  \dot{X}  }{2\, f\, H} \;, 
\label{genmodeqn}
\end{equation} 
where $\tau$ is the conformal time, which during inflation is related to the scale factor $a$ by $a = - \frac{1}{H \, \tau}$ at zeroth order in slow roll. 

Depending on the sign of $\xi$, one of the two helicity modes is unstable next to horizon crossing  (as we shall see, phenomenology requires $\xi \la {\rm O } \left( {\mathrm {few}} \right)$). We assume $\xi > 0$, so that the unstable polarization is the $+$ one (recall that $\tau < 0 $). This polarization can receive a substantial amplification (this is not the case for the negative helicity mode). The amplification is controlled by $\xi$, which has a very different behavior in the two different cases mentioned above that we consider in this work.

\subsection{Nearly constant gauge field from  $\frac{\alpha}{f} \, \phi \, F \, {\tilde F}$ } 
\label{sec:consxiAsol}

If $X$ is the inflaton field $\phi$, which is slowly rolling during inflation, we have $\xi \simeq \frac{\alpha \, \sqrt{2 \, \epsilon_\phi} \, M_p}{2 f}$, where $\epsilon_\phi$ is the standard slow roll inflaton parameter. This quantity is changing only at second order in slow roll, and therefore $\xi$  can be approximated as a constant while any single mode (a mode with a given value of $k$) has a size comparable to the horizon. As we shall see, this is the time range during  which a mode is produced, and can lead to potentially observable effects (before its energy is diluted away by the expansion of the universe). However, two different modes $k_1 \neq k_2$ leave the horizon at different moments, so they probe a possibly different value of $\xi$, depending on how much  $\sqrt{\epsilon_{\phi}}$ has changed between the two different times at which the two modes left the horizon. In this case we treat $\xi$ as an adiabatically evolving parameter, denoting by $\xi_k$ the value of $\xi$ assumed when a given mode $k$ left the horizon (this is the constant value of $\xi$  that we take in computing the evolution of that specific mode).~\footnote{Improving over this will provide corrections proportional to the slow roll parameters.} 
 
Under the assumption of constant $\xi$ (for a given mode $k$), eq.~(\ref{genmodeqn}) can be solved analytically. The normalized solution satisfying adiabatic vacuum  initial conditions is given in terms of the irregular Coulomb function
\begin{equation}
A_\pm \simeq \frac{1}{\sqrt{2 k}} \, H_0^{\pm} \left( \pm \xi ,\, - k \tau \right) \;. 
\label{colsol}
\end{equation}
(were the approximated equality is due to the fact that $\xi$ is not exactly constant). In the limit $\xi \gg -k \tau $, this  solution  is very-well approximated by
\begin{equation}
A_+ \simeq \sqrt{\frac{-\tau}{2}} \left[ 2 \, {\rm e}^{\pi \xi} \, \pi^{-1/2} \, K_1 \left( \sqrt{-8 \, \xi \, k \, \tau } \right) 
+  i \, {\rm e}^{-\pi \xi} \, \pi^{1/2} \, I_1 \left( \sqrt{-8 \, \xi \, k \, \tau } \right) \right] \;, 
\label{bessol}
\end{equation} 
where, $K_1$ and $I_1$ are modified Bessel functions of second and first types, respectively.  Since, as we shall see,  $\xi \ga O(1)$ in all interesting cases, the field amplification occurs around horizon crossing. One can further simplify this result by taking the large argument limit of Bessel functions
\begin{equation}
A_+ \simeq \frac{1}{\sqrt{2 k}} \left( \frac{-k \tau}{2 \xi} \right)^{1/4} \, {\rm e}^{\pi \xi - 2 \sqrt{-2 \xi k \tau}} 
+  \frac{i}{\sqrt{2 k}} \left( \frac{-k \tau}{2^5 \xi} \right)^{1/4} \, {\rm e}^{-\pi \xi + 2 \sqrt{-2 \xi k \tau}} 
\;\;\;,\;\;\; \frac{1}{8 \xi} \ll - k \, \tau \ll 2 \, \xi \;, 
\label{A-simple}
\end{equation} 
and
\begin{equation}
A_+' \simeq \sqrt{\frac{k}{2}} \,  \left( \frac{2 \xi}{-k \tau} \right)^{1/4} \, {\rm e}^{\pi \xi - 2 \sqrt{-2 \xi k \tau}} 
- i \, \sqrt{\frac{k}{2}} \,  \left( \frac{\xi}{-8 k \tau} \right)^{1/4} \, {\rm e}^{-\pi \xi + 2 \sqrt{-2 \xi k \tau}} 
\;\;\;,\;\;\; \frac{1}{8 \xi} \ll - k \, \tau \ll 2 \, \xi \;. 
\label{Ader-simple}
\end{equation} 

\begin{figure}[tbp]
\centering 
\includegraphics[width=0.5\textwidth,angle=0]{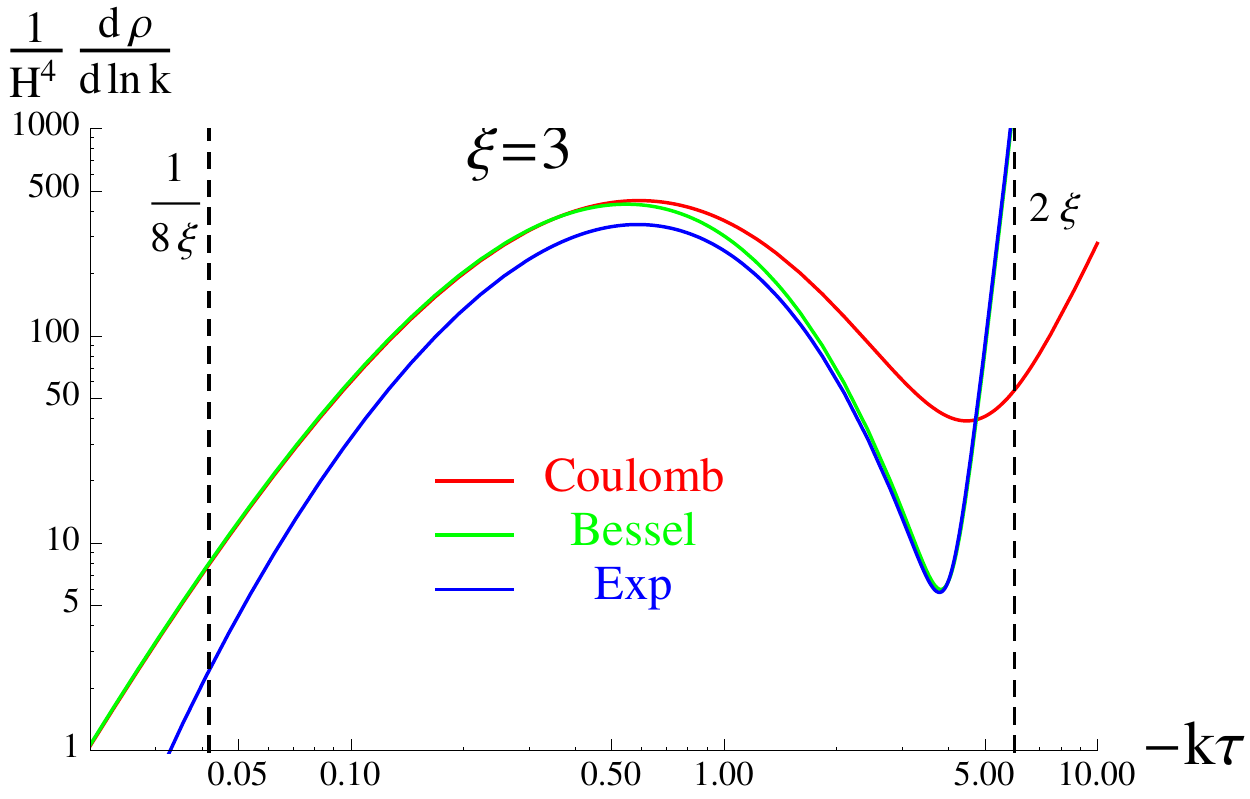}
\hfill
\caption{Time evolution of the contribution to the gauge field physical energy density from modes with a given comoving momentum $k$. The three different curves correspond to the three approximated solutions    (\ref{colsol}), (\ref{bessol}) and (\ref{A-simple}). From early to late times (from right to left), the figure shows the UV-divergent vacuum energy density, the gauge field amplification due its interaction with $X \left( t \right)$, and the dilution due to the expansion of the universe. For definiteness, the constant parameter $\xi = 3$  is assumed. 
}
\label{fig:rhoti}
\end{figure}

The real part of the approximations (\ref{A-simple})  and (\ref{Ader-simple}) encodes the amplification of the positive helicity gauge mode. The imaginary part guarantees that  the Wronskian condition   $A_+ A_+^{'*} - c.c. = i$ is satisfied. We note that these expressions are related by 
\begin{equation}
\frac{d A_+'\left( k ,\, \tau \right)}{d \tau}  =  \sqrt{\frac{2 \, k \, \xi }{- \tau}} \, A_+^* \left( k ,\, \tau \right) \;. 
\label{Ader}
\end{equation}
To appreciate the difference among the three approximations  (\ref{colsol}), (\ref{bessol}) and (\ref{A-simple}), and to understand the timescale of the gauge field amplification and subsequent dilution, we show in Figure \ref{fig:rhoti} the time evolution of the physical energy density of the gauge field modes with a given comoving momentum $k$. Details of the computation are given in Appendix \ref{app:rhoA}. 

Time flows from right to left in the figure, with $- k \tau = \frac{p}{H} = 1$ denoting horizon crossing ($p$ is the physical momentum of the mode, while $H$ is the Hubble rate). At the earliest times shown, the mode is deep inside the horizon, and the figure shows the energy density associated to the vacuum mode solution. Namely, at very large $\frac{k}{a}$, the term proportional to $\xi$ can be disregarded in eq. (\ref{genmodeqn}), and the mode has the standard (comoving) dispersion relation $\omega^2 = k^2$. The energy density of the vacuum mode is UV-divergent, and it needs to be renormalized away (we stress that this has nothing to do with the gauge field amplification studied in this work). As done in the literature, we simply cut-off this UV regime when we compute the observable effects of the gauge modes. Following the time evolution of the curves in the figure, we observe a decrease of this vacuum energy contribution, and then a growth of the energy density. For $\xi = {\rm O } \left( 1 \right)$, this growth takes place near horizon crossing (for definiteness, $\xi = 3$ was assumed in the evolutions shown in the figure). This growth is then followed by a decrease at the latest times shown in the figure, leading to a peak of the physical energy density close to horizon crossing. We stress that we are showing only the energy density of modes with a given comoving momentum $k$. At any times during inflation, there is a nearly constant energy density in gauge fields, due to the modes that have size comparable to the horizon at that given moment. 

The growth visible in the figure shows  the gauge field amplification due to its coupling to $X \left( t \right)$. The dilution is due to the expansion of the universe. The resulting peak is well separated from the UV-diverging part (we note that the vertical axis of the figure is in log scale), leading to a clear distinction between the physical field amplification, and the standard divergence associated with the empty vacuum state. 

The produced gauge field, before being diluted away,  sources scalar perturbations and gravitational waves. The phenomenological implications have been studied in a number of works in the literature that have used the approximate solution (\ref{A-simple}). The goal of this work is to study whether these results are stable under quantum correction and backreaction considerations. To do this, we need to consistently use the same approximation also in this work.~\footnote{In Appendix \ref{app:comparison} we show that the results do not change significantly if one instead uses the Coulomb functions (\ref{colsol}).} 

The condition $\frac{1}{8 \xi} \ll - k \tau \ll 2 \xi$ are mathematical conditions for (\ref{bessol}) to reproduce  (\ref{colsol}), and for  (\ref{A-simple}) to reproduce  (\ref{bessol}). However, we can see from the figure that the specific values $ - k \tau = \frac{1}{8 \xi} ,\, 2 \, \xi$ do not have an immediate physical meaning. In setting a UV cut-off, we rather use $- k \tau \vert_{\rm max} = \xi$, which, as visible in the figure, well approximates the position of the minimum between the unphysical vacuum energy density, and the physical bump in the energy density due to the gauge field amplification
(this is true in the $\xi \sim 3-5$ range we are interested in). We also see from the figure that we do not need to require that each individual mode is in the perturbative regime at arbitrary late times, when the physical energy density in that mode has become negligible due to the expansion of the universe. We want to ensure that this mode is the perturbative regime while it is contributing to a physical observable. Specifically, we do not care to ensure perturbativity when the physical energy density in a mode has decreased to less than a few percent of its peak value (we will study how the constraints change when we change this threshold).

\subsection{Bump in gauge field production from  $\frac{\alpha}{f} \, \sigma \, F \, {\tilde F}$ } 
\label{sec:bumpxiAsol}

We now discuss the gauge amplification in the case in which the parameter $\xi$ is significant for only a few e-folds during inflation. We assume that the field $X$ has a momentary faster roll in that period. In principle, we could still identify $X$ with the inflaton field, and the momentary speed-up with a feature in the inflaton potential. However, the inclusion of this case in the present work is done mostly to check the validity of the results of~\cite{Namba:2015gja}, in which the field $X$ was associated with a pseudo-scalar $\sigma$ which was not the inflaton field. For this reason, we make this assumption also here. 

We assume the simplest and most natural potential for a pseudo-scalar field  \cite{Namba:2015gja} 
\begin{equation}
V = \frac{\Lambda^4}{2} \left[ \cos \left( \frac{\sigma}{f} \right) + 1 \right] \;. 
\label{V-sigma}
\end{equation}
This term is added  to the inflation potential; therefore we assume no direct coupling between the inflaton and $\sigma$. The field $\sigma$ rolls for a few e-folds if the curvature of the potential is tuned to be comparable to $H$. Specifically, in terms of the parameter 
\begin{equation}
\delta \equiv \frac{\Lambda^4}{6 H^2 f^2} \;, 
\label{delta}
\end{equation} 
the field $\sigma$ acquires the mass $m_\sigma = \sqrt{3 \delta} \, H$ at the minimum of (\ref{V-sigma}). The equation of motion of $\sigma$ is solved by   \cite{Namba:2015gja} 
\begin{equation}
\sigma =2 \, f \, {\rm Arctan} \left[{\rm e}^{\delta \, H \left( t - t_* \right)} \right]  \;\;\; \Rightarrow \;\;\; \dot{\sigma} = \frac{f \, H \, \delta}{\cosh \left[ \delta \, H \left( t - t_* \right) \right] } \;, 
\label{dotsigma}
\end{equation}
under the assumption that $H$ is constant and that (\ref{V-sigma}) is much smaller than the inflaton potential. In this solution, $t_*$ denotes the time at which $\sigma$ evaluates to $\frac{\pi \, f}{2}$ and reaches its maximum speed $\dot{\sigma}_* \equiv \dot{\sigma} \left( t_* \right)$. We immediately see that $\dot{\sigma}$ is significant only for a number of e-folds $\Delta N \simeq \frac{1}{\delta}$ around $t=t_*$. 

In this model $\xi = \frac{\alpha \, \dot{\sigma}}{2 H f}$ is varying rapidly when most of the field amplification is taking place ($\dot \sigma \simeq \dot{\sigma}_*$). So the solutions obtained in the previous subsection are no longer valid.

Following  \cite{Namba:2015gja}, we compute this solution in Appendix \ref{app:A-bump}, where we find 
\begin{eqnarray}
A_+ \left( \tau > \tau_* \right) &\simeq&  N \left[ \xi_* ,\, x_* ,\, \delta \right] \, \left( \frac{-\tau}{8 \, k \, \xi \left( \tau \right)} \right)^{1/4}  \, {\rm exp} \left[ - \frac{4 \xi_*^{1/2}}{1+\delta} \, \left( \frac{-\tau}{-\tau_*} \right)^{\delta/2} \, \left( - k \tau \right)^{1/2} \right] 
\nonumber\\ 
&&+ \frac{i}{N \left[ \xi_* ,\, x_* ,\, \delta \right]} \, \left( \frac{-\tau}{2^7 \, \xi \left( \tau \right) \, k} \right)^{1/4}  \, {\rm exp} \left[  \frac{4 \xi_*^{1/2}}{1+\delta} \, \left( \frac{-\tau}{-\tau_*} \right)^{\delta/2} \, \left( - k \tau \right)^{1/2} \right] \;, \nonumber\\ 
\label{mainvarysigma-ARI}
\end{eqnarray}
while the mode is not amplified (and therefore, negligibly small) for $\tau < \tau_*$. In this expression, $\tau_*$ is the conformal time corresponding to the physical time $t_*$, while $x_*$ and $\xi_*$ denote the values assumed by $x \equiv - k \, \tau$ and by $\xi$ at this moment. (We note that 
$\xi_* = \frac{\alpha \, \delta}{2}$.) The normalization factor can be well fitted by a Gaussian shape  \cite{Namba:2015gja} 
\begin{equation}
N \left[ \xi_* ,\, x_* ,\, \delta \right] \approx N^c \left[ \xi_* ,\, \delta \right] \, {\rm exp} \left( - \frac{1}{2 \, \sigma^2 \left[ \xi_* ,\, \delta \right] } \, \ln^2 \left( \frac{x_*}{q^c \left[ \xi_* ,\, \delta \right] } \right) \right) \;. 
\label{N-fit}
\end{equation}
The coefficients $ N^c ,\; \sigma ,$ and $q^c$ control, respectively, the amplitude, the width, and the position of the bump. They can be evaluated 
 numerically, and in Appendix  \ref{app:A-bump} we provide their functional dependence on $\xi_*$ for the two choices $\delta = 0.2 ,\, 0.5$ considered in ref. \cite{Namba:2015gja}. Since $x_* = - k \, \tau_*$, we see that the position of the peak is at  $k_{\rm peak} = \frac{q^c}{-\tau_*}$. 
  The mode with $x_* =1$ is the mode that leaves the horizon when $\tau = \tau_*$, namely when $\xi$ is greatest. We expect that  modes with $x_* = {\rm O } \left( 1 \right)$ are the maximally amplified ones, and indeed we find $q^c \ga 1$ in the range of parameters we have studied (see Appendix \ref{app:A-bump}). 

In Figure \ref{fig:tdrho} we show the contribution to the energy density of three different modes of the gauge field (the result has been obtained by inserting  (\ref{mainvarysigma-ARI}) in eq. (\ref{endenpmode})). As in the previous figure, we note the decrease with time of the unphysical UV-divergent vacuum energy density, followed by the physical amplification, followed by the dilution due to the expansion of the universe. We note that, among the modes shown, the one with $x_* = 5$ is the one with greatest amplification. We verified that this is the case also among the modes that we do not show here, and that the amplification becomes progressively  smaller at values of $x_*$ greater or smaller than those shown here, in agreement with the Gaussian profile (\ref{N-fit}). We see from the figure that, for the maximally amplified $x_* = 5$ mode, the value $-k \, \tau = \xi_*$ provides a good position to separate between the unphysical vacuum energy density and the physical amplification. In our computation, we use the UV cut-off $- k \tau \vert_{\rm max} = \xi_*$ for all modes.

\begin{figure}[tbp]
\centering 
\includegraphics[width=0.5\textwidth,angle=0]{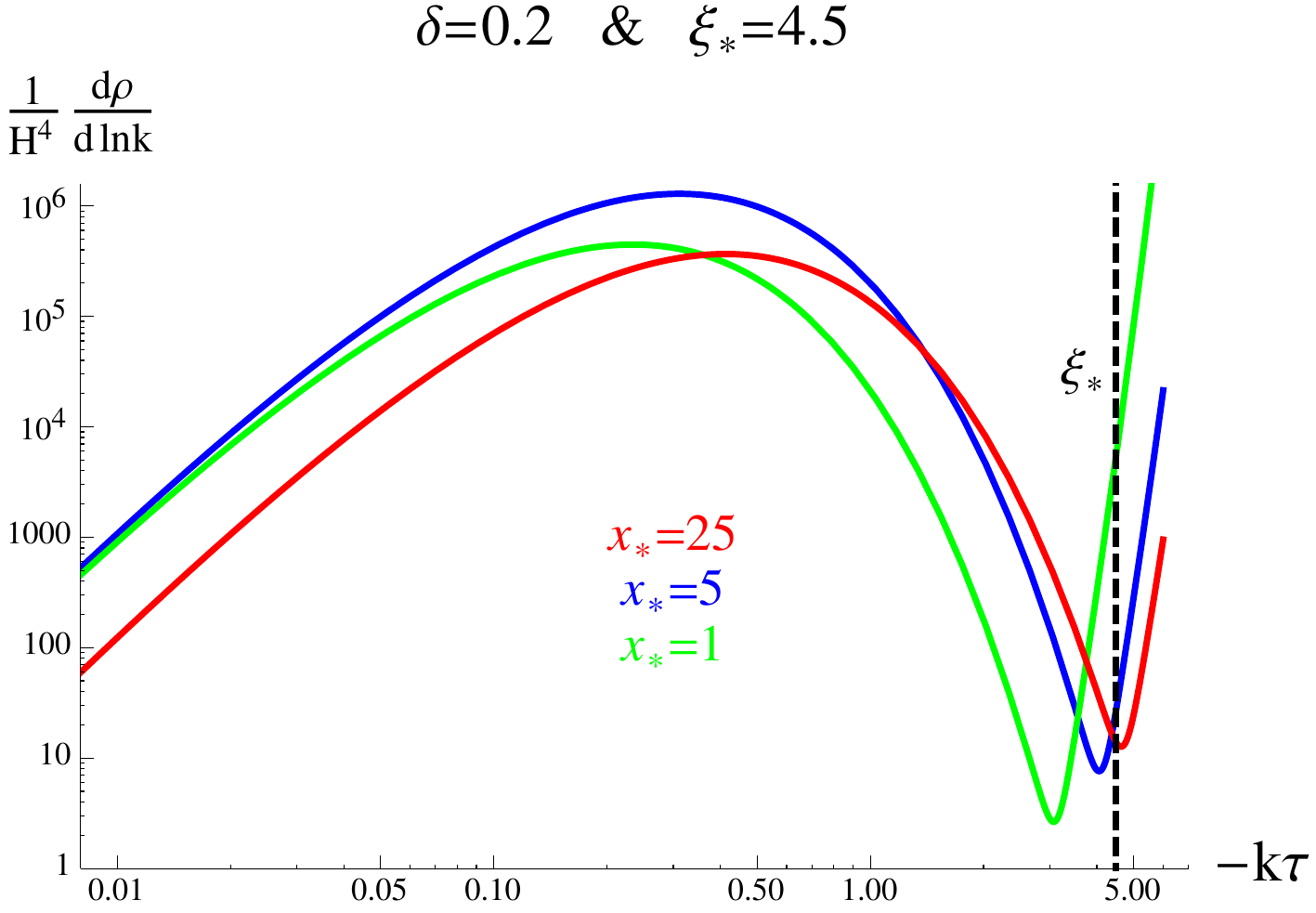}
\hfill
\caption{ 
Time evolution of the contribution to the gauge field physical energy density from modes with three different comoving momenta. The quantity $x_*$ is the ratio between the physical momentum of the mode and the Hubble rate at the time when $\sigma = \sigma_*$. 
}
\label{fig:tdrho}
\end{figure}

\section{Phenomenological signatures of the gauge field amplification}
\label{sec:phenomenology}

In this Section we briefly summarize the phenomenological signatures associated to the gauge field production discussed in the previous section. The main goal is to summarize which values of field amplification (controlled by the value of $\xi$) are needed  to produce those signatures. In the following sections, we  then study whether the required amount of field amplification is compatible with limits from backreaction and perturbativity. 

\subsection{Signatures for $X=\phi$}
\label{sec:Xphi-signatures}

We start with the model considered in Subsection \ref{sec:consxiAsol}, in which $X = \phi$ is the inflaton, whose speed is adiabatically evolving. 
The phenomenological signatures of this model studied in the literature are  primordial non-gaussianity  \cite{Barnaby:2010vf,Barnaby:2011vw,Ade:2015lrj}, growth of power-spectrum \cite{Meerburg:2012id,Ade:2015lrj}, primordial chiral gravity waves \cite{Sorbo:2011rz} at interferometer scales \cite{Cook:2011hg,Barnaby:2011qe,Crowder:2012ik,Domcke:2016bkh} and primordial black holes \cite{Linde:2012bt}. From these analysis, the value of $\xi$ required to obtained a visible signal is approximately 
\begin{eqnarray} 
\xi \left( N \simeq 60 \right) &\simeq& 2.5 \;\;\;\; {\rm from \; CMB \; measurement} \;, \nonumber\\ 
\xi \left( N \simeq 15 \right) &\simeq& 5 \;\;\;\; {\rm from \; GW \; at \; interferometers} \;, \nonumber\\ 
\xi \left( N \simeq 8 \right) &\simeq& 5 \;\;\;\; {\rm from \; primordial \; black \; holes } \;. 
\label{xi-phi-pheno}
\end{eqnarray} 
The value of $N$ in these expression is the number of e-folds before the end of inflation at which those limits apply. Specifically, these are the values assumed by $\xi$ when the mode leading to that specific signature left the horizon. The last two limits are obtained from refs.  \cite{Crowder:2012ik} and  \cite{Linde:2012bt}, respectively. Those works cast the limit in terms of the value assumed by $\xi$ at $N= 60$. We obtain the values written in (\ref{xi-phi-pheno}) by computing the evolution of $\xi \left( t \right)$ in the monomial inflaton potentials used in those works.

\subsection{Signatures for $X=\sigma$}
\label{sec:Xsigma-signatures}

Now, consider the model with $X = \sigma$ experiencing a momentary fast evolution studied in Subsection \ref{sec:bumpxiAsol}. This model was proposed in ref. \cite{Namba:2015gja} to source a potentially observable bump of GW at CMB scales, without significantly producing scalar perturbations, $P_{\zeta ,{\rm sourced}} \ll P_{\zeta,{\rm vacuum}}$. 

As seen from eq. (\ref{N-fit}), the spectrum of amplified gauge modes in this mechanism exhibits a peak at the scales that exited the horizon when $\sigma$ reaches its maximum speed. Correspondingly, the scalar perturbations $\zeta$, and the GW sourced by these gauge fields also present an analogous peak at these scales. The total power spectrum of scalar and tensor perturbations is a sum of the vacuum and the sourced contribution, where the latter  acquires the form  \cite{Namba:2015gja} 
\begin{eqnarray} 
P_{j,{\rm sourced}} \left( k \right) & = & \left[ \epsilon_\phi \, {\cal P}_\zeta^{(0)} \left( k \right) \right]^2 \, f_{2,j} \left( \frac{k}{k_*} ,\, \xi_* ,\, \delta \right) \;, \nonumber\\ 
f_{2,j} \left( \frac{k}{k_*} ,\, \xi_* ,\, \delta \right) & \simeq & f_{2,j}^c \left[ \xi_* ,\, \delta \right] \, {\rm exp} \left[ - \frac{1}{2 \, \sigma_{2,j}^2 \left[ \xi_* ,\, \delta \right] } \, {\rm ln}^2 \left( \frac{k}{k_* \, x_{2,j}^c \left[ \xi_* ,\, \delta \right] } \right) \right] \;, 
\label{f2}
\end{eqnarray} 
where $j = \left\{ \zeta ,\, + ,\, - \right\}$, with $\pm$ referring to the two GW helicities. In this expression,  $\epsilon_\phi$ is the slow roll inflaton parameter, and $P_\zeta^{(0)} \simeq 2.2 \cdot 10^{-9}$ \cite{Ade:2015lrj} is the amplitude of the scalar power spectrum,  
\begin{equation} 
P_\zeta \simeq P_\zeta^{(0)} \simeq \frac{H_{\rm inflation}^2}{8 \pi^2 \epsilon_\phi M_p^2} \simeq 2.2 \cdot 10^{-9} \;. 
\label{P-zeta}
\end{equation}

The three parameters $f_{2,+}^{\rm c} ,\, \sigma_{2,+} ,\,$ and $x_{2,+}^c$ control, respectively, the amplitude, the width, and the position of the peak of the sourced signal, and their functional dependence on $\xi_*$ is studied in  \cite{Namba:2015gja}. Analogous expressions apply for the sourced bispectra \cite{Namba:2015gja}.  
 
As shown in  \cite{Namba:2015gja}, this mechanisms admits a large region of parameters for which the vacuum scalar modes dominate over the sourced ones, leading to the tensor-to-scalar ratio 
\begin{equation}
r \left( k \right) \simeq r_{\rm vacuum} + \epsilon_\phi^2\, P_\zeta^{(0)} \, f_{2,+} \left( k \right) \;,  
\label{r-tot}
\end{equation}
where $r_{\rm vacuum} = 16 \, \epsilon_\phi$ is the ratio between the vacuum gravitational waves and scalar modes (we only included the sourced gravity waves of $+$ helicity in (\ref{r-tot}), as those of the other helicity are sourced in a negligible amount). 

Ref.~\cite{Namba:2015gja} studied the model for  $\delta = 0.2$ and $\delta = 0.5$. We also focus  our discussion on these two cases. Using the numerical fits of  \cite{Namba:2015gja},~\footnote{Specifically, we linearized in the $3 < \xi_* <6$ region the expression for $f_{2,+}$ given in Tables 1 and 2 of  ref.  \cite{Namba:2015gja}.}  we obtain the following result  
\begin{eqnarray}
&& \delta = 0.2 \;\;\Rightarrow\;\;  r_{\rm peak}^{1/2} \simeq 2.81 \cdot 10^{-7} \, \epsilon_\phi \, {\rm e}^{1.74 \, \pi \, \xi_*} \;,  \nonumber\\ 
&& \delta = 0.5 \;\;\Rightarrow\;\;  r_{\rm peak}^{1/2} \simeq 7.69 \cdot 10^{-7} \, \epsilon_\phi \, {\rm e}^{1.54 \, \pi \, \xi_*} \;, 
\label{tsrandepsphi}
\end{eqnarray} 
for the value of the tensor-to-scalar ratio at the peak of the sourced GW  signal, in the range of parameters for which this peak is above the vacuum GW signal  \cite{Namba:2015gja}, 
\begin{equation}
r_{\rm peak} \gg r_{\rm vacuum} \simeq 16 \, \epsilon_\phi \simeq \, \left( \frac{H_{\rm inflation}}{10^{-4} M_p} \right)^2 \;, 
\label{r-peak-par}
\end{equation} 
where in the last expression  eq. (\ref{P-zeta}) has been used. 

Combining eqs. (\ref{tsrandepsphi}) and (\ref{r-peak-par}), we obtain 
\begin{equation}
r_{\rm peak} \simeq  \left\{ \begin{array}{l} 
0.006 \left( \frac{H_{\rm inflation}}{10^{-6} \, M_p} \right)^4  \, {\rm e}^{3.48 \, \pi \, \left( \xi_* - 4.5 \right)} \;\;\;,\;\;\; \delta = 0.2 \;, \\ 
0.020 \left( \frac{H_{\rm inflation}}{10^{-6} \, M_p} \right)^4  \, {\rm e}^{3.08 \, \pi \, \left( \xi_* - 5 \right)} \;\;\;,\;\;\; \delta = 0.5 \;. 
\end{array} \right. 
\label{r-peak}
\end{equation}  
The result (\ref{r-peak})  shows that, provided $\xi_*$ is sufficiently large, a visible sourced GW at CMB scale can be obtained at arbitrary small scale of inflation. In the next Sections we  study what choice of parameters allows to achieve this and respect limits from backreaction and perturbativity. 

Such a sourced GW signal may be distinguished from the vacuum one by (i) its spectral dependence, (ii) its non-gaussian statistics, and (iii) its violation of parity. These aspects were already considered and studied in \cite{Namba:2015gja}. Here, we point out~\footnote{We thank Sarah Shandera for discussions on this point.} that the running of the tensor spectral tilt is likely the best parameter to quantify the spectral difference between the 
standard vacuum and the sourced GW signal. The sum of the vacuum mode and of (\ref{f2}) gives the total GW spectrum 
\begin{equation}
P_{\rm GW} \left( k \right) = 16 \, \epsilon_{\phi,0} \, P_\zeta^{(0)} \left( k_0 \right) \, \left( \frac{k}{k_0} \right)^{n_{t,0}} + 
\left[  \epsilon_{\phi,0} \, P_\zeta^{(0)} \left( k_0 \right) \, \left( \frac{k}{k_0} \right)^{n_{t,0}} \right]^2 \, f_{2,+}^c \, {\rm exp} \left[ - \frac{1}{2 \sigma_{2,+}^2} \, \ln^2 \left( \frac{k}{k_c} \right) \right] \;, 
\end{equation}
where $0$ indicates a quantity evaluated at the pivot scale $k_0$, where $n_t = - 2 \epsilon_\phi$ is the vacuum tensor spectral tilt, and where we redefined 
$k_* \, x_{2,+}^c \equiv k_c$ (as only this combination is observable) to indicate the location of the peak of the sourced signal. We disregard quantities that are second order in slow roll, and therefore, we disregard the running of the vacuum spectral tilt $n_t$. If we denote by 
\begin{equation}
{\cal F} \equiv \frac{P_{\rm GW,s}}{ P_{\rm GW,v}} \;, 
\end{equation} 
the fraction of the sourced divided by the vacuum tensor signal, where the total tensor spectral tilt $n_t$ and its running $\alpha_t$ are given by 
\begin{eqnarray} 
n_t \left( k \right) & \equiv & \frac{d \ln P_{\rm GW} \left( k \right)}{d \ln k} = n_{t,0} \, \frac{1+2\,{\cal F}}{1 + {\cal F}} - \frac{\ln \frac{k}{k_c}}{ \sigma_{2,+}^2} \, \frac{{\cal F}}{1+ {\cal F}} \;, \nonumber\\ 
\alpha_t \left( k \right) & \equiv & \frac{d n_t \left( k \right)}{d \ln k} = \left( n_{t,0} -  \frac{\ln \frac{k}{k_c}}{ \sigma_{2,+}^2} \right)^2 \, \frac{{\cal F}}{\left( 1 + {\cal F} \right)^2} - \frac{{\cal F}}{\sigma_{2,+}^2} \, \frac{1}{1+ {\cal F}} \;. 
\end{eqnarray}
We then have the two limits 
\begin{eqnarray}
&& {\rm vacuum \; GW \; dominate \;}: \; \left( {\cal F} \ll 1 \right)  \;\;\Rightarrow\;\;  n_t \simeq n_{t,0} \;\;,\;\; \alpha_t \simeq 0 \;, \nonumber\\ 
&& {\rm sourced \; GW \; dominates \;}: \; \left( {\cal F} \gg 1  \right)  \;\;\Rightarrow\;\; n_t \simeq 2 \, n_{t,0} -  \frac{\ln \frac{k}{k_c}}{ \sigma_{2,+}^2} \;\;,\;\; 
\alpha_t \simeq - \frac{1}{\sigma_{2,+}^2} \;.   
\end{eqnarray} 
Namely the presence of a peak of the sourced signal (as opposed to the nearly scale invariant vacuum signal) gives rise to a nonvanishing running of the spectral tilt, which is negative, and of magnitude inversely proportional to the square of the width of the peak. This running can be observed if the B mode of the CMB is measured for a sufficiently large window of multipoles, see Figure 6 of~\cite{Namba:2015gja}, for some specific examples. 
In these examples, the width $\sigma_{2,+}$ is of order one, see Tables 1 and 2 of~\cite{Namba:2015gja}.

\section{Backreaction}%
\label{sec:back}

In this section we discuss the effects of particle production on the background motion of $X$. We divide the discussion into two subsections, where we separately discuss the  $X = \phi$ (with adiabatic $\dot{\phi}$ evolution), and $X = \sigma$ (with a momentary speed up of this field) cases.

\subsection{Backreaction for $X=\phi$}
\label{sec:back-phi}

The gauge field enters in the evolution equations for the inflaton field $\phi$ and the scale factor as follows 
\begin{eqnarray}
&& \ddot{\phi} + 3 H \dot{\phi} + \partial_\phi V = \frac{\alpha}{f} \, \left\langle \vec{E} \cdot \vec{B} \right\rangle \;, \nonumber\\ 
&& 3 H^2 = \frac{1}{M_p^2} \left[ \frac{1}{2} \dot{\phi}^2 + V + \frac{1}{2} \left\langle \vec{E}^2 + \vec{B}^2 \right\rangle \right] \;, 
\end{eqnarray} 

Using (\ref{A-simple}) one finds \cite{Anber:2009ua}, in the $\xi \gg 1$ regime (namely, in the regime in which the gauge field amplification takes place), 
\begin{equation}
\left\langle \vec{E} \cdot \vec{B} \right\rangle \simeq - 2.4 \cdot 10^{-4} \, \frac{H^4}{\xi^4} \, {\rm e}^{2 \pi \xi} \;\;,\;\; 
\left\langle \frac{\vec{E}^2 + \vec{B}^2}{2} \right\rangle \simeq  1.4 \cdot 10^{-4} \, \frac{H^4}{\xi^3} \, {\rm e}^{2 \pi \xi} \;. 
\end{equation} 
Therefore \cite{Barnaby:2010vf,Barnaby:2011vw}
\begin{eqnarray}
&& \xi^{-3/2} \, {\rm e}^{\pi \xi} \ll 79 \frac{\vert \dot{\phi} \vert}{H^2} \;\Rightarrow\; {\rm negligible \; backreaction \; on \; } \phi {\rm \; eq.} \;, \nonumber\\ 
&& \xi^{-3/2} \, {\rm e}^{\pi \xi} \ll 146 \frac{M_p}{H} \;\Rightarrow\; {\rm negligible \; backreaction \; on \; Friedmann \; eq.} \;. 
\end{eqnarray} 
The first condition is more stringent (it is easier to modify the evolution of the inflaton field, that is moving slowly, than that of the scale factor), and using the normalization (\ref{P-zeta}), it leads to the bound 
\begin{equation}
{\rm Negligible \; backreaction \; on \; } \phi \;:\;\; \xi \ll 4.7 \;. 
\label{back-phi}
\end{equation} 
We discuss this condition in Section  \ref{sec:significance-phi}.

\subsection{Backreaction  for $X=\sigma$}
\label{sec:back-sig}

In eq. (\ref{r-peak}) we showed that, provided $\xi_*$ is sufficiently large, a visible sourced GW at CMB scale can be obtained at arbitrary small scale of inflation.  Here we study how large this field amplification can be, without violating bounds from backreaction on the dynamics of $\sigma$ and $\phi$. 

We first impose that $\sigma$  provides a negligible contribution to the energy density of the universe. 
To quantify this condition, in agreement with the slow roll relation $\dot{\phi} \simeq \sqrt{2 \, \epsilon_\phi} H \, M_p$, we define 
\begin{equation}
\epsilon_\sigma \equiv \frac{\dot{\sigma}^2}{2 \, H^2 \, M_p^2} \;. 
\label{eps-sig} 
\end{equation}
(as $\sigma$ is slowly rolling, we have $\epsilon_\sigma \simeq \frac{M_p^2}{2} \, \left( \frac{\partial_\sigma V}{V} \right)^2$). We denote by 
$\epsilon_{\sigma,*} = \frac{\delta^2}{2} \, \frac{f^2}{M_p^2}$ the maximum value acquired by this quantity, when $\sigma$ reaches its maximum speed $\dot{\sigma}_*$. Using eqs. (\ref{delta}) and (\ref{dotsigma}), we see that 
\begin{equation}
V_{\rm max} \left( \sigma \right) = \Lambda^4 = 3 H^2 M_p^2 \times \frac{4 \epsilon_{\sigma,*}}{\delta} \;\;,\;\; 
\left( \frac{\dot{\sigma}^2}{2} \right)_{\rm max} =  \frac{\dot{\sigma}_*^2}{2} = 3 H^2 M_p^2 \times \frac{\epsilon_{\sigma,*}}{3} \;. 
\end{equation} 
We are interested in $\delta \la 1$, so the potential energy dominates over the kinetic one, and we can write the condition
\begin{equation}
\rho_\sigma \ll 3 H^2 M_p^2 \;\;\; \Rightarrow \;\;\; \epsilon_{\sigma,*} \ll \frac{\delta}{4} \;. 
\label{back-sig-phi}
\end{equation} 

Secondly, we impose that the gauge field amplification does not significantly alter the motion of  $\sigma$.  The sourced gravity waves are proportional to the energy density $\rho_A$ of the sourcing gauge fields. The physical energy density in the gauge fields reaches its maximum when $\xi \la \xi_*$. As we show in eq. (\ref{rhoAmax}), the maximum value acquired by $\rho_A$ is 
\begin{equation}
\frac{\rho_{A,\rm max}}{\epsilon_\phi \, \rho_\phi} \sim \left\{ \begin{array}{l} 
2.75 \cdot 10^{-12} \, {\rm e}^{1.74 \, \pi \,\xi_*} \;\;\;,\;\;\; \delta = 0.2 \;, \\ 
8.86 \cdot 10^{-12} \, {\rm e}^{1.52 \, \pi \,\xi_*} \;\;\;,\;\;\; \delta = 0.5 \;. \end{array}  \right. 
\label{rhoAmaxtex}
\end{equation}  
We find that the maximum of $\rho_{A,{\rm max}}$ is achieved for $\tau \simeq \tau_* / 20$ in the $\delta = 0.2$ case, and for  $\tau \simeq \tau_* / 10$ in the $\delta = 0.5$ case. 

Both (\ref{tsrandepsphi}) and (\ref{rhoAmaxtex}) are the results of numerical fits. The two expressions show nearly the same  $\xi_*$ dependence.  This was expected since $\rho_A$ is the dominant GW source. For both $\delta$ values, we find 
\begin{equation}
\frac{\rho_{A,{\rm max}}}{\rho_\phi} \simeq 10^{-5} \, r_{\rm peak}^{1/2} \;. 
\label{rhoA-rhophi-1}
\end{equation} 

The gauge field is amplified at the expense of the kinetic energy of $\sigma$, so we need to impose that $\rho_{A,{\rm max}}$ (which  is reached short after $t_*$, see Figure \ref{fig:rho-varyA}) is smaller than the kinetic energy of $\sigma$ at this moment,~\footnote{We verified that, once this condition is verified, the right hand side of (\ref{eom-sigma}) can indeed be neglected.}
\begin{equation}
\rho_{A,{\rm max}} \ll \frac{\dot{\sigma}_*^2}{2} = \epsilon_{\sigma,*} \, H^2 \, M_p^2 \simeq  \epsilon_{\sigma,*} \, \frac{\rho_\phi}{3} \;. 
\label{rhoA-rhophi-2}
\end{equation}

Combining eqs. (\ref{rhoA-rhophi-1}) and (\ref{rhoA-rhophi-2}) we arrive to the limit 
\begin{equation}
\epsilon_{\sigma,*} \gg 3 \cdot 10^{-5} \, r_{\rm peak}^{1/2} \;. 
\end{equation}
Finally, this result can be combined with (\ref{tsrandepsphi}) to give 
\begin{eqnarray} 
&& \delta = 0.2 \;\;:\;\;\;\; \epsilon_{\sigma,*} \gg \epsilon_\phi \,  {\rm e}^{5.47 \, (\xi_* - 4.67)} \;, \nonumber\\ 
&& \delta = 0.5 \;\;:\;\;\;\; \epsilon_{\sigma,*} \gg \epsilon_\phi \,   {\rm e}^{4.84 \, (\xi_* - 5.06)}  \;.  
\label{back-sig}
\end{eqnarray} 
We discuss these conditions in Section \ref{sec:significance-sig}.

\section{Perturbativity}
\label{sec:pertlimits}

We now discuss the two criteria for perturbativity considered in  \cite{Ferreira:2015omg}, which we want to evaluate in the present work for the two cases  $X = \phi$ (with adiabatic $\dot{\phi}$ evolution), and $X = \sigma$ (with a momentary speed up of this field). 

The first criterion computed in   \cite{Ferreira:2015omg} is from the renormalization of the gauge field wave function: 
\begin{equation}
{\cal R}_A \equiv \left\vert \frac{\delta^{(1)} \langle {\hat A} \, {\hat A} \rangle'}{\langle {\hat A} \, {\hat A} \rangle'} \right\vert \ll 1 \;, 
\label{RA-def}
\end{equation}
(this ratio appears in eq. (15) of   \cite{Ferreira:2015omg}) where the numerator is the lowest order one loop contribution to the propagator and the denominator is the tree-level propagator.  The expectation values are taken in momentum space, and prime denotes the expectation value without the corresponding $\delta-$function.  The one (and higher) loop contributions are defined with respect to the interaction hamiltonian obtained from 
\begin{eqnarray}
{\cal L} &=& \left[ - \frac{1}{2} \left( \partial X \right)^2 - V \left( X \right) - \frac{1}{4} F^2 -  \frac{\alpha}{4\, f}  \;   X_{\rm background}  \, {\hat F} {\tilde {\hat F}} \right] + \left[  -  \frac{\alpha}{4\, f}  \;  \delta {\hat X} \, {\hat F} {\tilde {\hat F}} \right] \nonumber\\ 
&\equiv& {\cal L}_{\rm unperturbed} + {\cal L}_{\rm interaction} \;. 
\label{Hintconspa}
\end{eqnarray}
We decompose ${\hat X} = X_{\rm background} \left( t \right) + {\hat X}^{(0)} \left( t ,\, \vec{x} \right) +  {\hat X}^{(1)} \left( t ,\, \vec{x} \right) + \dots$ and ${\hat A} = {\hat A}_+^{(0)} \left( t ,\, \vec{x} \right) + {\hat A}_+^{(1)} \left( t ,\, \vec{x} \right) + \dots$, where the suffix indicates the order of the perturbations in the interaction defined in (\ref{Hintconspa}). 

The last term in ${\cal L}_{\rm unperturbed}$ encodes the backreaction of the produced gauge fields on the background dynamics. By definition, in the regime of small backreaction we can disregard the effects of this term on the background geometry, and so this term has the only effect of modifying the ``unperturbed'' gauge mode  ${\hat A}_+^{(0)}$ from the vacuum one to the one studied  in Section \ref{sec:consxiAsol}).~\footnote{This motivates the choice of ``unperturbed'' vs. interaction term made in ({Hintconspa}). Had we chosen to include the full  $-\frac{\alpha}{4 f} X F {\tilde F}$ interaction in ${\cal L}_{\rm int}$, then ${\hat A}_+^{(0)}$ would be the standard vacuum mode rather than the solution discussed in Section \ref{sec:consxiAsol}), and therefore the ratio (\ref{RA-def}) would not corresponds to the criterion studied in (\ref{RA-def}). Therefore, perturbativity as discussed in  \cite{Ferreira:2015omg} means perturbativity with respect to this splitting.}  With this understanding, and only in this limit, the expansion in (\ref{Hintconspa}) becomes an expansion in $\frac{\alpha}{f}$, or better, in the parameter $\xi = \frac{\alpha}{f} \, \frac{\dot{ X}_{\rm background} }{2  H}$ that controls the field amplification. In the case of strong backreaction, we cannot longer assume that the perturbations $ {\hat X}^{(0)} \left( t ,\, \vec{x} \right) $ and $ {\hat A}_+^{(0)} $ are a good approximation of the full perturbations of the model, simply because the last term in (\ref{Hintconspa}) breaks covariance (namely, both $ X_{\rm background} $ and $ {\hat X}^{(0)} $ are determined by ${\cal L}_{\rm unperturbed}$, but $F {\tilde F}$
only interacts with the former). Therefore, our perturbativity set-up is guaranteed to fail in the strong backreaction regime, and this is one of the reasons why backreaction has been studied in the previous Section. We deem our analysis meaningful as long as backreaction is negligible, or only marginally relevant (we expect that our  estimate for the range of $\xi$ under which perturbativity is under control will remain approximately valid). However, we cannot use it in the case of very strong backreaction, as the one of \cite{Anber:2009ua} (in that work, scalar perturbations have been studied with an equation   which is not a perturbative deformation of the one used here; our criterion cannot be used to determine whether the result of  \cite{Anber:2009ua} is out of perturbative control).

The one loop correction diagram can be  computed via the in-in formalism as
\begin{eqnarray}
\delta^{(1)} \langle AA \rangle  \simeq  - \int^\tau d \tau_1  \int^{\tau_1} d \tau_2 \, \left\langle \left[ \left[  {\hat A}_+^{(0)} \left( \vec{k}_1 ,\, \tau \right)  {\hat A}_+^{(0)} \left( \vec{k}_2 ,\, \tau \right) ,\, {\hat H}_{\rm int}^{(0)} \left( \tau_1 \right) \right] ,\, {\hat H}_{\rm int}^{(0)} \left( \tau_2 \right) \right] \right\rangle \;, 
\label{AA-inin}
\end{eqnarray}
To simplify the notation, in the following we omit the suffix $0$ from the zeroth-order vector mode functions. 

The second criterion studied in \cite{Ferreira:2015omg} is that the interaction (\ref{Hintconspa})  does not drive the amplitude of $\delta X$ out of the non-linear regime, 
\begin{equation}
{\cal R}_X \equiv  \frac{ \sqrt{ \langle \delta {\hat X}^{(1)} \left( \vec{x} ,\, \tau \right) \, \delta {\hat X}^{(1)} \left( \vec{x} ,\, \tau \right) \rangle }}{ f_{\rm period} } \ll 1 \;. 
\label{RX-def} 
\end{equation}
(This ratio encodes the discussion at the end of page 7 of  \cite{Ferreira:2015omg}.)  Contrary to (\ref{RA-def}), the expectation value appearing in this expression is in real space. Fourier transforming the field $X$ as we did for the gauge field (see eq. (\ref{A-deco})), we rewrite it as 
\begin{eqnarray} 
\left\langle \delta {\hat X}^{(1)}  \left( \vec{x} ,\, \tau \right) \,  \delta {\hat X}^{(1)}  \left( \vec{x} ,\, \tau \right) \right\rangle & = & \int \frac{d^3 k_1 \, d^3 k_2}{\left( 2 \pi \right)^3 } \, {\rm e}^{i \vec{x} \cdot \left( \vec{k}_1 + \vec{k}_2 \right)} \, \left\langle \delta {\hat X}^{(1)}  \left( \vec{k}_1 ,\, \tau \right) \,  \delta {\hat X}^{(1)}  \left( \vec{k}_2 ,\, \tau \right) \right\rangle \nonumber\\ 
& = & \int d \ln  k \; P_{\delta X}^{(1)} \left( k ,\, \tau \right) \;. 
\label{dXdX}
\end{eqnarray} 
To compute the correction to the power spectrum $P_{\delta X}^{(1)} \left( k \right) \equiv \frac{k^3}{2 \pi^2} \, \left\langle \delta X^{(1)} \left( \vec{k} \right)  \delta X^{(1)} \left( - \vec{k} \right) \right\rangle' $, for the inflaton case $X = \phi$ we employ results already given in the literature~\cite{Barnaby:2010vf,Barnaby:2011vw,Linde:2012bt}. For $X = \sigma$, we instead  make use of the in-in formalism 
\begin{eqnarray} 
&& \!\!\!\!\!\!\!\! \!\!\!\!\!\!\!\! \!\!\!\!\!\!\!\! 
\left\langle  \delta {\hat X}^{(1)} \left( \vec{k}_1 ,\, \tau \right)   \delta {\hat X}^{(1)}  \left( \vec{k}_2 ,\, \tau \right) \right\rangle \simeq - \int^\tau d \tau_1  \int^{\tau_1} d \tau_2 \, \left\langle \left[ \left[   \delta {\hat X}^{(0)} \left( \vec{k}_1 ,\, \tau \right)   \delta {\hat X}^{(0)} \left( \vec{k}_2 ,\, \tau \right) ,\, {\hat H}_{\rm int} \left( \tau_1 \right) \right] ,\, {\hat H}_{\rm int} \left( \tau_2 \right) \right] \right\rangle \;. \nonumber\\ 
\label{XX-inin}
\end{eqnarray} 

The scale $f_{\rm period}$ in eq. (\ref{RX-def})  is the periodicity of the potential of $X$. This has an immediate identification in the case $X = \sigma$, where the potential is (\ref{V-sigma}), and therefore $f_{\rm period} = f$. As pointed out in  \cite{Ferreira:2015omg}, if the interaction (\ref{Hintconspa}) leads to typical field displacements comparable to (or even greater than) $f$, then all operators obtained by Taylor expanding (\ref{V-sigma}) become strong, driving the system out of the perturbative regime. For the case $X=\phi$, the identification of $f_{\rm period}$ depends on the specific model under consideration. For instance, the potential of aligned natural inflation \cite{Kim:2004rp} has four axions scales $f_i \ll M_p$ ($i=1,\dots,4$), but two different terms for two axions are arranged so to produce a large periodicity for one linear combination of the two axions. Therefore, $f_{\rm period} \gg f_i$. The coupling between the axions and gauge field is controlled by the original axion scales in the model, leading to an interaction lagrangian (\ref{Hintconspa}) in terms of the original scales, $f = {\rm O } \left( f_i \right)$ (see section V of \cite{Peloso:2015dsa} for a detailed computation). However, a ${\rm O } \left( f_i \right)$ field displacement would have a very suppressed effect in the potential along the  $\phi$ direction, which has the periodicity $f_{\rm period} \gg f_i$. Therefore, if $X = \phi$ is the inflaton field, the best indicator for the periodicity of the inflation potential is the classical value assumed by the inflaton $\phi_{\rm cl} \la {\rm O } \left( f_{\rm period} \right)$ (under the assumption that the minimum of the inflaton potential is at $\phi = 0$). Therefore, we write the two different conditions 
\begin{equation}
{\cal R}_\phi \equiv  \frac{\sqrt{ \langle \delta {\hat \phi}^{(1)}  \, \delta {\hat \phi}^{(1)}  \rangle }}{ \phi_{\rm cl} } \ll 1 \;\;\;\;,\;\;\;\; 
{\cal R}_\sigma \equiv  \frac{ \sqrt{ \langle \delta {\hat \sigma}^{(1)} \, \delta {\hat \sigma }^{(1)} \rangle}}{ f } \ll 1 \;. 
\label{RX} 
\end{equation}

\begin{figure}[ht!]
\centerline{
\includegraphics[width=0.7\textwidth,angle=0]{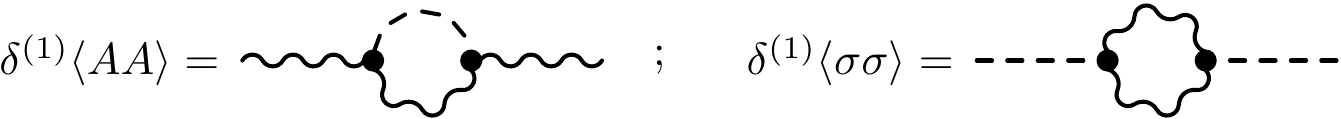}
}
\caption{Diagrammatic representation of the one loop terms (\ref{AA-inin}) and  (\ref{XX-inin}).  Solid lines denote $\delta X$ modes, while wiggly lines denote vector field $A_+$ modes. The small bullets denote the interaction (\ref{Hintconspa}). 
}
\label{fig:loop}
\end{figure}

Figure \ref{fig:loop} provides the diagrammatic expression of the one loop terms  (\ref{AA-inin}) and  (\ref{XX-inin}). 

Before concluding this section, we should emphasize that all the perturbativity constraints we have considered so far concern the amplitude of the classical modes that have been generated by amplification of vacuum fluctuations. However, we should also require perturbativity  in the more usual sense of sensitivity of the theory to UV quantum modes in a time-independent background. The presence of the coupling in the Lagrangian (2.1) tell us that this should be treated as an effective field theory below an energy scale $\sim 4\pi f/\alpha$, and we generally expect new states to exist at or below such energy scale. Since, as we have extensively discussed above, the typical (physical) energies of the system we are considering are of the order of $H\,\xi_*$, consistency of our analysis should require 
\begin{equation}
\xi_*\,H\ll \frac{4\pi f}{\alpha}\,,
\end{equation}
which is easily satisfied. This is particularly true in the case $X=\sigma$ considered in  \cite{Namba:2015gja}, where it is assumed that the scale of inflation is low, so that the vacuum GW are unobservable. 

\subsection{Perturbativity  for $X=\phi$}
\label{sec:pert-xi-phi}

Let us discuss the conditions ${\cal R}_A \ll 1$ and ${\cal R}_\phi \ll 1$ in the case in which $X$ is the inflaton field $\phi$. We start from the second condition, and we show that it is satisfied in all cases of interest. Using eq. (\ref{dXdX}), the invariant curvature (in spatially flat gauge) 
$\zeta = - \frac{H}{\dot{\phi}} \, \delta \phi$, and the slow roll relation $\dot{\phi} \simeq - \sqrt{2 \epsilon} \, H \, M_p$, the condition (\ref{RX})  rewrites 
\begin{equation}
{\cal R}_\phi \simeq \sqrt{2 \epsilon} \, \times \left(  \frac{M_p}{\phi_{\rm cl}} \right) \times \sqrt{ \int d \ln k \, P_\zeta^{(1)} } \ll 1 \;,   
\label{R-phi}
\end{equation} 
where we recall that $ P_\zeta^{(1)} $ is the power spectrum of the sourced scalar mode $\zeta$. This condition is satisfied since all the factors on the right hand side of eq.~(\ref{R-phi}) are smaller than one. Let us discuss these factors one by one. 

The first factor is $\ll 1$  due to slow roll. At CMB scales, the existing limit $\xi \la 2.5$ (see eq. (\ref{xi-phi-pheno})), implies negligible backreaction of the gauge fields on the inflaton background dynamics. For the GW signatures at interferometers, and for production of primordial black holes, we are in a regime in which the gauge field production slows down the motion of the inflaton (see for instance Figure 2 of \cite{Barnaby:2011qe}). However, $\dot{\phi} = {\rm O } \left(  \sqrt{2 \epsilon} \, H \, M_p \right)$ is still valid, and the fact that the particle production slows down the inflaton effectively decreases the value of $\epsilon$ in eq. (\ref{R-phi}), further decreasing this factor. 
 
The second factor is model dependent, but we recall that models of axion inflation realize a super-Planckian inflaton excursion. We already discussed this after eq. (\ref{XX-inin}). Therefore $\phi_{\rm cl} \ga M_p$. 

Finally, the third factor is $\ll 1$ because $P_\zeta^{(1)} \ll 1$. This is for sure true at CMB scales, where  $\xi \la 2.5$ forces  $P_\zeta^{(1)} \ll {\cal P}_\zeta^{(0)} = {\rm O } \left( 10^{-9} \right)$. The sourced power is significantly greater at progressively smaller scales. However, also when the primordial black hole limit is saturated, one finds  $P_\zeta^{(1)} = {\rm O } \left( 10^{-3} \right)$, see Figure 5 of \cite{Linde:2012bt}. 

Let us now discuss the condition ${\cal R}_A \ll 1$. We evaluated this ratio using eqs. (\ref{AAfinwithpre}) and (\ref{pref-AA}). The ratio is function of both comoving momentum $k$ of the mode in the propagator, and of time. Due to approximate scale invariance of the vacuum signal, the dependence on $k$ is extremely weak, so let us first focus our discussion on CMB scales ($k_1 = k_{\rm p}$ in eq.  (\ref{pref-AA})). For any fixed mode, the ratio ${\cal R}_A$ assumes a different value at different times. 

As discussed in Subsection \ref{sec:consxiAsol}, the  level of amplification of any given mode due to the $\phi F {\tilde F}$ interaction is a function of time. In particular, the fractional energy density $\frac{d \rho_{\rm gauge}}{d \ln k}$ of any given mode reaches a maximum value when the mode has a size comparable to the horizon, and it is then diluted away by the expansion of the universe. Most of the contribution of any given mode to a cosmological observable takes place when  $\frac{d \rho_{\rm gauge}}{d \ln k}$ is close to its peak value. This defines the time interval for which we need to make sure that  ${\cal R}_A \ll 1$ for that given mode. 

\begin{figure}[ht!]
\centerline{
\includegraphics[width=0.5\textwidth,angle=0]{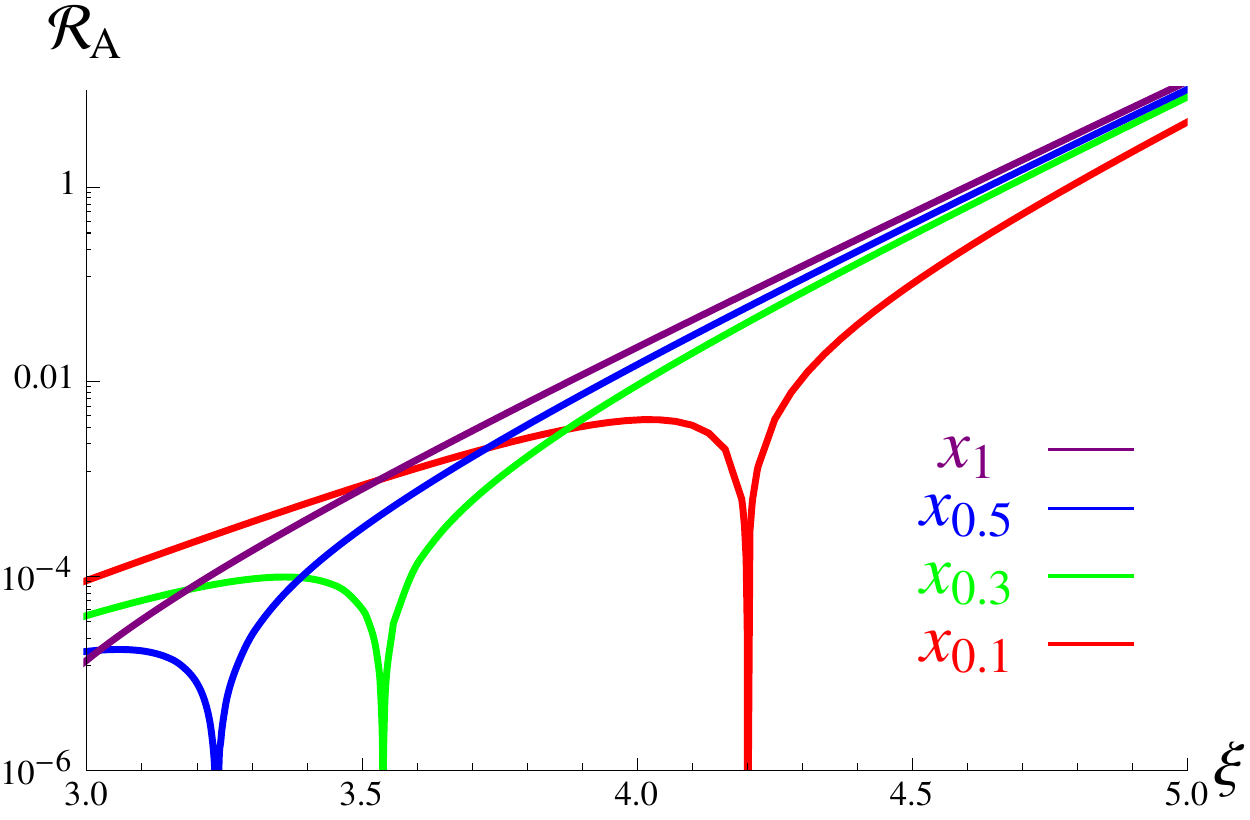}
}
\caption{Ratio ${\cal R}_A$ for the $X = \phi$ case, as a function of  $\xi$. The ratio is evaluated for a fixed mode (fixed $k$)  of the size of the Planck pivot scale (due to approximate scale invariance, nearly the same bound is obtained at smaller scales). The different curves shown correspond to  different values of the rescaled time $x \equiv - k \tau$ at which   ${\cal R}_A$ is evaluated. 
For instance $x_{0.1}$ indicates that ${\cal R}_A$ is evaluated when the energy density in that mode is $10\%$ of the peaked value that it had previously assumed  (as shown in Figure \ref{fig:rhoti}, the energy density in one given mode reaches a peak value, and it then decreases).
}
\label{fig:R-constant}
\end{figure}

In Figure \ref{fig:R-constant} we show the value of ${\cal R}_A$ for one given mode with wave number equal the Planck pivot scale ($k =  k_{\rm p}$) as a function of $\xi$. The value of $\xi$ should be understood as the value assumed by this quantity when the mode of our interest left the horizon. In general, a greater value of $\xi$ results in a greater gauge field amplification, and therefore in a greater value of  ${\cal R}_A$.~\footnote{We disregard the narrow spikes seen in the Figure, which take place when $\frac{\delta^{(1)} \langle {\hat A} \, {\hat A} \rangle'}{\langle {\hat A} \, {\hat A} \rangle'} $ changes sign.} Different lines in the figure show the value assume by ${\cal R}_A$ for $k = k_{\rm p}$ and for that specific value of $\xi$, but at different times. The values of time shown are chosen by evaluating the value of  $\frac{d \rho_{\rm gauge}}{d \ln k}$. For instance $x_{0.1}$ denotes the value of time ($x \equiv - k \, \tau$) when the energy density in that mode has been diluted to $10\%$ of the value it had at its peak (and analogously for the other values shown in the figure; therefore $x_{1} > x_{0.5} > x_{0.3} > x_{0.1}$, corresponding to $\tau_1 < \tau_{0.5} < \tau_{0.3} < \tau_{0.1}$). 

The strongest limit is obtained for $x_1$, when the energy density in that mode is maximum. In this case, a numerical fit of the curve shown in the figure gives 
\begin{equation} 
{\cal R}_A \simeq {\rm e}^{2.01 \, \pi \left( \xi - 4.60 \right)} \;. 
\label{RA-phi-res}
\end{equation} 
The fit well reproduces the $\propto {\rm e}^{2 \pi \xi}$ scaling expected from analytic considerations (the numerator of eq. (\ref{AAfinwithpre}) has two extra powers of $A_R$ than the denominator). 

We recall that the limit in  Figure \ref{fig:R-constant} assumes a mode with  wave number equal the Planck pivot scale. For signatures at smaller scales ($k > k_{\rm p}$), eq.  (\ref{pref-AA}) presents the extra factor  $\left( \frac{k_1}{k_{\rm p}} \right)^{n_s-1} \simeq {\rm e}^{-\left( 1- n_s  \right) \left( 60 - N_k  \right)}$, where $N_k$ is the number of e-folds before the end of inflation when the mode of wavenumber $k$ exited the horizon, while we have assumed that $k_{\rm p}$ exited the horizon $60$ e-folds before the end of inflation.  Taking this into account, the perturbativity limit can be cast in the form 
\begin{equation}
 {\rm e}^{2.01 \, \pi \left( \xi_k - 4.60 \right)} \, {\rm e}^{-\left( 1- n_s  \right) \left( 60 - N_k  \right)} \ll 1 \;. 
\label{pert-phi}
\end{equation} 
This condition is a function of wavenumber, since both $\xi_k$ and $N_k$ refer to the value assumed by $\xi$ and by $N$ when the mode of wave number $k$ exited the horizon during inflation. We discuss this condition in Section \ref{sec:significance-phi}. 

\subsection{Perturbativity  for $X=\sigma$}
\label{sec:pert-xi-sigma}

In the case in which $X = \sigma$ is a pseudo-scalar different from the inflaton, that has a nonvanishing speed only for a few e-folds during inflation, we find the two perturbativity conditions 
\begin{equation}
 {\cal R}_A \equiv {\rm R}_A \left[ x_* ,\,  \xi_* ,\, \delta ,\, \frac{\tau}{\tau_*} \right] \, \frac{\epsilon_{\phi}}{\epsilon_{\sigma_*}}  \ll 1  \quad \quad \quad ,\quad \quad \quad   {\cal R}_{\sigma} \equiv {\rm R}_{\sigma}  \left[ \xi_* ,\, \delta ,\, \frac{\tau}{\tau_*} \right] \frac{\epsilon_{\phi}}{\epsilon_{\sigma_*}}  \ll 1 \;, 
\label{conditions-sigma}
\end{equation}
where $R_A$ and $R_\sigma$ are given in eqs. (\ref{Xsig-RA}) and (\ref{Xsig-Rsig}), respectively. 

The ratio ${\cal R}_A$ is evaluated mode by mode, and it therefore depends on the comoving momentum $k$ of the mode through $x_* = - k \tau_*$. We recall that $\tau_*$ is the conformal time at which $\dot{\sigma}$ is maximum, and $-\frac{1}{\tau_*}$ is the comoving momentum of the mode that left the horizon at this time. For each mode, the quantity $R_A$ then depends on the time at which it is evaluated; we express this dependence as a dependence on the ratio $\frac{\tau}{\tau_*}$.   The ratio  ${\cal R}_\sigma$ is instead obtained after an integral over momentum (performed at any given fixed time) and so it  depends on the time variable  $\frac{\tau}{\tau_*}$.  Finally, both ratios depend on the parameter $\xi_*$ (the maximum value of acquired by $\xi$; this is the parameter that controls the amount of field amplification), and on $\delta$ (which controls the duration of the phase for which the evolution of $\sigma$ is significant; specifically, the field $\sigma$ has a non-negligible evolution for $N \simeq \frac{1}{\delta}$ e-folds of inflation).

\begin{figure}[ht!]
\centerline{
\includegraphics[width=0.5\textwidth,angle=0]{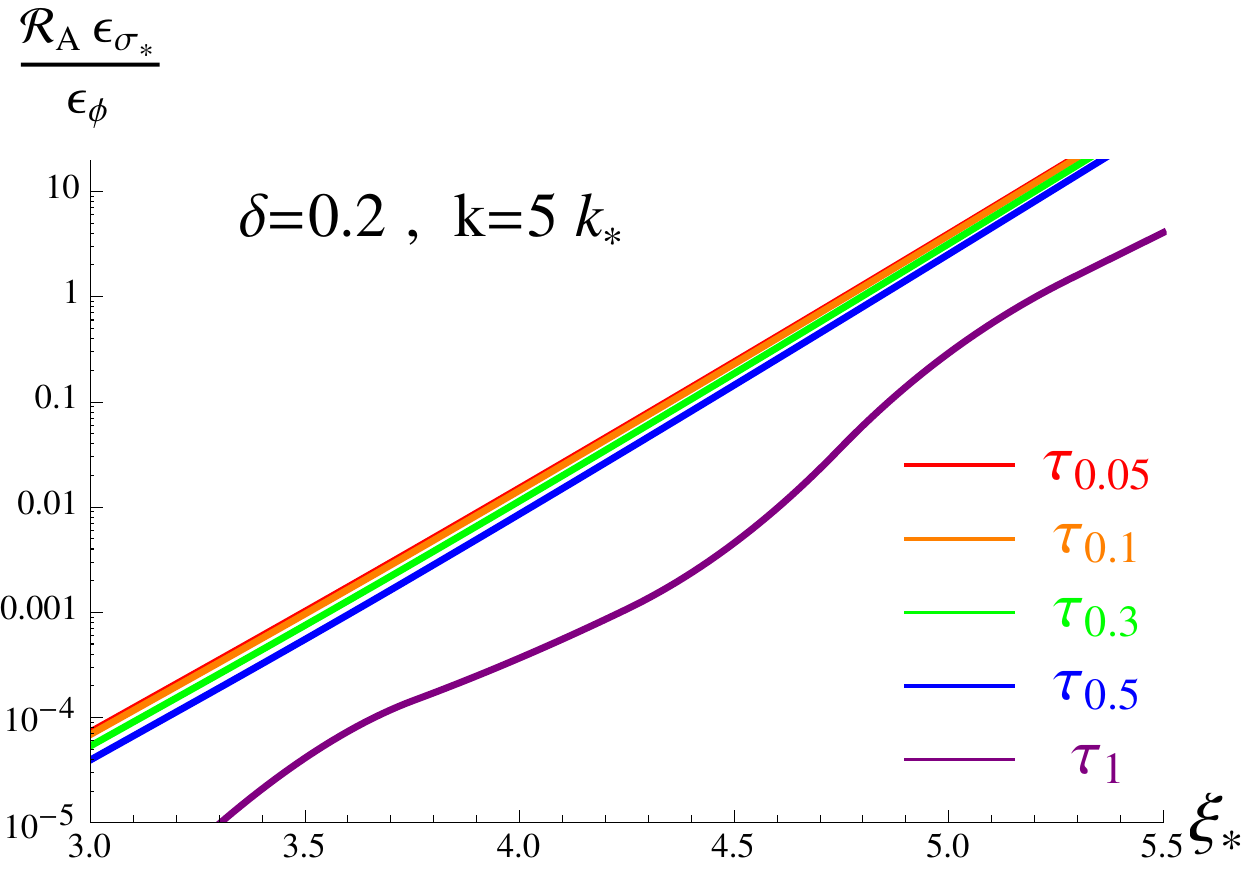}
\includegraphics[width=0.5\textwidth,angle=0]{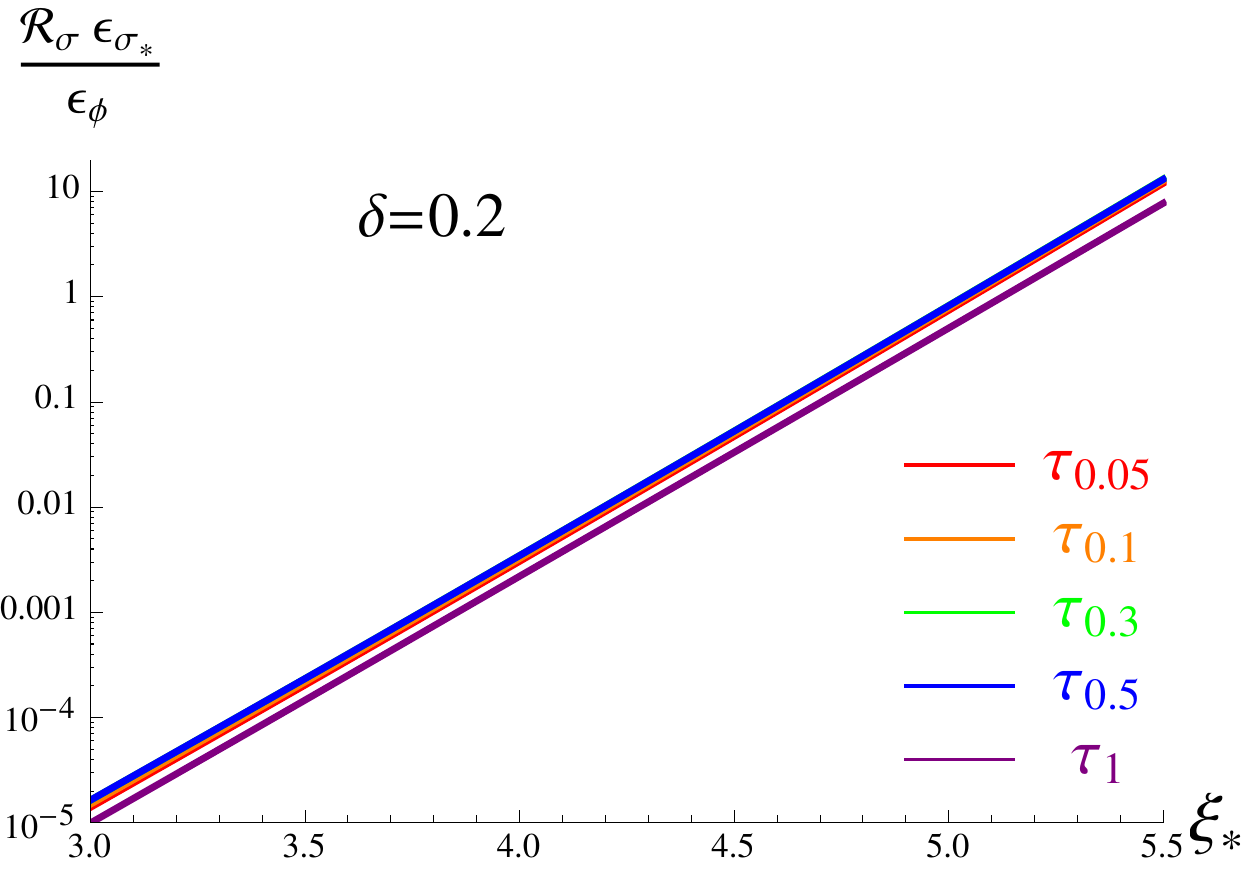}
}
\caption{Ratio $R_A$ (left panel) and  $R_A$ (right panel) for $\delta = 0.2$ and for varying $\xi_*$. The different lines correspond to different times at which the ratios are evaluated. 
}
\label{fig:R-delta02}
\end{figure}

In Figure \ref{fig:R-delta02} we show the two quantities $R_A = \frac{{\cal R}_A \, \epsilon_{\sigma,*}}{\epsilon_\phi}$ (left panel) and $R_\sigma = \frac{{\cal R}_\sigma \, \epsilon_{\sigma,*}}{\epsilon_\phi}$ (right panel), for $\delta = 0.2$ and varying $\xi_*$. In the left panel, we fixed $k = 5 \, k_*$. The quantity shown in the right panel is instead $k-$independent. The different lines in the figure correspond to different times. The times are chosen when the total energy density in the gauge field decreases (due to the expansion of the universe) by a given amount with respect to the value it had at its peak. For example $\tau_{0.5}$, is the time at which the energy density has decreased to $50\%$ of the peak value. We only studied the limits in the time interval between $\tau_1$ and $\tau_{0.05}$, since the gauge field has significantly decreased after this times, and therefore it has no significant impact on observables (in short, we do not need to care if the conditions (\ref{conditions-sigma}) are violated after  $\tau_{0.05}$). The $\xi_*$ dependence of the largest value for $R_A$ and $R_\sigma$ obtained among the lines shown can be well fitted by the exponential dependence 
\begin{equation}
\delta = 0.2 \;\;:\;\;\;\;\; {\rm R}_A  \simeq {\rm e}^{5.50 \, (\xi_* - 4.75)} \quad\quad,\quad\quad  {\rm R}_\sigma \simeq {\rm e}^{5.44 \left( \xi_* - 5.03 \right)}
\label{del02fit}
\end{equation}

\begin{figure}[ht!]
\centerline{
\includegraphics[width=0.5\textwidth,angle=0]{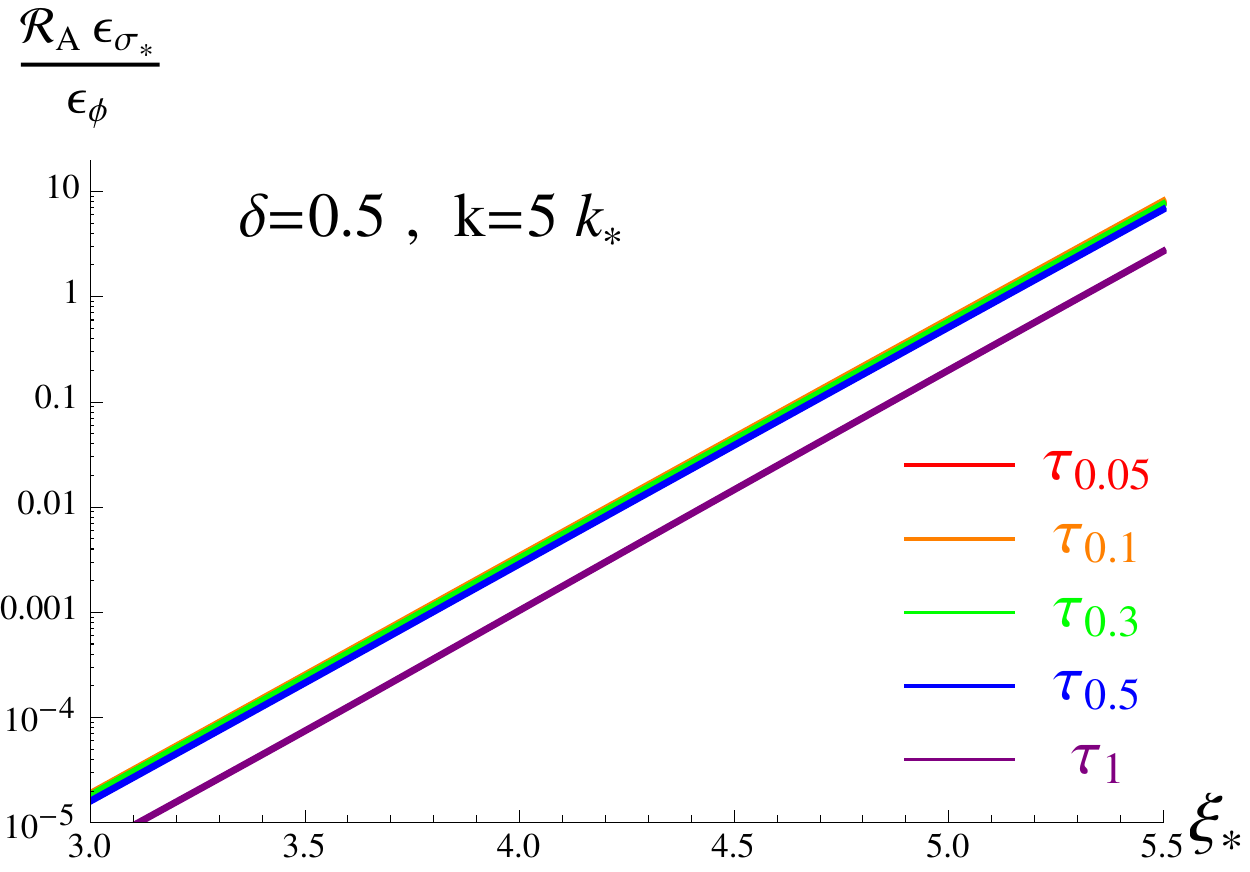}
\includegraphics[width=0.5\textwidth,angle=0]{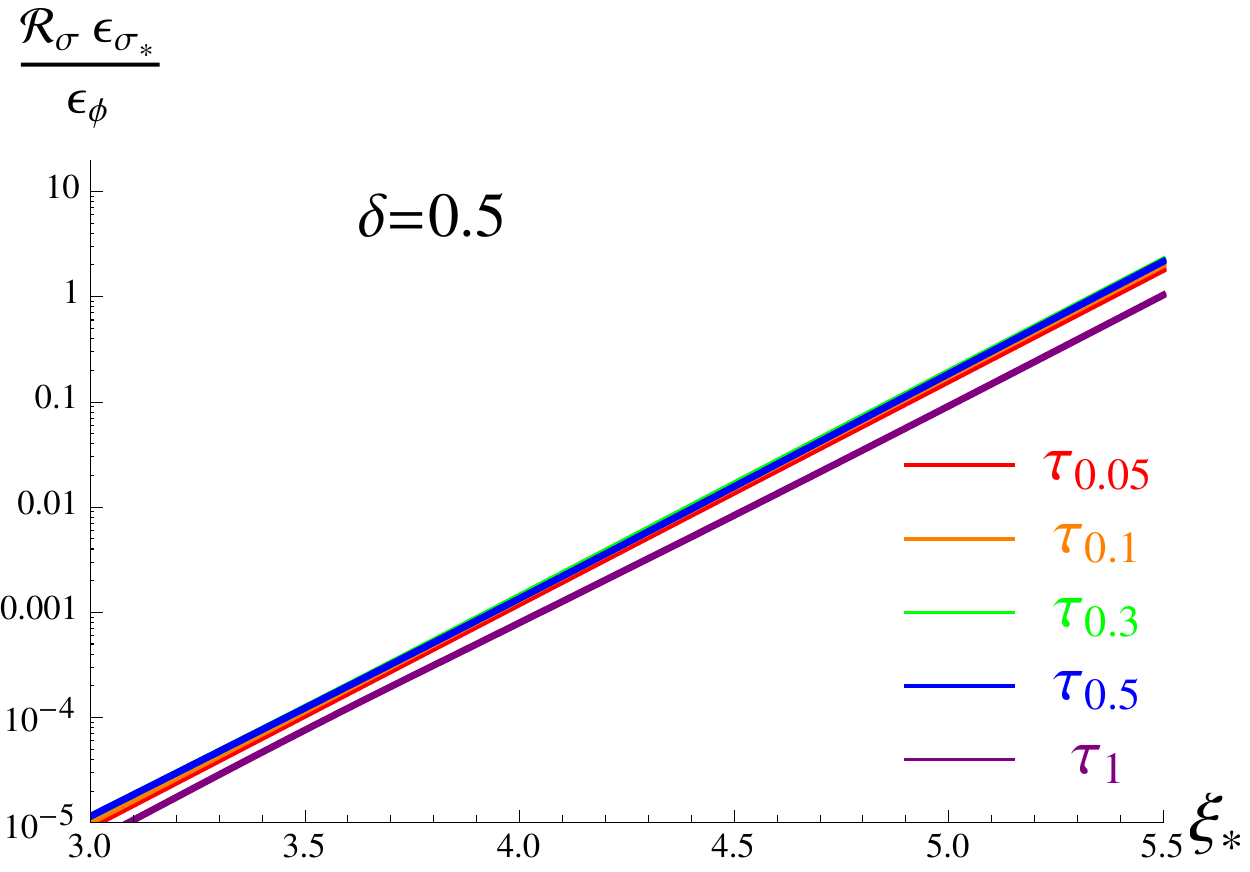}
}
\caption{Ratio $R_A$ (left panel) and  $R_A$ (right panel) for $\delta = 0.5$ and for varying $\xi_*$. The different lines correspond to different times at which the ratios are evaluated. 
}
\label{fig:R-delta05}
\end{figure}

In Figure  \ref{fig:R-delta05} we repeat the same study for $\delta = 0.5$. In this case, we obtain the accurate fits 
\begin{equation}
\delta = 0.5 \;\;:\;\;\;\;\; {\rm R}_A  \simeq {\rm e}^{5.19 \, (\xi_* - 5.10)} \quad\quad,\quad\quad  {\rm R}_\sigma \simeq {\rm e}^{4.88 \left( \xi_* - 5.34 \right)}
\label{del05fit}
\end{equation}

We find that $R_A > R_\sigma$  both for $\delta = 0.2$ and for $\delta = 0.5$. Therefore our final bounds from perturbativity, ${\cal R}_A \ll 1$, can be cast in the form 
\begin{eqnarray} 
&& \delta = 0.2 \;\;:\;\;\;\; \epsilon_{\sigma,*} \gg \epsilon_\phi \,  {\rm e}^{5.50 \, (\xi_* - 4.75)} \;, \nonumber\\ 
&& \delta = 0.5 \;\;:\;\;\;\; \epsilon_{\sigma,*} \gg \epsilon_\phi \,   {\rm e}^{5.19 \, (\xi_* - 5.10)}  \;.  
\label{pert-sig}
\end{eqnarray} 
We discuss this condition in Section \ref{sec:significance-sig}. 

\section{Discussion of the constraints}%
\label{sec:significance}

In this section we discuss the significance of the constraints on gauge field amplification imposed by backreaction on the background evolution (obtained in Section \ref{sec:back}) and by perturbativity (obtained in Section \ref{sec:pertlimits}). We then compare these constraints with the amplification required to obtain visible signatures (discussed in Section \ref{sec:phenomenology}).  We divide the discussion in two subsections, 
where we separately discuss the  $X = \phi$ (with adiabatic $\dot{\phi}$ evolution), and $X = \sigma$ (with a momentary speed up of this field) cases. 

\subsection{Discussion  for $X=\phi$}
\label{sec:significance-phi}

The values in (\ref{xi-phi-pheno}) indicate that $\xi \simeq 2.5$ is required to obtain phenomenological signatures at CMB scales, 
while $\xi \simeq 5$ is required for both significant GW at interferometer scales and primordial black holes. 

The condition (\ref{back-phi}) indicates that the  vector field amplification has a negligible backreaction on the inflaton dynamics for $\xi \ll 4.7$. 
Since the field amplification is exponentially sensitive to $\xi$, we interpret this condition (and the ones discussed below) as $\xi \la 4.7$.  This condition  is well satisfied when the phenomenological limits at CMB scales are respected,  but it is violated in the realizations that produce visible GW at interferometer scales, and primordial black-holes. In these two cases, the backreaction of the produced gauge fields slows down the motion of the inflaton, and this effect has been taken into account in those studies. This condition is also violated (by construction) in the mechanism of Ref.  \cite{Anber:2009ua}, in which the gauge field amplification is the main source of friction in the motion of the inflaton, and can lead to slow-roll inflation in a potential that would be otherwise too steep to drive inflation.~\footnote{In this mechanism, extra assumptions (as for instance the presence of $N \ga 100$ gauge fields, or the curvaton mechanism for the primordial perturbations) have to be made to ensure that the limits from CMB non-gaussianity are met.}

Finally, the relation (\ref{pert-phi}) provides a condition under which the computations of the gauge field and inflaton modes are under perturbative control. This condition evaluates to  $\xi \la 4.6$ for the value of $\xi$ assumed when the CMB modes left the horizon ($N=60$). This condition becomes slightly relaxed for smaller wavelength modes. Using $n_s \simeq 0.965$  \cite{Ade:2015lrj}, we find  $\xi \simeq 4.8$ for the LIGO scale relevant for the GW signature (corresponding to  $N\simeq 15$), and  $\xi \simeq 4.9$ for the scale relevant for the black hole limit. Since in both cases a visible signature is obtained for $\xi \simeq 5$, we see that this value is only marginally out of the perturbative and of the backreaction bounds. As a consequence, we do not think that this affects the conclusion that significant GW and primordial black holes will be generated for $\xi = {\rm O } \left( 5 \right)$, although order one corrections to the result present in the literature (as already mentioned in some of those works) can be expected.

\subsection{Discussion   for $X=\sigma$}
\label{sec:significance-sig}

Eq.~(\ref{r-peak}) shows that, for sufficiently high field amplification (sufficiently high $\xi_*$), it is possible to have a sourced GW signal at arbitrary low scale of inflation. For a large range of parameters, this GW amplification can take place without a corresponding increase of scalar perturbations beyond the bounds of cosmic variance, 
and it is therefore not ruled out by the TT and TTT observations  \cite{Namba:2015gja}. 

The condition Eq. (\ref{back-sig-phi}) guarantees that $\sigma$ affects in a negligible way the expansion of the universe, and (therefore) the motion of the inflaton. Eq. (\ref{back-sig}) ensures that the gauge field amplification has a negligible impact on the evolution of $\sigma$. The perturbativity considerations discussed in the previous section give instead the bound (\ref{pert-sig}). We note that both  bounds  (\ref{back-sig}) and  (\ref{pert-sig}) can be satisfied for arbitrary values of $\xi_*$,  provided that $\epsilon_{\sigma,*} / \epsilon_\phi$ is sufficiently large. Ref. \cite{Ozsoy:2014sba} claimed that a sourced GW signal can be observed only if the scale of inflation is not much smaller than that necessary to also generate a visible vacuum GW signal, so that, according to their claim, $r_{\rm sourced} \gg r_{\rm vacuum}$ is not possible. They only studied the case of a constant $\dot{\sigma}$ evolution, but the origin  of their claim can be discussed in the present context as well. This conclusion is based on the lines with positive $45^\circ$ slope in their Figure 1. Their ``allowed region'' at the right of these lines is the region with  $\epsilon_\sigma < \epsilon_\phi$ (in their notation, $\epsilon_\chi < \epsilon$). The basic idea is that, since $r_{\rm vacuum}$ is proportional to $\epsilon_\phi$, while the sourced GW signal is controlled by the speed of $\sigma$, the ratio $ \frac{r_{\rm peak}}{r_{\rm vacuum}}$, can be increased only if also the ratio $\frac{\epsilon_{\sigma}}{\epsilon_\phi}$ increases, eventually leading to a faster motion of $\sigma$ than of $\phi$. This statement agrees with our analysis. However, while ref.~\cite{Ozsoy:2014sba} imposes  $\epsilon_\sigma < \epsilon_\phi$, we now show that there is no reason to impose this requirement. 

First of all, we note that  $\epsilon_\sigma > \epsilon_\phi$ does not impact the Friedmann equation (given that with the condition  (\ref{back-sig-phi}) we impose that $\rho_\sigma \ll V \left( \phi \right)$), nor the background equation for the inflaton field. In principle, it may  affect the relation for the spectral tilt $n_s$ of the scalar perturbations.  The spectral tilt receives a contribution $\propto \frac{\dot{H}}{H^2} \simeq - \epsilon_\phi - \epsilon_\sigma$, resulting in 
\begin{equation}
n_s - 1 \simeq 2 \eta_\phi - 6 \epsilon_\phi - 4 \epsilon_\sigma \;. 
\label{ns} 
\end{equation} 
(Where $\eta_\phi \equiv M_p^2 \, \partial_\phi^2 V / V$ is the other standard slow roll parameter.)  Given the measured value of $n_s - 1 \sim - 0.03$  \cite{Ade:2015lrj}, we  require $ \epsilon_{\sigma,*} \la 10^{-2} $ to avoid fine tunings in (\ref{ns}), but we do not need to impose $\epsilon_{\sigma,*} < \epsilon_\phi$. If the vacuum GW signal is too small to be observed, it follows that $\epsilon_\phi \ll \vert n_s - 1 \vert$, so that eq. (\ref{ns}) should read $n_s - 1 \simeq 2 \eta_\phi  - 4 \epsilon_\sigma$. Having $\epsilon_{\sigma,*} > \epsilon_\phi$ simply implies that, for the few e-folds in which the motion of $\sigma$ is non negligible, $\dot{H}$ is controlled be the motion of $\sigma$. However, this has no phenomenological consequence as long as  $\epsilon_{\sigma,*} \la 10^{-2}$, since in this case the scalar spectral tilt will simply be controlled by $\eta_\phi$. We are not aware of any other problems that could be induced by $\epsilon_{\sigma,*} > \epsilon_\phi$, and therefore we do not impose this condition. 

Therefore all backreaction and perturbativity constraints can be summarized as 
\begin{eqnarray}
&& \delta = 0.2 \;\;:\;\;\;\; {\rm Max} \left[ \epsilon_\phi \,  {\rm e}^{5.47 \, (\xi_* - 4.67)} ,\,  \epsilon_\phi \,  {\rm e}^{5.50 \, (\xi_* - 4.75)} \right] \la \epsilon_{\sigma,*} \la 10^{-2} \;, \nonumber\\ 
&& \delta = 0.5 \;\;:\;\;\;\; {\rm Max} \left[ \epsilon_\phi \,  {\rm e}^{4.84 \, (\xi_* - 5.06)} ,\,  \epsilon_\phi \,  {\rm e}^{5.19 \, (\xi_* - 5.10)} \right] \la \epsilon_{\sigma,*} \la 10^{-2} \;.  
\end{eqnarray} 
For $\delta =0.2$, the first condition in the square parenthesis (the one obtained from backreaction) dominates over the second one for $\xi_* \la 19$, so we disregard the second condition. For $\delta =0.5$, instead, the first condition (from backreaction) dominates for $\xi_* \la 5.7$. Therefore, we will keep both conditions. Using  (\ref{eps-sig}), we can rewrite the conditions as 
\begin{eqnarray}
&& \delta = 0.2 \;\;:\;\;\;\;  2 \cdot 10^{-5} \, {\rm e}^{2.74 \, \xi_*} \, \sqrt{\epsilon_\phi}  \la \frac{f}{M_p} \la 0.71 \;, \nonumber\\ 
&& \delta = 0.5 \;\;:\;\;\;\; {\rm Max} \left[ 1.4 \cdot 10^{-5} \, {\rm e}^{2.42 \, \xi_*} \, \sqrt{\epsilon_\phi} ,\, 5.1 \cdot 10^{-6} \, {\rm e}^{2.60 \, \xi_*} 
\, \sqrt{\epsilon_\phi} \right] \la \frac{f}{M_p} \la 0.28 \;. 
\end{eqnarray} 

Finally, using the relation (\ref{tsrandepsphi}), which is valid when the peak of the sourced GW signal is much greater than the vacuum GW, 
we can rewrite these limits as 
\begin{eqnarray}
&& \delta = 0.2 \;\;:\;\;\;\;    0.038 \, {\rm e}^{0.0018 \, \xi_*} \, r_{\rm peak}^{1/4} \la \frac{f}{M_p} \la 0.71 \;, \nonumber\\ 
&& \delta = 0.5 \;\;:\;\;\;\; {\rm Max} \left[ 0.016 \, {\rm e}^{0.00097 \, \xi_*} ,\,  0.0058 \, {\rm e}^{0.18 \, \xi_*}   \right] r_{\rm peak}^{1/4} \la \frac{f}{M_p} \la 0.28 \;.  
\end{eqnarray} 
where $r_{\rm peak}$ is the tensor-to-scalar ratio at the peak of the GW signal. We see that there is a large region of parameter space in which all these limits can be satisfied.

\section{Conclusions}
\label{sec:conclusions}

Axion fields are very natural candidates for the inflaton or for light extra degrees of freedom during inflation, since their potential is protected by a shift symmetry, that can be broken in a controllable way \cite{Pajer:2013fsa}. The main decay of a pseudoscalar $X$ is through the least irrelevant operator $X F {\tilde F}$ (the other dimension-five operator $\partial_\mu X \, {\bar \psi} \gamma^\mu \gamma_5 \psi$ leads to a helicity-suppressed decay in the limit of small fermion mass $m_\psi \ll m_X$). This operator can lead to interesting gauge field amplification already during inflation: in its presence, one polarization of the gauge field becomes unstable at horizon crossing; the energy density of any given gauge mode can be highly amplified by this effect, and it is eventually diluted away by the expansion of the universe. In the few e-folds between the amplification and the dilution, the mode can give rise to a number of signatures that we have outlined in the Introduction. 

In our discussion we have distinguished between the case in which $X$ rolls at an approximately constant velocity~\cite{Sorbo:2011rz,Barnaby:2012xt}, or it experiences a transient of relatively fast roll~\cite{Namba:2015gja} (we assume that, even at its fastest roll, the field continues to satisfy $\vert \dot{X} \vert \ll H M_p$). We have reviewed this second case at length, as  it is one of the very few mechanisms that can produce sourced GW at CMB scales (therefore violating the condition (\ref{standard_pt})), and at the same time avoids overproducing scalar perturbations, so that it is not ruled out by the strong limits on the latter. In our discussion we have provided analytic expressions for the peak of the sourced GW signal, and for the needed energy density in the gauge field, which were not given in  \cite{Namba:2015gja}. We have also remarked that this mechanism can source visible GW even if the scale of inflation is arbitrarily below what is required by  (\ref{standard_pt}) to produce a visible vacuum signal.  The sourced GW signal has clear properties (it is localized in $\ell$ space, it is chiral, and it is highly non-gaussian) that could allow us to distinguish it from the vacuum one. These properties will allow to distinguish a vacuum vs non-vacuum origin of the GW background, once it will be observed. 

The gauge field amplification needs to be sufficiently strong to lead to observable effects. One should therefore include its backreaction on the background dynamics, as consistently done in many applications, and also make sure that the computations remain in the perturbative regime. A strong motivation for the present work has been the study of two criteria pointed out in \cite{Ferreira:2015omg} for the validity of the perturbative regime. Following their definitions, and their starting point, we have checked their computation in the case in which $\dot{X}$ is adiabatically rolling, and we have extended it to the case in which $\dot{X}$ has a momentary  roll. In this second case, we have found that there exists a large region of parameters in which the two criteria are satisfied (we find that the perturbativity criteria are satisfied in most of the region in which the produced gauge field does not backreact on the background evolution). In the first case, we have found that the criteria are satisfied in a greater region than what was obtained in~\cite{Ferreira:2015omg}.  As a consequence,  the mechanism is generally in better shape than what concluded in~\cite{Ferreira:2015omg}. We have outlined the reasons for this difference  in Appendix \ref{app:comparison}. 

From our results, it emerges that the criteria are well satisfied in the applications that produce signatures at CMB scales. One of the two criteria is instead marginally violated when GW and primordial black holes are produced at smaller scales. In this case, $\xi \simeq 5$ is needed to produce a visible effect (where $\xi$ is the parameter that controls the gauge field amplification), while the perturbativity criterion fails at $\xi \geq 4.8$. This result is not surprising, given that several of the original works already pointed out that the scalar perturbations are in a strong coupling regime for those values of $\xi$, and order one corrections can be expected (we are comforted in the statement by the fact that the $\xi \la 4.8$ region is not very far from $\xi = 5$).  We do not believe that the conclusion that significant GW and primordial black holes  will be generated for $\xi = {\rm O } \left( 5 \right)$ is in question. An exact computation (performed perhaps through lattice simulation) would certainly allow to refine the precise bound on $\xi$, but not the presence of these physical effects.

\vskip.25cm
\noindent{\bf Acknowledgements:} 

We thank Raphael Flauger, Matthew Kleban, and  Sarah Shandera  for useful discussions. The work of M.P. and C.U. partially supported from the DOE grant DE-SC0011842  at the University of Minnesota. The work of L.S. is partially supported by the NSF grant PHY-1205986.

\appendix

\section{Energy density in the gauge field} 
\label{app:rhoA}

For brevity, we use electromagnetic notation, although we are not assuming that the vector field studied in this work is the Standard Model  electromagnetic field. In this notation, the  energy density of the gauge field acquires the form 
\begin{equation}
\rho_A = \left\langle \frac{E^2+B^2}{2} \right\rangle \;, 
\end{equation} 
where $E$ and $B$ fields are 
\begin{eqnarray}
\vec{E} & = &  - \frac{1}{a^2} \, \int \frac{d^3 k}{\left( 2 \pi \right)^{3/2}} \, {\rm e}^{i \vec{k} \cdot \vec{x}} \, \vec{\epsilon}_+  \left( \vec{k} \right) \, {\hat A}_+' \left( \vec{k} \right) \;, \nonumber\\ 
\vec{B} & = &   \frac{1}{a^2} \, \int \frac{d^3 k}{\left( 2 \pi \right)^{3/2}} \, {\rm e}^{i \vec{k} \cdot \vec{x}} \, \vec{\epsilon}_+  \left( \vec{k} \right) \, k \,  {\hat A}_+ \left( \vec{k} \right) \;. 
\label{E-B}
\end{eqnarray}  
From these expressions, we see that  the energy density per mode reduces to the following simple form:
\begin{equation}
\frac{d \rho_k}{d k} = \frac{1}{4 \pi^2 a^4} \left\{ k^2 \, \vert A_+' \left( k \right) \vert^2 +  k^4 \, \vert A_+ \left( k \right) \vert^2 \right\} 
=  \frac{   k^3}{8 \pi^2 a^4} \left\{   \left \vert \frac{d {\tilde A}}{d  x } \right \vert^2 + \vert {\tilde A} \vert^2 \right\} \;, 
\label{endenpermode}
\end{equation} 
where we have defined 
\begin{equation}
A_+ \left( \tau ,\, k \right) \equiv \frac{1}{\sqrt{2 k}} \, {\tilde A} \left( x \equiv - \, k \, \tau  \right) \;. 
\end{equation} 
A dimensionless expression is obtained by dividing the energy density by the  $4^{th}$ power of the Hubble parameter, leading to 
\begin{equation}
\frac{d \, \left( \rho_k / H^4 \right)}{d \ln k}  = \frac{  x^4 }{8 \pi^2 } \left\{  \left \vert \frac{d {\tilde A}}{d  x } \right \vert^2 + \vert {\tilde A} \vert^2 \right\} \;. 
\label{endenpmode}
\end{equation} 
This is the expression plotted in Figure \ref{fig:rhoti}, for the three approximations (\ref{colsol}), (\ref{bessol}) and (\ref{A-simple}) for the gauge field amplitude. 

We also need to evaluate $\langle E \cdot B \rangle$, as it affects the evolution equation for the field $\sigma$ 
\begin{equation}
\ddot{\sigma} + 3 H \dot{\sigma} + \partial_\sigma V = \frac{\alpha}{f} \, \langle \vec{E} \cdot \vec{B} \rangle \;. 
\label{eom-sigma}
\end{equation}
(we want to ensure that this terms can be neglected). Using (\ref{E-B}), we find 
\begin{equation}
\frac{d \, \langle \vec{E} \cdot \vec{B} \rangle / H^4 }{d \ln k}  = \frac{  x^4 }{8 \pi^2 } \, \frac{d}{d x} \, 
 \vert {\tilde A}  \vert^2 \;. 
\label{endenpmode2}
\end{equation}

\section{Gauge field amplitude in the $X = \sigma$ case} 
\label{app:A-bump}

In this appendix, we provide some details on the evolution of the gauge field mode function and energy density for the  $X=\sigma$ case. Performing a WKB approximation of the evolution equation, 
Ref.  \cite{Namba:2015gja} obtained the following accurate expression for the real part of the gauge field amplitude: 
\begin{eqnarray}
A_R \left( \tau ,\, k \right) &\simeq&  N \left[ \xi_* ,\, x_* ,\, \delta \right] \, \left( \frac{-\tau}{8 \, k \, \xi \left( \tau \right)} \right)^{1/4}  \, {\rm exp} \left[ - \frac{4 \xi_*^{1/2}}{1+\delta} \, \left( \frac{-\tau}{-\tau_*} \right)^{\delta/2} \, \left( - k \tau \right)^{1/2} \right] \;, \nonumber\\ 
 \frac{d A_R \left( \tau ,\, k \right)}{d \tau} & \simeq & \sqrt{\frac{2 \, k \, \xi \left( \tau \right)}{-\tau}} \, A_R \left( \tau ,\, k \right) , \quad\quad\quad\quad    \tau  >  \tau_*  \;, 
\label{varysig-AR}
\end{eqnarray}
where, combining the definition of $\xi$ with eq. (\ref{dotsigma}), we can see that 
\begin{equation}
\xi \left( \tau \right) = \frac{2 \, \xi_*}{\left(  \frac{-\tau}{-\tau_*} \right)^\delta + \left(  \frac{-\tau_*}{-\tau} \right)^\delta } \;. 
\label{varysig}
\end{equation}

We recall that $x = - k \tau$, where $k$ is the comoving momentum of the mode, and $\tau$ is conformal time. We also have $x_* = - k \, \tau_*$, where $\tau_*$ denotes the time at which $\sigma$ reaches its maximum speed. The expression (\ref{varysig-AR}) is valid only for $\tau > \tau_*$. We verified numerically that the amplitude of the gauge field is much smaller that  (\ref{varysig-AR}) at earlier times, and therefore we can disregard these earlier times in our computations. 

We provide  the imaginary part of $A_+$ using the  Wronskian condition $A_+ \, A_+^{'*} - {\rm c.c.} = i$ :
\begin{eqnarray}
A_I \left( \tau ,\, k \right) &\simeq&  \frac{1}{N \left[ \xi_* ,\, x_* ,\, \delta \right]} \, \left( \frac{-\tau}{2^7 \, \xi \, k} \right)^{1/4}  \, {\rm exp} \left[  \frac{4 \xi_*^{1/2}}{1+\delta} \, \left( \frac{-\tau}{-\tau_*} \right)^{\delta/2} \, \left( - k \tau \right)^{1/2} \right]  \;, \nonumber\\ 
\frac{d A_I \left( \tau ,\, k \right)}{d \tau} &\simeq& - \sqrt{\frac{2 \, k \, \xi \left( \tau \right)}{-\tau}} \, A_I \left( \tau ,\, k \right) \;. 
\label{varysig-AI}
\end{eqnarray}
We note that $A_+ = A_R + i \, A_I$ and its derivative also satisfy the relation (\ref{Ader}). 

As the analogous quantities for the sourced power spectra and bispectra, we find that the  normalization factor $N \left[ \xi_* ,\, x_* ,\, \delta \right]$ is well approximated by the shape 
\begin{equation}
N \left[ \xi_* ,\, q ,\, \delta \right] \approx N^c \left[ \xi_* ,\, \delta \right] \, {\rm exp} \left( - \frac{1}{2 \, \sigma^2 \left[ \xi_* ,\, \delta \right] } \, \ln^2 \left( \frac{q}{q^c \left[ \xi_* ,\, \delta \right] } \right) \right) \;. 
\label{N-fit-app}
\end{equation} 
The three parameters $N^c ,\, q^c ,\,$ and $\sigma$ depend on $\xi_*$ and $\delta$. In agreement with the computations of \cite{Namba:2015gja}, we evaluate this dependence for the two specific cases $\delta = 0.2 ,\, 0.5$, and we then fit the $\xi_*$ dependence numerically in the $3 \leq \xi_* \leq 7$ interval. We obtain 
\begin{eqnarray}
N^c  &=& {\rm exp} \left[ 0.437 + 2.97 \, \xi_* + 0.00105 \, \xi_*^2 \right] \;\;\;,\;\;\; \delta = 0.2 \;\;\;,\;\;\; 3 \leq \xi_* \leq 7 \;, \nonumber\\ 
q^c &=& -0.150 + 0.594 \, \xi_* - 0.00105 \, \xi_*^2 \;, \nonumber\\ 
\sigma &=& 2.78 - 0.387 \, \xi_* + 0.0229 \, \xi_*^2 \;, 
\end{eqnarray}  
and 
\begin{eqnarray}
N^c  &=& {\rm exp} \left[ 0.117 + 2.54 \, \xi_* + 0.000525 \, \xi_*^2 \right] \;\;\;,\;\;\; \delta = 0.5 \;\;\;,\;\;\; 3 \leq \xi_* \leq 7 \;, \nonumber\\ 
q^c &=& -0.0500 + 0.683 \, \xi_* - 0.000716 \, \xi_*^2 \;, \nonumber\\ 
\sigma &=& 1.51 - 0.220 \, \xi_* + 0.0137 \, \xi_*^2 \;. 
\end{eqnarray} 

\begin{figure}[ht!]
\centerline{
\includegraphics[width=0.5\textwidth,angle=0]{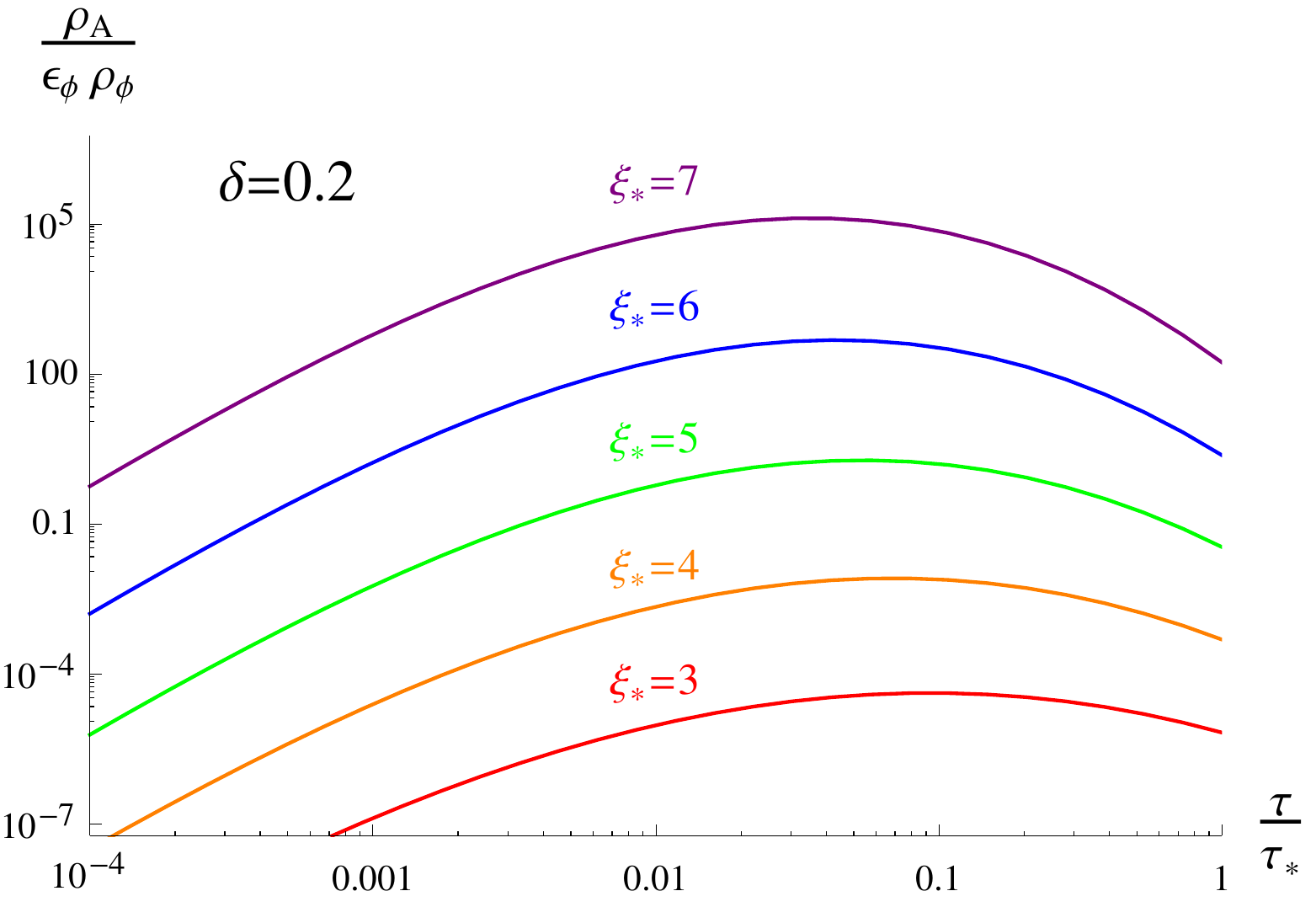}
\includegraphics[width=0.5\textwidth,angle=0]{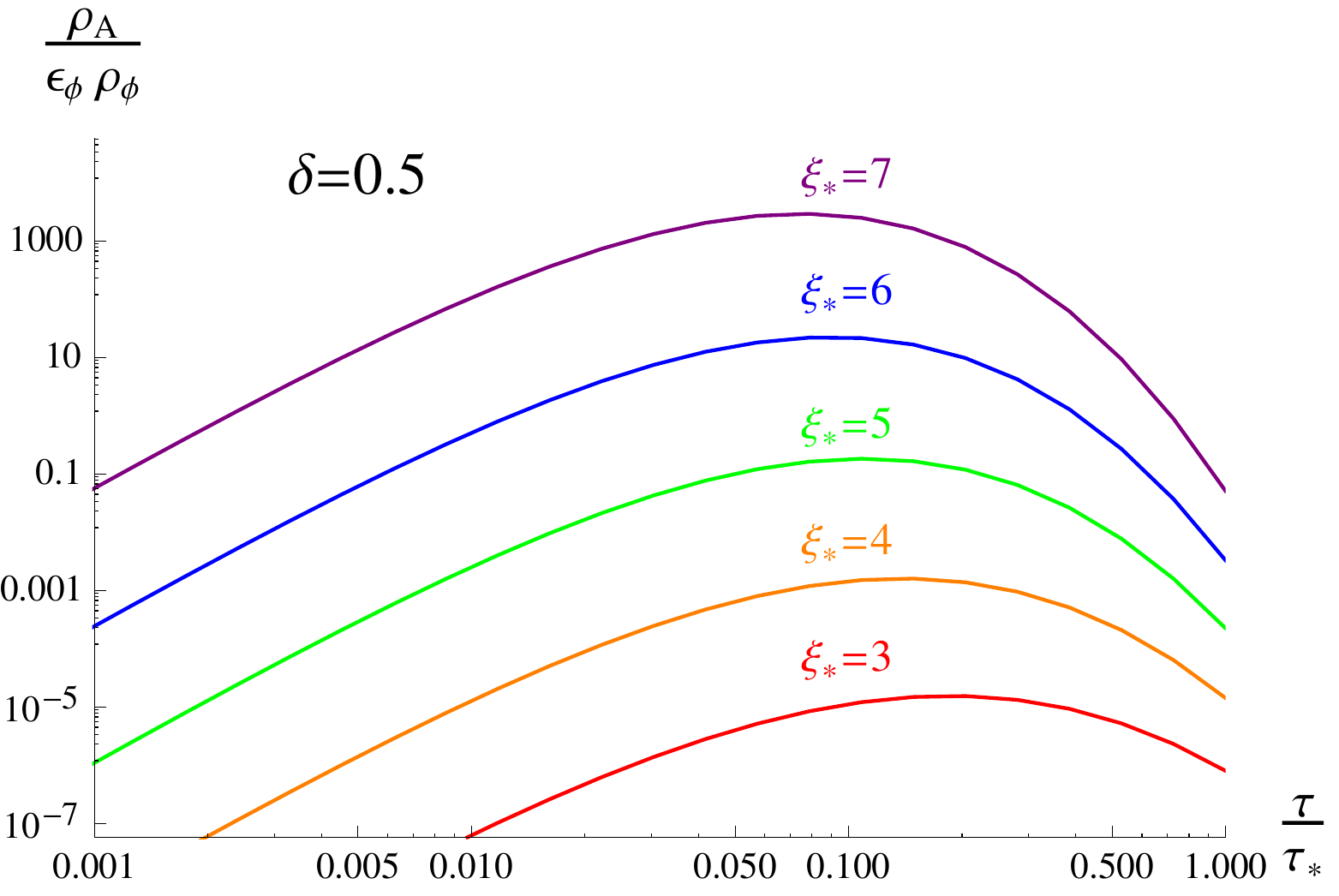}
}
\caption{Total Energy density in the gauge field, rescaled by $\epsilon_{\phi}\, \rho_{\phi}$, as a function of time, for fixed values of $\xi_* $ 
}
\label{fig:rho-varyA}
\end{figure}

Inserting the expression for the vector field amplitude into (\ref{endenpmode}), using  $\rho_{\phi} \simeq 3 \, H^2 \, M_p^2$, and eliminating $H$ through eq. (\ref{P-zeta}), we obtain  the following expression for the energy in the vector field 
\begin{eqnarray} 
\frac{\rho_A}{\epsilon_\phi \, \rho_\phi} &=& \frac{N_c^2 P_\zeta^{(0)} \, y^{7/2}}{6 \sqrt{\xi_*} \sqrt{y^\delta + y^{-\delta}}} 
\int_0^{\infty} \frac{d x_*}{x_*} \, x_*^{7/2} \, {\rm e}^{-\frac{8 \sqrt{\xi_*} \sqrt{x_*} y^{\frac{1}{2} + \frac{\delta}{2}}}{1+\delta}-\frac{\ln^2 
\left( \frac{x_*}{q_c} \right)}{\sigma^2} }  \,  \left[ 4 \xi_* + x_* y \left( y^\delta + y^{-\delta} \right) \right] \;, \nonumber\\
\end{eqnarray}  
where we have defined  $y \equiv \tau / \tau_*$ (therefore, our expression are valid for $0 \leq y \leq 1$).  We show the result for different values of $\xi_*$ and for different times in Figure \ref{fig:rho-varyA}. 

We observe that the energy density in the vector field reaches a maximum value at $y = {\rm O } \left( 0.1 \right)$, and it is then diluted away by the expansion of the universe. We fitted numerically the dependence on $\xi_*$ of the maximum value assumed by $\rho_A$, obtaining the very accurate expression 
\begin{equation}
\frac{\rho_{A,\rm max}}{\epsilon_\phi \, \rho_\phi} \sim \left\{ \begin{array}{l} 
2.75 \cdot 10^{-12} \, {\rm e}^{1.74 \, \pi \,\xi_*} \;\;\;,\;\;\; \delta = 0.2 \;, \\ 
7.48 \cdot 10^{-12} \, {\rm e}^{1.52 \, \pi \,\xi_*} \;\;\;,\;\;\; \delta = 0.5 \;. \end{array} \right. 
\label{rhoAmax}
\end{equation}  

\section{Computation of the perturbativity limits for  $X=\phi$ }
\label{app:Xphi}

In this appendix we derive the limits discussed in Section \ref{sec:pert-xi-phi} for the  $X=\phi$ case. 

We compute the gauge field two-point correlator at leading (zeroth) and first subleading (one loop) order in the interaction (\ref{Hintconspa}), where $X$ is the inflaton field $\phi$.  We note that  this interaction term involves the inflaton perturbation $\delta \phi$, and not the inflaton vev. The inflation zeroth mode modifies the vector amplitude from the vacuum mode to (\ref{colsol}).~\footnote{More accurately, we  use the approximation (\ref{A-simple}) in most of our explicit computations, for the reasons explained at the end of Section \ref{sec:consxiAsol}. } The resulting wave function is used as the ``unperturbed'' wave function at the right hand side of (\ref{AA-inin}) (namely, we are perturbing in $\delta \phi F {\tilde F}$, not in $\phi  F {\tilde F}$; this is the same starting point as the computations of \cite{Ferreira:2015omg}). 

The unperturbed two point correlator is 
\begin{equation} 
\left\langle {\hat A}_+^{(0)} \left( \vec{k}_1 ,\, \tau \right)  {\hat A}_+^{(0)} \left( \vec{k}_2 ,\, \tau \right) \right\rangle = \vert A_+ \left( k_1 ,\, \tau \right) \vert^2 \delta^{(3)} \left( \vec{k}_1 + \vec{k}_2 \right) \;. \nonumber\\ 
\end{equation} 
For the one loop expression, using eqs. (\ref{AA-inin}) and (\ref{Hintconspa}), we obtain 
\begin{eqnarray}
&&   \!\!\!\!\!\!\!\!   \!\!\!\!\!\!\!\!   \!\!\!\!\!\!\!\!   \!\!\!\!\!\!\!\! 
\delta^{(1)} \left\langle {\hat A}_+ \left( \vec{k}_1 ,\, \tau \right)  {\hat A}_+ \left( \vec{k}_2 ,\, \tau \right) \right\rangle = 
- \frac{\alpha^2}{f^2} \int \frac{d^3 k d^3 p d^3 k' d^3 p'}{\left( 2 \pi \right)^3} \, \vert \vec{k} + \vec{p} \vert \,  \vert \vec{k}' + \vec{p}' \vert \, \nonumber\\
&& \!\!\!\!\!\!\!\! 
\times \left[ \vec{\epsilon}^{(+)} \left( \vec{p} \right) \cdot  \vec{\epsilon}^{(+)} \left( - \vec{k} - \vec{p} \right) \right] \,  
\left[ \vec{\epsilon}^{(+)} \left( \vec{p}' \right) \cdot  \vec{\epsilon}^{(+)} \left( - \vec{k}' - \vec{p}' \right) \right] \, 
\int^{\tau} d \tau_1 \int^{\tau_1} d \tau_2 \, \,\, {\cal C} \;,  
\end{eqnarray} 
where the integrand is 
\begin{eqnarray}
&&  \!\!\!\!\!\!\!\! \!\!\!\!\!\!\!\! \!\!\!\!\!\!\!\! \!\!\!\!\!\!\!\! 
{\cal C} = \Bigg\langle \Bigg[ \left[ {\hat A}_+^{(0)} \left(  \vec{k}_1 ,\, \tau \right)  {\hat A}_+^{(0)} \left( \vec{k}_2 ,\, \tau \right) ,\, 
\delta {\hat \phi} \left( \vec{k} ,\, \tau_1 \right) \, \frac{d {\hat A}_+^{(0)} \left( \vec{p} ,\, \tau_1 \right)}{d \tau_1} \,   {\hat A}_+^{(0)} \left( - \vec{k} - \vec{p} ,\, \tau_1 \right) \right] ,\, \nonumber\\
&&  \quad\quad\quad\quad\quad\quad\quad\quad \quad\quad  \quad\quad 
\delta {\hat \phi} \left( \vec{k}' ,\, \tau_2 \right)  \, \frac{d {\hat A}_+^{(0)} \left( \vec{p}\, ' ,\, \tau_2 \right)}{d \tau_2} \,   {\hat A}_+^{(0)} \left( - \vec{k}' - \vec{p}\,' ,\, \tau_2 \right) \Bigg] \Bigg\rangle \;. 
\end{eqnarray}

The explicit evaluation of the commutators and the expectation values gives 
\begin{eqnarray}
&&  \!\!\!\!\!\!\!\! 
 \delta^{(1)} \left\langle {\hat A}_+ \left( \vec{k}_1 ,\, \tau \right)  {\hat A}_+ \left( \vec{k}_2 ,\, \tau \right) \right\rangle = 
  \frac{2 \alpha^2 \,\, \delta^{(3)} \left( \vec{k}_1 + \vec{k}_2  \right)  }{f^2} \, \int^{\tau} d \tau_1 \int^{\tau_1} d \tau_2 \;   \int \frac{d^3 p}{\left( 2 \pi \right)^3} \,   \left[ 1 - \frac{\vec{k}_1 \cdot \vec{p}}{k_1 \, p} \right]^2 
       \nonumber\\ 
&& \quad\quad\quad \times
\Bigg\{ k_1^2 \, 
{\rm Im }  \left[ A_+ \left( k_1 ,\, \tau \right) \,  A_+^* \left( k_1 ,\, \tau_1 \right) \right] \, {\cal C}_1 
+ k_1 \, p \, {\rm Im }  \left[ A_+ \left( k_1 ,\, \tau \right)   A_+^{'*} \left( k_1 ,\, \tau_1 \right) \right]    \, 
{\cal C}_2 \nonumber\\ 
&&  \quad\quad\quad\quad 
+ k_1 \, p \,  {\rm Im }  \left[ A_+ \left( k_1 ,\, \tau \right) \,  A_+^* \left( k_1 ,\, \tau_1 \right) \right]   \, 
{\cal C}_3 
+ p^2 \, {\rm Im }  \left[ A_+ \left( k_1 ,\, \tau \right)   A_+^{*'} \left( k_1 ,\, \tau_1 \right) \right]  \, 
{\cal C}_4   \Bigg\} \;, 
\label{AAcorgen}
\end{eqnarray} 
where
\begin{eqnarray}
{\cal C}_1 &=& {\rm Im}  \left[  A_+ \left( k_1 ,\, \tau \right) \,  A_+^{'*} \left( p ,\, \tau_2 \right)  \, A_+' \left( p ,\, \tau_1 \right) \,  A_+^* \left( k_1 ,\, \tau_2 \right)  \, \delta \phi \left( \vert \vec{k}_1 + \vec{p} \vert ,\, \tau_1 \right) \, \delta \phi^* \left( \vert \vec{k}_1 + \vec{p} \vert ,\, \tau_2 \right)  \right] \;, \nonumber\\ 
{\cal C}_2 &=& {\rm Im}  \left[  A_+ \left( k_1 ,\, \tau \right)   A_+^{'*} \left( p ,\, \tau_2 \right)  A_+ \left( p  ,\, \tau_1 \right)   A_+^* \left( k_1 ,\, \tau_2 \right)  \delta \phi \left( \vert \vec{k}_1 + \vec{p} \vert ,\, \tau_1 \right)   \delta \phi^* \left( \vert \vec{k}_1 + \vec{p} \vert ,\, \tau_2 \right) \right] \;, \nonumber\\ 
{\cal C}_3 &=&  {\rm Im}  \left[  A_+ \left( k_1 ,\, \tau \right) \,  A_+^* \left( p ,\, \tau_2 \right)  \,  A_+' \left( p ,\, \tau_1 \right) \,  A_+^{'*} \left( k_1 ,\, \tau_2 \right)  \,  \delta \phi \left( \vert \vec{k}_1 + \vec{p} \vert ,\, \tau_1 \right) \,  \delta \phi^* \left( \vert \vec{k}_1 + \vec{p} \vert ,\, \tau_2 \right) \right] \;, \nonumber\\  
{\cal C}_4 &=& {\rm Im}  \left[  A_+ \left( k_1 ,\, \tau \right)   A_+^* \left( p ,\, \tau_2 \right)  A_+ \left( p  ,\, \tau_1 \right)   A_+^{'*} \left( k_1 ,\, \tau_2 \right)  \delta \phi \left( \vert \vec{k}_1 + \vec{p} \vert ,\, \tau_1 \right)   \delta \phi^* \left( \vert \vec{k}_1 + \vec{p} \vert ,\, \tau_2 \right) \right] \;. \nonumber\\ 
\end{eqnarray} 

This expression is exact. We now note that it is highly dominated by terms which have the highest possible powers of $A_R$. These are terms $\propto A_R^5 \, A_I \propto {\rm e}^{4 \pi \xi} $. Keeping only these terms, we have the very accurate approximation 
\begin{eqnarray}
&&  \!\!\!\!\!\!\!\! \!\!\!\
 \delta^{(1)} \left\langle {\hat A}_+ \left( \vec{k}_1 ,\, \tau \right)  {\hat A}_+ \left( \vec{k}_2 ,\, \tau \right) \right\rangle \simeq  
  \frac{ 2 \, \alpha^2\,\,  \delta^{(3)} \left( \vec{k}_1 + \vec{k}_2  \right)  }{f^2} \, \int^{\tau} d \tau_1 \int^{\tau_1} d \tau_2 \;   \int \frac{d^3 p}{\left( 2 \pi \right)^3} \,   \left[ 1 - \frac{\vec{k}_1 \cdot \vec{p}}{k_1 \, p} \right]^2 
       \nonumber\\ 
&&  \!\!\!\!\!\!\!\! \!\!\!\!\!\!\!\! \!\!\!\!\!\!\!\! 
\times {\rm Im } \left[  \delta \phi \left( \vert \vec{k}_1 + \vec{p} \vert ,\, \tau_1 \right)   \delta \phi^* \left( \vert \vec{k}_1 + \vec{p} \vert ,\, \tau_2 \right) \right] \; A_R \left( k_1 ,\, \tau \right) 
\times \Bigg[ p \, A_R \left( p ,\, \tau_2 \right) \, A_R' \left( k_1 ,\, \tau_2 \right) + k_1 \, A_R \left( k_1 ,\, \tau_2 \right) \, A_R' \left( p ,\, \tau_2 \right)   \Bigg] \nonumber\\ 
&&  
\quad\quad \quad \quad  \times \Bigg\{ p \, A_R \left( p ,\, \tau_1 \right) 
\left[ A_I \left( k_1 ,\, \tau \right) \, A_R' \left( k_1 ,\, \tau_1 \right) -  A_R \left( k_1 ,\, \tau \right) \, A_I' \left( k_1 ,\, \tau_1 \right) \right] 
\nonumber\\ 
&& \quad \quad \quad\quad\quad \quad \quad \quad 
+ k_1 \, A_R' \left( p ,\, \tau_1 \right) 
\left[ A_I \left( k_1 ,\, \tau \right) \, A_R \left( k_1 ,\, \tau_1 \right) -  A_R \left( k_1 ,\, \tau \right) \, A_I \left( k_1 ,\, \tau_1 \right) \right] \Bigg\} \;. 
\label{AAcordom}
\end{eqnarray} 

As we already remarked, the expressions for $A_+$ and for $\delta \phi$ entering at the right hand side are the unperturbed ones, namely those obtained without the interaction (\ref{Hintconspa}). For the inflaton field we have 
\begin{equation}
\delta \phi \left( k ,\, \tau \right) = \frac{H_k \left( 1 + i \, k \, \tau \right) {\rm e}^{-i \, k \, \tau}}{\sqrt{2} \, k^{3/2} } \;, 
\end{equation} 
where $H_k$ denotes the value of the Hubble rate when the $\delta \phi \left( k \right)$ left the horizon

Defining the dimensionless quantities 
\begin{equation}
x \equiv - k_1 \, \tau \;\;,\;\; 
x_1 \equiv - k_1 \, \tau_1 \;\;,\;\; 
x_2 \equiv - k_2 \, \tau_2 \;\;,\;\; 
\vec{q} \equiv \frac{\vec{p}}{k_1} \;\;,\;\; 
{\tilde A} \left( x \right) \equiv \sqrt{2 k_1} \, A \left( k_1 ,\, \tau \right) \;, 
\label{dimensionless} 
\end{equation} 
we arrive to 
\begin{eqnarray}
\!\!\!\!\!\!\!\! \!\!\!\! \!\!\!\!\!\!\!\! \!\!\!\!\!\!\!\!\!\!\!\!  \left\langle {\hat A}_+^{(0)} \left( \vec{k}_1 ,\, \tau \right)  {\hat A}_+^{(0)} \left( \vec{k}_2 ,\, \tau \right) \right\rangle' 
\simeq \frac{{\tilde A}_R^2 \left( x \right)}{2 \, k_1} \; ,  
\end{eqnarray}
for the tree level correlator (we recall that the prime  denotes the expectation value without the corresponding $\delta-$function), and, after some algebra, to 
\begin{eqnarray}
&&  \!\!\!\!\!\!\!\!   \!\!\!\!\!\!\!\!   \!\!\!\!\!\!\!\! 
{\cal R}_A = \left\vert \frac{\delta^{(1)} \langle AA \rangle'}{\langle AA \rangle'} \right\vert \simeq  \frac{  \alpha^2\,H_{k_1}^2   }{4 \, f^2 } \Bigg\vert
\int \frac{d^3 q}{\left( 2 \pi \right)^3} \, \frac{ q \,  \left[ 1 - {\hat k}_1 \cdot {\hat q}  \right]^2 }{ \vert {\hat k}_1 + \vec{q} \vert^3  } \, 
 \int_x d x_1 \, \int_{x_1} d x_2 
 \left[ {\tilde A}_R \left( q \, x_2 \right)  {\tilde A}'_R \left( x_2 \right) +  {\tilde A}_R \left( x_2 \right)  {\tilde A}'_R \left( q \, x_2 \right) 
\right] \nonumber\\ 
&&  \!\!\!\!\!\!\!\! \!\!\!\!
\times \Bigg\{ \frac{{\tilde A}_I \left( x \right)}{{\tilde A}_R \left( x \right)} \left[ 
{\tilde A}_R \left( q \, x_1 \right)  {\tilde A}'_R \left( x_1 \right) +  {\tilde A}_R \left( x_1 \right)  {\tilde A}'_R \left( q \, x_1 \right) 
 \right]  -  \left[
{\tilde A}_R \left( q \, x_1 \right)  {\tilde A}'_I \left( x_1 \right) +  {\tilde A}_I \left( x_1 \right)  {\tilde A}'_R \left( q \, x_1 \right) \right] \Bigg\} 
 \nonumber\\ 
&&  \!\!\!\!\!\!\!\! \!\!\!   \times    
\Bigg\{ \vert {\hat k}_1 + \vec{q} \vert \left( x_2 - x_1 \right) \, \cos \left[ \vert {\hat k}_1 + \vec{q} \vert \left( x_1 - x_2 \right) \right] 
 + \left( 1 + \vert {\hat k}_1 + \vec{q} \vert^2 \, x_1 \, x_2 \right) 
\sin \left[ \vert {\hat k}_1 + \vec{q} \vert \left( x_1 - x_2 \right) \right]   \Bigg\}  \Bigg\vert \;, \nonumber\\ 
\label{AAfinwithpre}
\end{eqnarray} 
where prime on the function denotes derivative with respect to its argument.  We stress that this is the leading term for ${\cal R}_A$ in an expansion series in $A_I / A_R$. This is very accurate, since $A_I / A_R = {\rm O } \left( {\rm e}^{-2 \pi \xi} \right) \ll 1$.

In eq. (\ref{AAfinwithpre}) we approximated $H_{k_1  q} \simeq H_{k_1 \vert \vec{q} + {\hat k}_1 \vert } \simeq H_{k_1}$. This is a good approximation since $H$ is slowly evolving, and since most of the support of the integral is at $q = {\rm O } \left( 1 \right)$. We can then evaluate the prefactor using 
\begin{equation} 
\frac{\alpha^2 H_{k_1}^2}{4 f^2}
=\,  \frac{ \xi^2 H_{k_1}^4}{\dot{\phi}^2} = 
4 \pi^2 \xi^2 P_\zeta^{(0)} \left( k_1 \right) \, 
\simeq 8.7 \cdot 10^{-8} \, \xi^2 \,  \left( \frac{k_1}{k_{\rm p}} \right)^{n_s-1} \;. 
\label{pref-AA}
\end{equation} 
where $k_{\rm p}$ indicates the pivot scale (0.05 Mpc$^{-1}$), and $n_s-1$ the scalar spectral tilt.

\section{Computation of the perturbativity limits for  $X=\sigma$ }
\label{app:Xsig}

Here we derive the limits stated in Section \ref{sec:pert-xi-sigma} in the case in which $X=\sigma$. In evaluating the condition ${\cal R}_A \ll 1$, the formal expression (\ref{AAcordom}) still applies, 
where we need to use the unperturbed gauge functions appropriate for this case. In the next subsection we provide an approximate solution (not given in  \cite{Namba:2015gja})  for the unperturbed fluctuations of $\sigma$. In Subsection \ref{app:Xsig-AA2} we then provide the expression for ${\cal R}_A$ for this case. Finally, in Subsection \ref{app:Xsig-ss2} we compute the expression for ${\cal R}_\sigma$.

\subsection{ $\delta \sigma$ vacuum solutions }
\label{app:Xsig-dsig}

As remarked, in the in-in expression (\ref{AAcordom}) the unperturbed operators for the gauge and pseudo-scalar field must be used (where by unperturbed we mean the mode at zeroth order in the interaction  (\ref{Hintconspa}); we remark that this is the same starting point of the computations of ref. \cite{Ferreira:2015omg}).  The unperturbed pseudo-scalar modes satisfy the equation 
\begin{equation}
\delta \ddot{\sigma} + 3 H \delta \dot{\sigma} + \left( \frac{k^2}{a^2} + \partial_\sigma^2 \, V \right) \delta \sigma = 0 \;. 
\label{ds0-eq}
\end{equation} 
Using the background solution (\ref{dotsigma}), we find that the mass term  evolves with time as 
\begin{equation}
 \partial_\sigma^2 \, V  \left( t \right) = - 3 \, \delta \, H^2 \, \frac{1-{\rm e}^{2 \, \delta \, H \, \left( t - t_* \right)} }{ 1+{\rm e}^{2 \, \delta \, H \, \left( t - t_* \right)} } =  - 3 \, \delta \, H^2 \, \frac{1-\left( \frac{\tau_*}{\tau} \right)^{2 \delta} }{1+\left( \frac{\tau_*}{\tau} \right)^{2 \delta} } \;. 
\end{equation} 
In terms of  $y \equiv a \, \delta \sigma $,  $x \equiv - k \tau$, and  $x_* \equiv - k \tau_*$, eq. (\ref{ds0-eq}) rewrites 
\begin{equation} 
\frac{d^2 y}{d x^2} +  \left[ 1  - \frac{2}{x^2}  \left( 1 + 
 \frac{  3 \, \delta  }{2} \, \frac{1-\left( \frac{x_*}{x} \right)^{2 \delta} }{1+\left( \frac{x_*}{x} \right)^{2 \delta} } \right) 
\right] y = 0 \;. 
\end{equation} 

The last fraction in the round parenthesis varies from $\simeq 1$ at $x > x_*$ to $\simeq -1$ at $x < x_*$. 
This transition is very fast for $\delta < 1$. We can obtain an approximate analytic solution for the unperturbed $\delta \sigma$ modes in the limit in which the transition is replaced by a step function 
\begin{equation}
\left\{ 
\begin{array}{l}
\frac{d^2 y}{d x^2} +  \left[ 1  - \frac{2}{x^2}  \left( 1 + 
 \frac{  3 \, \delta  }{2} \right) \right] y = 0 \;\;\;,\;\;\; x > x_* \\ \\ 
\frac{d^2 y}{d x^2} +  \left[ 1  - \frac{2}{x^2}  \left( 1 - 
 \frac{  3 \, \delta  }{2} \right) \right] y = 0 \;\;\;,\;\;\; x < x_* 
 \end{array} \right. 
\end{equation} 
The early time ($x > x_*$) solution that reduces to the adiabatic vacuum mode in the deep UV is 
\begin{equation}
y \left( x \right) =   \frac{1}{2} \, \sqrt{\frac{\pi}{k}} \, \sqrt{x} \, H^{(1)}_{\frac{3}{2} \sqrt{1+ \frac{4 \delta}{3}}} \left( x \right) \;\;\;,\;\;\; x > x_* 
\end{equation} 
We then solve the late time  ($x < x_*$) equation, and impose continuity of $y$ and $\frac{d y}{d x}$ at $x=x_*$, to obtain 
\begin{eqnarray}
&& \delta \sigma =  \frac{H}{k^{3/2}} x^{3/2} \left\{ C_1 \left[ \, \delta, \,  x_* \right]  \, J_{\Delta_-  } \left( x \right) 
+ C_2 \left[ \, \delta, \, x_* \right]   \, Y_{\Delta_-  } \left( x \right) \right\}   \quad , \quad x < x_*  \;, 
\label{dsig0}
\end{eqnarray} 
where 
\begin{eqnarray} 
&& \!\!\!\!\!\!\!\! \!\!\!\!\!\!\!\!  \!\!\!\!\!\!\!\!  \!\!\!\!\!\!\!\! 
C_1 \left[ \delta ,\, x_* \right] = \frac{\pi^{3/2}}{4} \Bigg[ x_* 
\left( Y_{\Delta_-} \left( x_* \right) H^{(1)}_{1+\Delta_+} \left( x_* \right) -  Y_{1+\Delta_-} \left( x_* \right) H^{(1)}_{\Delta_+} \left( x_* \right) \right) \nonumber\\ 
&& \quad\quad  \quad\quad \quad\quad \quad\quad + \left( \Delta_- - \Delta_+ \right) Y_{\Delta_-} \left( x_* \right) H^{(1)}_{\Delta_+} \left( x_* \right) \Bigg] \;, \nonumber\\ 
&& \!\!\!\!\!\!\!\! \!\!\!\!\!\!\!\!  \!\!\!\!\!\!\!\!  \!\!\!\!\!\!\!\! 
C_2  \left[ \delta ,\, x_* \right] = - \frac{\pi^{3/2}}{4} \Bigg[ x_* 
\left( J_{\Delta_-} \left( x_* \right) H^{(1)}_{1+\Delta_+} \left( x_* \right) -  J_{1+\Delta_-} \left( x_* \right) H^{(1)}_{\Delta_+} \left( x_* \right) \right) \nonumber\\ 
&& \quad\quad  \quad\quad \quad\quad \quad\quad + \left( \Delta_- - \Delta_+ \right) J_{\Delta_-} \left( x_* \right) H^{(1)}_{\Delta_+} \left( x_* \right) \Bigg] \;, 
\label{C1C2}
\end{eqnarray} 
and where 

\begin{equation}
 \Delta_+  \equiv  \frac{3}{2} \sqrt{1+ \frac{4 \delta}{3}} \quad , \quad \Delta_-  \equiv  \frac{3}{2} \sqrt{1- \frac{4 \delta}{3}} \;. 
\end{equation}

\subsection{Vector field renormalization}
\label{app:Xsig-AA2}

As discussed in Appendix \ref{app:A-bump}, also in this case the gauge modes satisfy the approximate relation (\ref{Ader}). Using this in eq.  (\ref{AAcordom}), which, as remarked above, continues to hold also in the present case, we can write 
\begin{eqnarray}
&&  \!\!\!\!\!\!\!\! \!\!\!\!\!\!\!\! \!\!\!\!\!\!\!\! \!\!\!\!\!\!\!\!  
 \delta^{(1)} \left\langle {\hat A}_+ \left( \vec{k}_1 ,\, \tau \right)  {\hat A}_+ \left( \vec{k}_2 ,\, \tau \right) \right\rangle'  \simeq 
  \frac{ 4 \,\alpha^2   }{f^2} \,   \int_{\tau_*}^{\tau} \frac{d \tau_1 \, \sqrt{\xi \left( \tau_1 \right)} }{\sqrt{-\tau_1}}  \int_{\tau_*}^{\tau_1} \frac{d \tau_2 \, \sqrt{\xi \left( \tau_2 \right)}}{\sqrt{-\tau_2}}   \nonumber\\ 
&&  \!\!\!\!\!\!\!\! 
\int \frac{d^3 p}{\left( 2 \pi \right)^3} \, p \, k_1 \,  \left[ 1 - \frac{\vec{k}_1 \cdot \vec{p}}{k_1 \, p} \right]^2 
\; A_R \left( k_1 ,\, \tau \right) A_R \left( p ,\, \tau_1 \right) A_R \left( k_1 ,\, \tau_2 \right)  A_R \left( p ,\, \tau_2 \right) \nonumber\\ 
&&  
\times \left[ 
\left( - k_1 + p \right)  A_{R} \left( k_1 ,\, \tau \right) A_{I} \left( k_1 ,\, \tau_1 \right) 
+ \left( \sqrt{ k_1 } + \sqrt{ p } \right)^2  A_{R} \left( k_1 ,\, \tau_1 \right) A_{I} \left( k_1 ,\, \tau \right) 
\right] \nonumber\\ 
&& 
\times 
{\rm Im } \left[   \delta \sigma \left( \vert \vec{k}_1 + \vec{p} \vert ,\, \tau_1 \right) \,  \delta \sigma^* \left( \vert \vec{k}_1 + \vec{p} \vert ,\, \tau_2 \right) \right]   \;. 
\end{eqnarray} 

To evaluate this expression, we again introduce rescaled dimensionless times and momenta 
\begin{equation}
x \equiv - k_1 \, \tau \;\;,\;\; 
x_1 \equiv - k_1 \, \tau_1 \;\;,\;\; 
x_2 \equiv - k_1 \, \tau_2 \;\;,\;\; 
\vec{q} \equiv \frac{\vec{p}}{k_1} \;\;,\;\; 
\label{dimensionless2}
\end{equation} 
and we rescale the pseudo-scalar and vector functions according to 
\begin{eqnarray}
{\tilde A}_R  \left( x \right) &\equiv& \left( \frac{8 k_1 \xi \left( \tau \right)}{-\tau} \right)^{1/4} \, A_R \left( \tau ,\, k_1 \right) \simeq N \left[ \xi_* ,\, x_* ,\, \delta \right] \, {\rm exp} \left[ - \frac{4 \xi_*^{1/2}}{1+\delta} \, \left( \frac{x}{x_*} \right)^{\delta/2} \, x^{1/2} \right] \,, \nonumber\\ 
{\tilde A}_I  \left( x \right) &\equiv& \left( \frac{8 k_1 \xi \left( \tau \right)}{-\tau} \right)^{1/4} \, A_I \left( \tau ,\, k_1 \right) \simeq \frac{1}{2 N \left[ \xi_* ,\, x_* ,\, \delta \right]} \, {\rm exp} \left[  \frac{4 \xi_*^{1/2}}{1+\delta} \, \left( \frac{x}{x_*} \right)^{\delta/2} \, x^{1/2} \right] \,, \nonumber\\ 
\delta {\tilde \sigma} \left( x \right) &\equiv& \frac{k^{3/2}}{H} \, \delta \sigma  \left( \tau ,\, k_1 \right) \simeq x^{3/2}
\, \left\{ C_1 \left[ \, \delta, \,  x_* \right]  \, J_{\Delta_-  } \left( x \right) + C_2 \left[ \, \delta, \, x_* \right]   \, Y_{\Delta_-  } \left( x \right) \right\}  \,. 
\label{A-ds-til}
\end{eqnarray} 
(We only specify the $x-$dependence of the rescaled quantities; their dependence on $x_* ,\, \xi_* ,\, \delta$ is left understood). In this notation, we can rewrite 
\begin{eqnarray}
&&  \!\!\!\!\!\!\!\! 
{\cal R}_A = \frac{ \delta^{(1)} \left\langle {\hat A}_+ \left( \vec{k}_1 ,\, \tau \right)  {\hat A}_+ \left( - \vec{k}_1 ,\, \tau \right) \right\rangle' }{  \left\langle {\hat A}_+ \left( \vec{k}_1 ,\, \tau \right)  {\hat A}_+ \left( - \vec{k}_1 ,\, \tau \right) \right\rangle' } \simeq 
  \frac{  \,\alpha^2  H^2 }{f^2} \,   \int_{x}^{x_*}  d x_1  \int_{x_1}^{x_*} d x_2    \nonumber\\ 
&&  \times 
  \int \frac{d^3 q}{\left( 2 \pi \right)^3}   \frac{q^{1/2}  \left[ 1 - {\hat k}_1 \cdot {\hat q}  \right]^2 }{2 \vert {\hat k}_1 + \vec{q} \vert^3 } 
\;   {\tilde A}_R \left( q \, x_1 \right)  \; {\rm Im } \left[   \delta {\tilde \sigma } \left(  \vert {\hat k}_1 + \vec{q} \vert \, x_1 \right) \,  \delta {\tilde \sigma}^* \left(  \vert {\hat k}_1 + \vec{q} \vert \, x_2 \right) \right]  \nonumber\\ 
&&  \quad\quad 
\times  \left[ 
\left( - 1 + q \right)   {\tilde A}_{I} \left( x_1  \right) 
+ \left( 1 + \sqrt{ q } \right)^2 
\frac{  {\tilde A}_{R} \left( x_1 \right) }{  {\tilde A}_R \left( x  \right) } \,   
  {\tilde A}_{I} \left( x   \right) 
\right]   \,  {\tilde A}_R \left(    x_2 \right)   {\tilde A}_R \left(  q \, x_2 \right) \;. 
\end{eqnarray} 

Let us rewrite the prefactor $  \frac{  \,\alpha^2  H^2 }{f^2} $. In succession, we use the definition (\ref{genmodeqn}) of $\xi$, evaluated at $\tau = \tau_*$, to eliminate $\frac{\alpha}{f}$; then, the relation (\ref{eps-sig}) to eliminate $\dot{\sigma}_*$; finally, the relation (\ref{P-zeta}) to eliminate $H$. We find 
\begin{equation}
\frac{\alpha^2 \, H^2}{f^2} \simeq 16 \pi^2 {\cal P}_\zeta^{(0)} \, \xi_*^2 \, \frac{\epsilon_\phi} {\epsilon_{\sigma,*}}
\simeq 3.5 \cdot 10^{-7} \, \xi_*^2 \,  \frac{\epsilon_\phi}{\epsilon_{\sigma,*}}  \;. 
\label{resc-aHf} 
\end{equation} 

Combining the last two expressions, we can write 
\begin{eqnarray}
{\cal R}_A &\equiv& R_A \left[ x_* ,\, \xi_* ,\, \delta ,\, x \right] \,  \frac{\epsilon_\phi}{\epsilon_{\sigma,*}}  \;, \nonumber\\ 
R_A &\simeq&  3.5 \cdot 10^{-7} \, \xi_*^2 \,   \int_{x}^{x_*} d x_1   \int_{x_1}^{x_*} d x_2   \int \frac{d^3 q}{\left( 2 \pi \right)^3}   \frac{ q^{1/2} \left[ 1 - {\hat k}_1 \cdot {\hat q}  \right]^2 }{2 \vert {\hat k}_1 + \vec{q} \vert^3 }  \nonumber\\ 
&&  \times 
\;  {\tilde A}_R \left( q \, x_1 \right) 
  \, {\tilde A}_R \left(   x_2 \right)  {\tilde A}_R \left(  q ,\, x_2 \right) 
 \; {\rm Im } \left[   \delta {\tilde \sigma } \left(  \vert {\hat k}_1 + \vec{q} \vert \, x_1 \right) \,  \delta {\tilde \sigma}^* \left(  \vert {\hat k}_1 + \vec{q} \vert \, x_2 \right) \right]  \nonumber\\ 
&&  \quad\quad 
\times  \left[ 
\left( - 1 + q \right)  {\tilde A}_{I} \left( x_1  \right) 
+ \left( 1 + \sqrt{ q } \right)^2 
\frac{ {\tilde A}_{R} \left( x_1 \right) }{ {\tilde A}_R \left(  x \right) } \,   
 {\tilde A}_{I} \left( x \right) 
\right]  \;,  
\label{Xsig-RA}
\end{eqnarray} 
were we remarked the parametric dependence of $R_A$, and where we recall that the gauge and pseudo-scalar mode functions are given in eq. (\ref{A-ds-til}). We note that, since $x_* = - k \, \tau_*$, and $x= - k \, \tau$, we can express the functional dependence of $R_A$ also as 
$ R_A \left[ x_* ,\, \xi_* ,\, \delta ,\, \frac{\tau}{\tau_*} \right]$. This is the form that appears in the main text.

\subsection{Pseudo-scalar field renormalization}
\label{app:Xsig-ss2}

We now evaluate the condition (\ref{RX}) for $R_\sigma$. From eqs.  (\ref{Hintconspa}) and  (\ref{XX-inin})  we obtain 
\begin{eqnarray}
&& \!\!\!\!\!\!\!\!  \!\!\!\!\!\!\!\!  \!\!\!\!\!\!\!\! 
 \langle \delta \hat{\sigma}^{(1)} (\tau, \vec{k_1})  \delta \hat{\sigma}^{(1)} (\tau, \vec{k_2})  \rangle  \, = \,- \, \frac{\alpha^2}{f^2} \int  \frac{ d^3 k\, d^3 p  \, d^3 k' \, d^3 p'}  {(2\pi)^{3}} \,    \int^{\tau}  d\tau_1 \,\,  \int^{\tau_1}  d\tau_2  \nonumber\\ 
&& \times \vert \,   \vec{k} + \vec{p}      \, \vert   \, \vert \,   \vec{k}' + \vec{p}'      \, \vert  \, 
 \left( \vec{\epsilon}^{\,+}  (\vec p) \cdot \vec{\epsilon}^{\, +} (  -  \vec {k} -  \vec {p})  \right)      \left( \vec{\epsilon}^{\,+}  (\vec p\,') \cdot \vec{\epsilon}^{\, +} (  -  \vec {k}\,' -  \vec {p}\,' )  \right)  \nonumber\\ 
&&   \!\!\!\!\!\!\!\!  \times \left\langle \,\, \left[ \left[  \delta \hat{\sigma}  (\tau, \vec{k_1})  \, \delta \hat{\sigma}  (\tau, \vec{k_2}) \, , \, \delta \hat{\sigma}  (\tau_1, \vec{k})\, \hat{A' } (\tau_1, \vec{p}) \,  \hat{A} (\tau_1,  - \vec{k} - \vec{p})  \right]  ,  \, \delta \hat{\sigma}  (\tau_2, \vec{k} \,' ) \,  \hat{A' } (\tau_2, \vec{p} \, ' )  \,  \hat{A} (\tau_2,  - \vec{k}\,' - \vec{p}\,')  \right] \, \right\rangle \,. \nonumber\\
\end{eqnarray}
We decompose this expression in several terms, each containing propagators of two different fields.  Each commutator of two fields is proportional to the imaginary part of the product of the wave functions of the two fields. The real $A_R$ and imaginary $A_I$ parts of the gauge field amplitude are given by   eqs. (\ref{varysig-AR}) and (\ref{varysig-AI}), respectively. With the phase convention that we have chosen, only the real part is amplified. 
For this reason the expression that we have just written is strongly dominated by the terms that contain the largest powers of $A_R$. These are the terms in which the vector fields are not commuted over, and are therefore proportional to $A^2 \, A^{'2} \simeq A_R^2 \, A_R^{'2}$. These terms give 
 \begin{eqnarray}
\!\!\!\!\!\!\!\!\!\!\!\!\!\!\!\!\!\!\!\!\!\!\!\! \!\!\!    \left\langle \delta \hat{\sigma}^{(1)} (\tau, \vec{k_1})  \delta \hat{\sigma}^{(1)} (\tau, \vec{k_2})  \right\rangle & & \, \simeq \, \,\delta^3(\vec k_1 + \vec k_2) \frac{4 \alpha^2}{f^2} \int  \frac{ \, d^3 p }  {(2\pi)^{3}}  \, \,\, \left\vert  \vec{\epsilon}^{\,+}  (\vec p) \cdot \vec{\epsilon}^{\, +} (  \vec {k_1} -  \vec {p})   \right\vert^2   \nonumber\\ \nonumber\\ 
 &&    \!\!\!\!\!\!\!\!\!\!\!\!\!\!\!\!\!\!\!\!\!\!\!\!\!\!\!\!\!\!\!\!\!\!\!\!\!\!\!\!\!\!\!\!\!\!\!\!\!\!\!\!\!\!\!\!\!\!\!   \int_{\tau_*}^{\tau}  d\tau_1  \,\,  \int_{\tau_*}^{\tau_1}  d\tau_2 \,\,\, 
\bigg\{  \vert \,   \vec{k_1} - \vec{p}    \, \vert^2  A_R' (\tau_1, p) \, A_R' (\tau_2, p\, ) \, A_R (\tau_1, \vert  \vec{k_1}  - \vec{p} \vert \,) \, A_R  (\tau_2,   \vert \vec{k_1}\,  - \vec{p}\, \vert \,) \,\, \nonumber\\   \nonumber\\ 
 && \!\!\!\!\!\!\!\!\!\!\!\!\!\!\!\!\!\!\!\!\!\!\!\! \!\!\!   +\,\, p\,  \vert \,   \vec{k_1} - \vec{p}    \, \vert  A_R' (\tau_1, p) \,  A_R (\tau_2, p \,)\,  A_R (\tau_1, \vert \vec{k_1}  - \vec{p}  \vert \,) \, A_R' (\tau_2,  \vert \vec{k_1} - \vec{p} \vert \, )  \bigg\}  \nonumber\\   \nonumber\\ 
&&\!\!\!\!\!\!\!\!\!\!\!\!\!\!\!\!\!\!\!\!\!\!\!\! \!\!\!  \times  \left(  {\rm Im} \left[ \delta \sigma(\tau, k_1)\, \delta \sigma ^* (\tau_1, k_1)  \right] \,\,  {\rm Im} \left[ \delta \sigma(\tau, k_1)\, \delta \sigma ^* (\tau_2, k_1)  \right]  \,\, + \vec{k_1} \leftrightarrow k_2  \right) \,.  
\end{eqnarray}

Working out the product of the polarization operators, and symmetrizing the time integration, we obtain, after some algebra, 
\begin{eqnarray} 
&& \!\!\!\!\!\!\!\! \!\!\!\!\!\!\!\! 
\left\langle  \delta \hat{\sigma}^{(1)} \left( \tau, \vec{k}_1 \right) \,    \delta \hat{\sigma}^{(1)} \left( \tau, \vec{k}_2 \right) \right\rangle'  \simeq 
 \frac{ \alpha^2}{2 f^2} \int  \frac{ \, d^3 p }  {\left( 2\pi \right)^{3}}  \, 
\frac{\left( p^{1/2} + \vert \vec{k}_1 - \vec{p} \vert^{1/2} \right)^2 \left[ \left( p + \vert \vec{k}_1 - \vec{p} \vert \right)^2 - k_1^2 \right]^2}{4 \, p \, \vert \vec{k}_1 - \vec{p} \vert}  \nonumber\\ 
&& \quad 
\times \left[ \int_{\tau_*}^\tau d \tau_1  \sqrt{\frac{2 \xi \left( \tau_1 \right)}{-\tau_1}}  {\rm Im} \left[ \, \delta \sigma \left(\tau, k_1 \right) \, \delta \sigma ^* \left( \tau_1, k_1 \right)\, \right]  A_R \left( \tau_1 ,\, p \right)  A_R \left( \tau_1 ,\, \vert \vec{k}_1 - \vec{p} \vert  \right) 
 \right]^2 \;. 
\end{eqnarray}
In terms of the dimensionless quantities introduced in (\ref{dimensionless2}) and (\ref{A-ds-til}), and recalling the definition of the power spectrum given after eq. (\ref{dXdX}), we can then write  
\begin{eqnarray}
&& \!\!\!\!\!\!\!\! 
P_{\delta \sigma}^{(1)} \left( k ,\, \tau \right)    \simeq 
\frac{\alpha^2 H^4}{64 \pi^2 f^2} \,  \int  \frac{ \, d^3 q }  {\left( 2\pi \right)^{3}}  \, \frac{\left( q^{1/2} + \vert {\hat k} - \vec{q} \vert^{1/2} \right)^2 \left[ \left( q + \vert {\hat k} - \vec{q} \vert \right)^2 - 1 \right]^2}{ q^{3/2} \, \vert {\hat k} - \vec{q} \vert^{3/2}}  \nonumber\\ 
 && \quad\quad \quad\quad  \quad\quad 
  \times \left[ \int_x^{x_*} d x_1 \,  {\rm Im} \left[ \, \delta {\tilde \sigma} \left( x \right) \, \delta {\tilde \sigma}^* \left( x_1 \right)\, \right]  {\tilde A}_R \left( q \, x_1 \right)  {\tilde A}_R \left(  \vert {\hat k} - \vec{q} \vert \, x_1 \right) \right]^2 \;, 
\label{Pdsds}
\end{eqnarray}  
where we have relabeled as $\vec{k}_1 \rightarrow \vec{k}$ the momentum of the generic mode that we are considering. 

\begin{figure}[ht!]
\centerline{
\includegraphics[width=0.5\textwidth,angle=0]{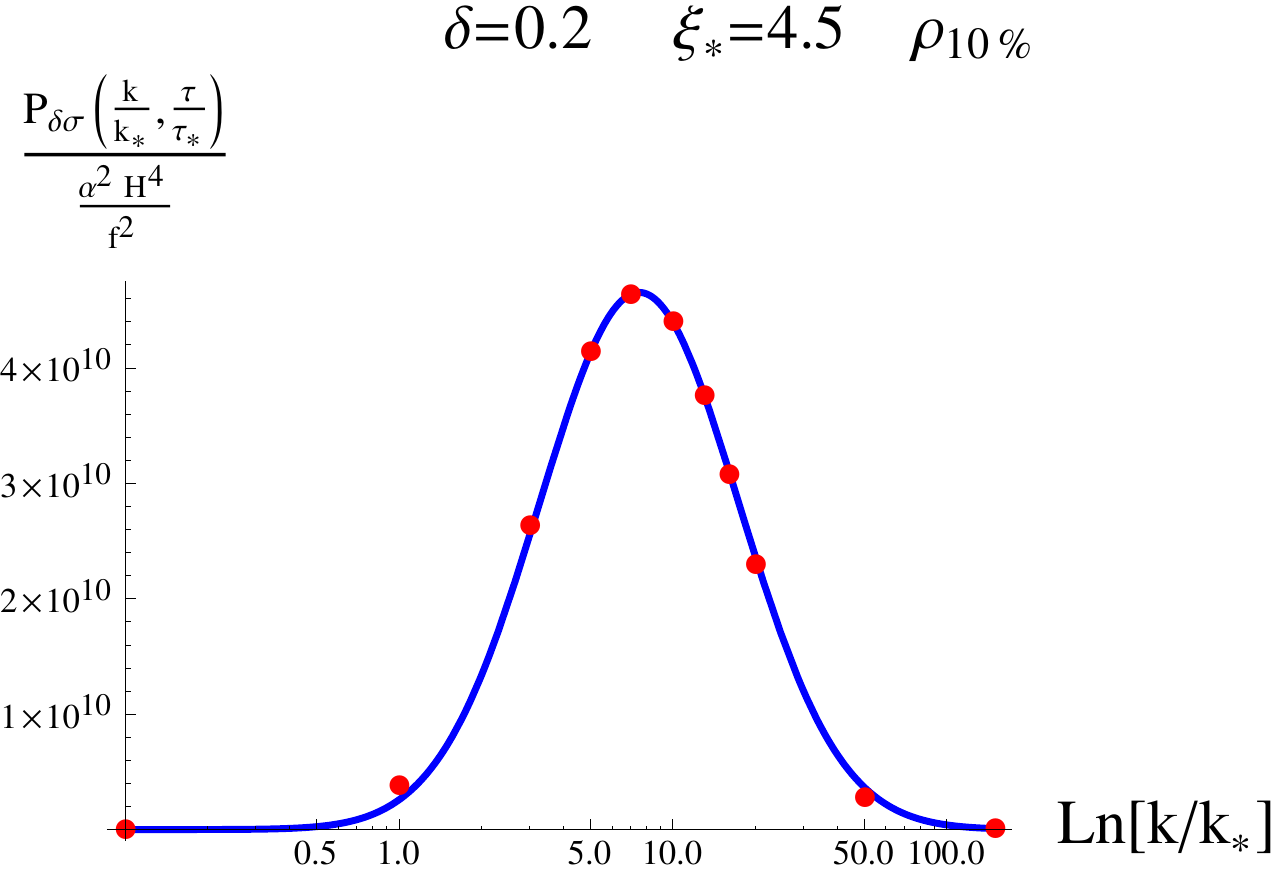}
\includegraphics[width=0.5\textwidth,angle=0]{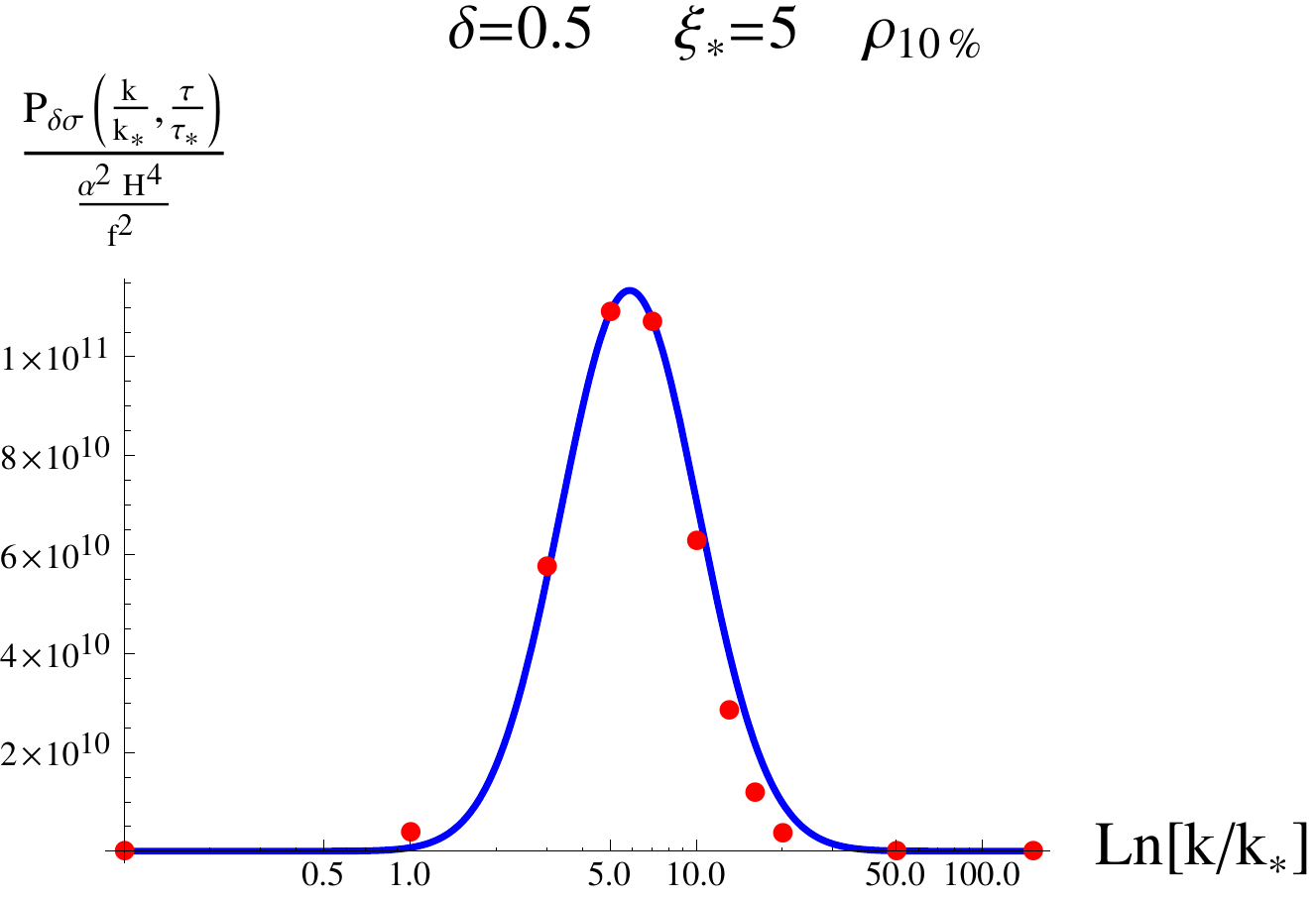}
}
\caption{Power spectrum of the pseudo-scalar $\delta \sigma$ sourced by the gauge field, for two specific values of $\delta$ and $\xi_*$. The time $\tau/\tau_*$ is chosen to be the moment at which the total energy density in the gauge field has decreased to $10 \%$ than the value it had at its peak.
The red dots are the values obtained from a numerical evolution of eq. (\ref{Pdsds}). The blue solid line is the fitting function (\ref{Pdsds-param}). 
} 
\label{fig:fit-ds}
\end{figure}

As can be seen from the rhs of  (\ref{Pdsds}), the power spectrum dependence on the momentum and on time can be written as a dependence on the dimensionless quantities $-k \tau_*$ and $\frac{\tau}{\tau_*}$. We therefore write the power spectrum as $P_{\delta \sigma}^{(1)} \left( x_* = - k \tau_* ,\, \frac{\tau}{\tau_*} \right)$.  In \cite{Namba:2015gja}, it was shown that the power spectra of the sourced GW and inflation perturbations in this model are well fitted by a Gaussian peak. Not surprisingly, the same is true for the power spectrum (\ref{Pdsds}) (the reason is that the amplitude of the sourcing gauge field is well fitted by this parametrization, cf. eq. (\ref{N-fit-app})). Indeed, we evaluated the expression (\ref{Pdsds}) numerically, and we found that it is well parametrized by  
\begin{equation}
P_{\delta \sigma}^{(1)}  \left(  x_* ,\, \frac{\tau}{\tau_*} \right) \simeq \frac{\alpha^2 H^4}{f^2} \, {\cal M} \left[ \xi_* ,\, \delta ,\, \frac{\tau}{\tau_*} \right] \, {\rm exp} \left[ - \frac{\ln^2 \left( \frac{x_*}{  x_{c,{\cal M}} \left[ \xi_* ,\, \delta ,\, \frac{\tau}{\tau_*} \right]} \right)}{2 \sigma_{\cal M}^2 \left[ \xi_* ,\, \delta ,\, \frac{\tau}{\tau_*} \right]} \right] \;. 
\label{Pdsds-param}
\end{equation} 
As an example of the goodness of this fit, in Figure \ref{fig:fit-ds} we show the power spectrum for two specific values of $\delta$ and $\xi_*$, and for a specific time $\tau$. Equally good fits are obtained in the other cases we have studied. 

The amplitude ${\cal M}$, the central position $x_c$, and the width of the peak $\sigma_{\cal M}$, evolve with time, as they parametrize the growth of the power spectrum sourced by the gauge fields $A_+$.~\footnote{The three parameters ${\cal M} ,\, x_{c,{\cal M}}  ,\, \sigma_{\cal M},\, $ also depend on the two parameters $\xi_*$ and $\delta$ that control the motion of the field $\sigma$.}  Since the amplitude of $A_+$ is negligible at $\tau < \tau_*$, all our results, and in particular the expression (\ref{Pdsds-param}), are valid for  $\frac{\tau}{\tau_*} < 1$, while we can simply set $P_{\delta \sigma}^{(1)} \left( x_* ,\, \frac{\tau}{\tau_*} \right) \simeq 0$ at  $\frac{\tau}{\tau_*} > 1$. The power spectrum then experiences a fast growth at  $\vert \tau  \vert \la \vert \tau_* \vert$, and it eventually saturates to a constant value at  $\vert \tau  \vert \ll \vert \tau_* \vert$. The parameter $x_{c,{\cal M}} = {\rm O } \left( 1 \right)$. So, expression (\ref{Pdsds-param}) has a peak at $x_* = {\rm O } \left( 1 \right)$, namely for the modes that left the horizon close to the $\dot{\sigma}$ was maximum.  

Using the fit (\ref{Pdsds-param}), we can perform the integral (\ref{dXdX}) analytically. We can then express the condition (\ref{RX}) as 
\begin{equation}
{\cal R}_\sigma =  \frac{\sqrt{\int d \ln x_* \, P_{\delta \sigma}^{(1)} } }{f}  = \frac{\alpha \, H^2}{f^2}  \, \left( 2 \pi \right)^{1/4} \, \sqrt{ {\cal M}  \left[ \xi_* ,\, \delta ,\, \frac{\tau}{\tau_*} \right] \,  \sigma_{\cal M}  \left[ \xi_* ,\, \delta ,\, \frac{\tau}{\tau_*} \right] } \ll1 \,. 
\end{equation} 

Using the relation (\ref{resc-aHf}), as well as $\alpha =  \frac{2 \xi_*}{\delta}$ (as can be obtained from eqs. (\ref{genmodeqn}) and (\ref{dotsigma})), we can rewrite this as 
\begin{eqnarray}
{\cal R}_{\sigma}  &\equiv& \frac{\epsilon_\phi}{\epsilon_{\sigma_*}} \, {\rm R}_\sigma \left[ \xi_* ,\, \delta ,\, \frac{\tau}{\tau_*} \right] \ll 1 \;, \nonumber\\ 
{\rm R}_\sigma &\simeq& 2.8 \cdot 10^{-7} \, \delta \, \xi_* \, \sqrt{ {\cal M}  \left[ \xi_* ,\, \delta ,\, \frac{\tau}{\tau_*} \right] \,  \sigma_{\cal M}  \left[ \xi_* ,\, \delta ,\, \frac{\tau}{\tau_*} \right]  } \;. 
\label{Xsig-Rsig}
\end{eqnarray} 
We recall that the parameters ${\cal M}$ and $\sigma_{\cal M}$ parametrize, respectively, the amplitude and the width of the peak in the power spectrum of sourced $\delta \sigma$ modes, see eq. (\ref{Pdsds-param}).

\section{Comparison with previous results} 
\label{app:comparison}

\begin{figure}[ht!]
\centerline{
\includegraphics[width=0.5\textwidth,angle=0]{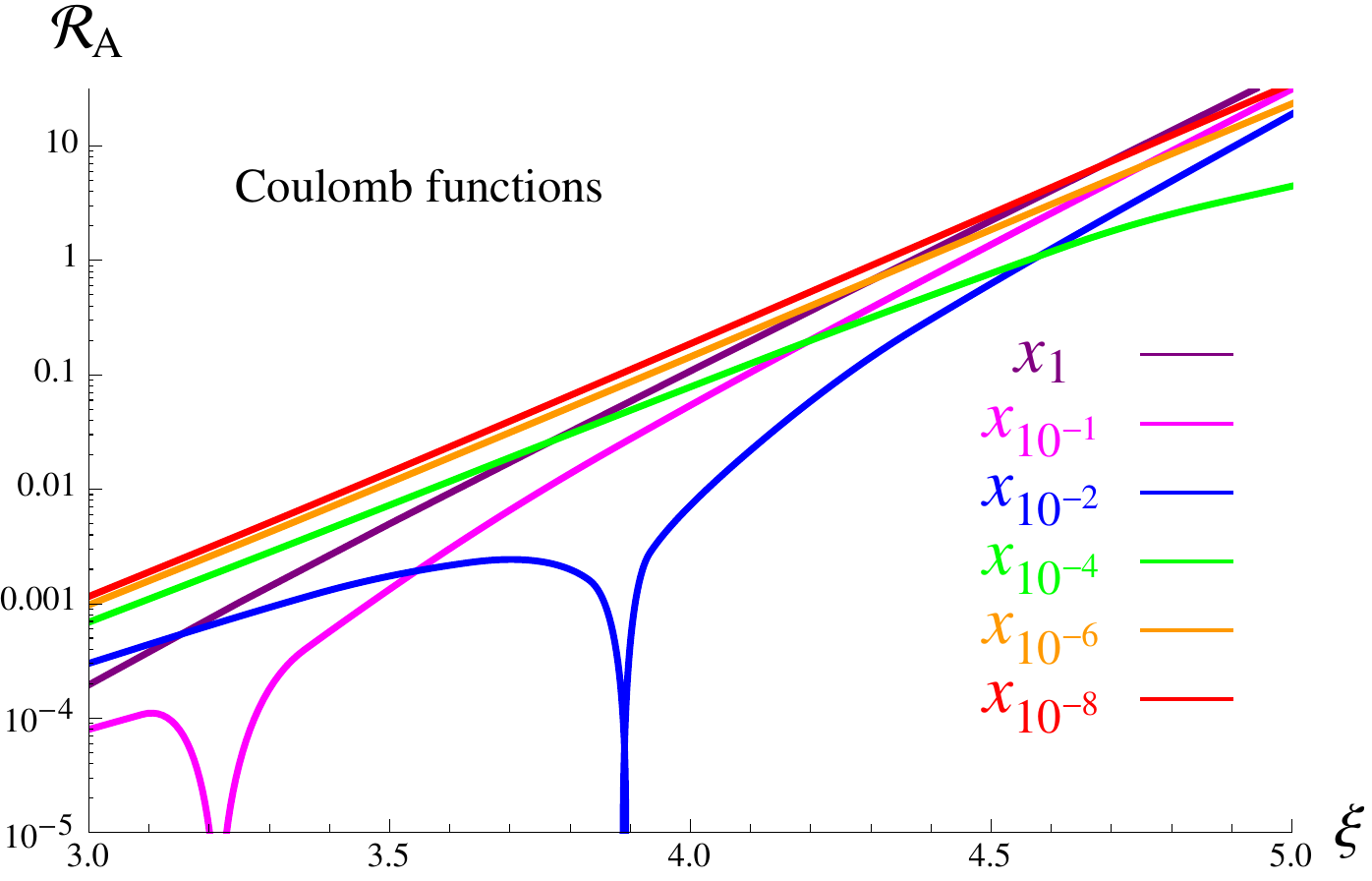}
}
\caption{Perturbativity criterion for $X=\phi$. This is nearly the same quantity shown in Figure \ref{fig:R-constant}, with the only difference that 
we are here using the expression (\ref{colsol}) for the $A+$ modes. From this figure, we obtain $\xi \la 4.4$ for CMB-scale modes, which is close to the limit ($\xi \la 4.6$) obtained in the main text.  
} 
\label{fig:Ra-coulomb}
\end{figure}

In Section \ref{sec:significance} we found that, for the $X = \phi$ case, perturbativity at the CMB scales is ensured for $\xi \la 4.6$. This is the case also studied in ref.  \cite{Ferreira:2015omg}, so we can compare our result with theirs.  Ref. \cite{Ferreira:2015omg}  first provides the analytic estimate $\xi \la 3.5$. It then computes ${\cal R}_A$ numerically, and it reports the result in Figure 5.  Their numerical computation gives $\xi \la 3.7$. Given that ${\cal R}_A \propto {\rm e}^{2 \pi \xi}$, we see that such a limit is significantly stronger than ours.  In this Appendix we discuss the reasons of this discrepancy.

We first of all note that the analytic result of  \cite{Ferreira:2015omg} is based on their analytic expression ${\cal R}_A \simeq \frac{\xi^2}{100} {\rm e}^{2 \pi \xi} {\cal P}_\zeta^{(0)}$.  It is claimed in   \cite{Ferreira:2015omg} that the $100$ factor is a  ``loop factor'', namely this analytic estimate  assumes that the loop integral (see the left diagram of our Figure \ref{fig:loop}) gives a $\sim 10^{-2}$ suppression to the result. No justification for this is given in  \cite{Ferreira:2015omg}. In fact, the results given in the literature for these kinds of loops are typically $\sim 10^{-4}$, about two orders of magnitude smaller than the estimate of  \cite{Ferreira:2015omg}, see for example eq. (8) of \cite{Barnaby:2010vf} (the  $\sim 10^{-4}$ figure agrees with our numerical result). Let us therefore disregard the analytic estimate of   \cite{Ferreira:2015omg}, and move to the comparison between the numerical results. 

There are $4$ different aspects that contribute to the discrepancy between the numerical results: (i) a different formal expression is integrated numerically; (ii) different approximations are used  for the mode functions $A_+$; (iii) the quantity ${\cal R}_A$ is evaluated at different times. Once we account for all these effects, we still find a significant discrepancy. We therefore conclude that, in addition to this, (iv) the integration performed in  \cite{Ferreira:2015omg} has some numerical inaccuracy.

Let us address (i): let us compare our expression (\ref{AAcordom}) for the numerator of ${\cal R}_A$ with their expression (B.4). One difference between the two expressions is a $\left( 2 \pi \right)^3$ factor. This factor is just due to different $2 \pi$ convention, and it cancels in the ratio ${\cal R}_A$. A difference that instead remains is the fact that \cite{Ferreira:2015omg} has $\left( 1 + \cos \theta \right)^2$ (where $\theta$ is the angle between the external momentum and the loop momentum), while we have instead  $\left( 1 - \cos \theta \right)^2$. Our expression originates from the identity $\left\vert \left[  \vec{\epsilon}_+ \left( \hat k \right) \times \vec{\epsilon}_+ \left( \hat q \right) \right] \cdot {\hat q} \right\vert^2 = \frac{\left( 1 - {\hat k} \cdot {\hat q} \right)^2}{4}$, while ref.  \cite{Ferreira:2015omg}  has an incorrect plus sign at the right hand side of this identity. We verified that the change of this sign affects the result only at the $\sim 10 \%$ level, and therefore it is not the main origin of the discrepancy. 

Let us address (ii):  in Section \ref{sec:consxiAsol}, three approximate expressions for $A_+$ are given. As we remarked there, the expression (\ref{A-simple}) is used in our numerical computations. The reason for this is that most of the computations of the phenomenology associated with this mechanism uses this expression. Since we want to verify whether those results are under perturbative control, we should use the same expression in the computation of ${\cal R}_A$. On the other hand, ref.~\cite{Ferreira:2015omg} uses the expression (\ref{colsol}). To compare with the results of that paper, we performed again our evaluation of ${\cal R}_A$ (using the correct  $\left( 1 - \cos \theta \right)^2$ factor) using the Coulomb function  (\ref{colsol}), and we present the result in Figure  \ref{fig:Ra-coulomb} (we use the cut-off $k < 2 \, \xi$ in this computation). If we evaluate ${\cal R}_A$ when the physical energy density in the mode has its maximum value ($x_1$ in the figure), we find the limit $\xi \la 4.4$. This value is not very far from the $\xi \la 4.6$ result that we obtained with the exponential functions (\ref{A-simple}), which is due to the fact that both expressions are good expressions for $A_+$ in the region where the gauge field is relevant. Therefore, it still does not explain the stronger result claimed by  \cite{Ferreira:2015omg}. 

Let us address (iii). In figure  \ref{fig:Ra-coulomb} we showed several lines, corresponding to the same momentum $k$, but to a different time $x_N = - k \tau_N$. We recall that the times are defined as follows: $\tau_{0.1}$ is the time at which the physical energy density in that mode has decreased to $10\%$ of the maximum value it had (see Figure \ref{fig:rhoti}). Ref.  \cite{Ferreira:2015omg} does not explicitly indicates at which time the numerical result shown in their Figure 5 is evaluated, but states that the result ``is the same for a wide range of values [times]''. We see from our Figure that the result becomes time-independent only at the latest times shown. At this time the wavelength of the mode has become much greater than the horizon, and everything, including ${\cal R}_A$, freezes. However, we stress that at this late time the physical energy density of the gauge mode has decreased to a very small value, and the mode is no longer contributing to any observable. So, this is not where the ${\cal R}_A$ should be evaluated. If we do so, we find the slightly more stringent limit $\xi \la 4.3$. 

To conclude, even when we changed our integrand and our method to reproduce that of ref.  \cite{Ferreira:2015omg}, we still do not get a limit as strong as theirs. This leads us to think that the result of \cite{Ferreira:2015omg} is affected by some numerical inaccuracy.


\begin{thebibliography}{99}


\bibitem{Kamionkowski:2015yta} 
  M.~Kamionkowski and E.~D.~Kovetz,
  arXiv:1510.06042 [astro-ph.CO].


\bibitem{Cook:2011hg} 
  J.~L.~Cook and L.~Sorbo,
  Phys.\ Rev.\ D {\bf 85}, 023534 (2012)
  [Phys.\ Rev.\ D {\bf 86}, 069901 (2012)]
  [arXiv:1109.0022 [astro-ph.CO]].

\bibitem{Senatore:2011sp} 
  L.~Senatore, E.~Silverstein and M.~Zaldarriaga,
  JCAP {\bf 1408}, 016 (2014)
  [arXiv:1109.0542 [hep-th]].



\bibitem{Barnaby:2012xt} 
  N.~Barnaby, J.~Moxon, R.~Namba, M.~Peloso, G.~Shiu and P.~Zhou,
  Phys.\ Rev.\ D {\bf 86}, 103508 (2012)
  [arXiv:1206.6117 [astro-ph.CO]].


\bibitem{Barnaby:2010vf} 
  N.~Barnaby and M.~Peloso,
  Phys.\ Rev.\ Lett.\  {\bf 106}, 181301 (2011)
  [arXiv:1011.1500 [hep-ph]].

\bibitem{Mirbabayi:2014jqa} 
  M.~Mirbabayi, L.~Senatore, E.~Silverstein and M.~Zaldarriaga,
  Phys.\ Rev.\ D {\bf 91}, 063518 (2015)
  doi:10.1103/PhysRevD.91.063518
  [arXiv:1412.0665 [hep-th]].

\bibitem{spectator} 
  M.~Biagetti, M.~Fasiello and A.~Riotto,
  Phys.\ Rev.\ D {\bf 88}, 103518 (2013)
  doi:10.1103/PhysRevD.88.103518
  [arXiv:1305.7241 [astro-ph.CO]]; 
%
  M.~Biagetti, E.~Dimastrogiovanni, M.~Fasiello and M.~Peloso,
  JCAP {\bf 1504}, 011 (2015)
  doi:10.1088/1475-7516/2015/04/011
  [arXiv:1411.3029 [astro-ph.CO]]; 
%
  T.~Fujita, J.~Yokoyama and S.~Yokoyama,
  PTEP {\bf 2015}, 043E01 (2015)
  doi:10.1093/ptep/ptv037
  [arXiv:1411.3658 [astro-ph.CO]]; 
%
  K.~Choi, K.~Y.~Choi, H.~Kim and C.~S.~Shin,
  JCAP {\bf 1510}, no. 10, 046 (2015)
  doi:10.1088/1475-7516/2015/10/046
  [arXiv:1507.04977 [astro-ph.CO]].


\bibitem{Carney:2012pk} 
  D.~Carney, W.~Fischler, E.~D.~Kovetz, D.~Lorshbough and S.~Paban,
  JHEP {\bf 1211}, 042 (2012)
  [arXiv:1209.3848 [hep-th]].


\bibitem{Cannone:2014uqa} 
  D.~Cannone, G.~Tasinato and D.~Wands,
  JCAP {\bf 1501}, no. 01, 029 (2015)
  [arXiv:1409.6568 [astro-ph.CO]].

\bibitem{Cannone:2015rra} 
  D.~Cannone, J.~O.~Gong and G.~Tasinato,
  arXiv:1505.05773 [hep-th].


\bibitem{Bartolo:2015qvr} 
  N.~Bartolo, D.~Cannone, A.~Ricciardone and G.~Tasinato,
  JCAP {\bf 1603}, no. 03, 044 (2016)
  doi:10.1088/1475-7516/2016/03/044
  [arXiv:1511.07414 [astro-ph.CO]].

\bibitem{Cai:2015dta} 
  Y.~Cai, Y.~T.~Wang and Y.~S.~Piao,
  Phys.\ Rev.\ D {\bf 91}, 103001 (2015)
  [arXiv:1501.06345 [astro-ph.CO]].


\bibitem{Cai:2016ldn} 
  Y.~Cai, Y.~T.~Wang and Y.~S.~Piao,
  arXiv:1602.05431 [astro-ph.CO].


\bibitem{non-abelian}
  E.~Dimastrogiovanni and M.~Peloso,
  Phys.\ Rev.\ D {\bf 87}, no. 10, 103501 (2013)
  doi:10.1103/PhysRevD.87.103501
  [arXiv:1212.5184 [astro-ph.CO]]; 
%
  P.~Adshead, E.~Martinec and M.~Wyman,
  Phys.\ Rev.\ D {\bf 88}, no. 2, 021302 (2013)
  doi:10.1103/PhysRevD.88.021302
  [arXiv:1301.2598 [hep-th]];  
%
  R.~Namba, E.~Dimastrogiovanni and M.~Peloso,
  JCAP {\bf 1311}, 045 (2013)
  [arXiv:1308.1366 [astro-ph.CO]];
%
  I.~Obata, T.~Miura and J.~Soda,
  Phys.\ Rev.\ D {\bf 92}, no. 6, 063516 (2015)
  doi:10.1103/PhysRevD.92.063516
  [arXiv:1412.7620 [hep-ph]];
%
  I.~Obata and J.~Soda,
  arXiv:1602.06024 [hep-th];
%
  A.~Maleknejad,
  arXiv:1604.03327 [hep-ph].
%


\bibitem{Dimastrogiovanni:2014ina} 
  E.~Dimastrogiovanni, M.~Fasiello, D.~Jeong and M.~Kamionkowski,
  JCAP {\bf 1412}, 050 (2014)
  doi:10.1088/1475-7516/2014/12/050
  [arXiv:1407.8204 [astro-ph.CO]].


\bibitem{preheating}
  S.~Y.~Khlebnikov and I.~I.~Tkachev,
  Phys.\ Rev.\ D {\bf 56}, 653 (1997)
  [hep-ph/9701423];
%
  R.~Easther and E.~A.~Lim,
  JCAP {\bf 0604}, 010 (2006)
  [astro-ph/0601617];
%
  J.~Garcia-Bellido and D.~G.~Figueroa,
  Phys.\ Rev.\ Lett.\  {\bf 98}, 061302 (2007)
  [astro-ph/0701014];
%
  J.~F.~Dufaux, A.~Bergman, G.~N.~Felder, L.~Kofman and J.~P.~Uzan,
  Phys.\ Rev.\ D {\bf 76}, 123517 (2007)
  [arXiv:0707.0875 [astro-ph]];
%
  J.~F.~Dufaux, G.~Felder, L.~Kofman and O.~Navros,
  JCAP {\bf 0903}, 001 (2009)
  [arXiv:0812.2917 [astro-ph]];
%
  J.~F.~Dufaux,
  Phys.\ Rev.\ Lett.\  {\bf 103}, 041301 (2009)
  [arXiv:0902.2574 [astro-ph.CO]];
%
  J.~F.~Dufaux, D.~G.~Figueroa and J.~Garcia-Bellido,
  Phys.\ Rev.\ D {\bf 82}, 083518 (2010)
  [arXiv:1006.0217 [astro-ph.CO]];
%
  D.~G.~Figueroa and T.~Meriniemi,
  JHEP {\bf 1310}, 101 (2013)
  [arXiv:1306.6911 [astro-ph.CO]].



\bibitem{Antoniadis:2014xva} 
  I.~Antoniadis and S.~P.~Patil,
  Eur.\ Phys.\ J.\ C {\bf 75}, 182 (2015)
  doi:10.1140/epjc/s10052-015-3411-z
  [arXiv:1410.8845 [hep-th]].


\bibitem{Kleban:2015daa} 
  M.~Kleban, M.~Mirbabayi and M.~Porrati,
  JCAP {\bf 1601}, no. 01, 017 (2016)
  doi:10.1088/1475-7516/2016/01/017
  [arXiv:1508.01527 [hep-th]].


\bibitem{Antoniadis:2015txa} 
  I.~Antoniadis and S.~P.~Patil,
  arXiv:1510.06759 [hep-th].




\bibitem{Guzzetti:2016mkm} 
  M.~C.~Guzzetti, N.~Bartolo, M.~Liguori and S.~Matarrese,
  arXiv:1605.01615 [astro-ph.CO].


\bibitem{Sorbo:2011rz} 
  L.~Sorbo,
  JCAP {\bf 1106}, 003 (2011)
  [arXiv:1101.1525 [astro-ph.CO]].





\bibitem{Namba:2015gja} 
  R.~Namba, M.~Peloso, M.~Shiraishi, L.~Sorbo and C.~Unal,
  JCAP {\bf 1601}, no. 01, 041 (2016)
  doi:10.1088/1475-7516/2016/01/041
  [arXiv:1509.07521 [astro-ph.CO]].


\bibitem{Ferreira:2014zia} 
  R.~Z.~Ferreira and M.~S.~Sloth,
  JHEP {\bf 1412}, 139 (2014)
  [arXiv:1409.5799 [hep-ph]].



\bibitem{Ferreira:2015omg} 
  R.~Z.~Ferreira, J.~Ganc, J.~Noreña and M.~S.~Sloth,
  arXiv:1512.06116 [astro-ph.CO].
  

\bibitem{Ozsoy:2014sba} 
  O.~…zsoy, K.~Sinha and S.~Watson,
  Phys.\ Rev.\ D {\bf 91}, no. 10, 103509 (2015)
  doi:10.1103/PhysRevD.91.103509
  [arXiv:1410.0016 [hep-th]].


\bibitem{Anber:2006xt} 
  M.~M.~Anber and L.~Sorbo,
  JCAP {\bf 0610}, 018 (2006)
  [astro-ph/0606534].


\bibitem{Caprini:2014mja} 
 C.~Caprini and L.~Sorbo,
 JCAP {\bf 1410}, no. 10, 056 (2014)
 doi:10.1088/1475-7516/2014/10/056
 [arXiv:1407.2809 [astro-ph.CO]].


\bibitem{Bamba:2014vda} 
  K.~Bamba,
  Phys.\ Rev.\ D {\bf 91}, 043509 (2015)
  doi:10.1103/PhysRevD.91.043509
  [arXiv:1411.4335 [astro-ph.CO]].


\bibitem{Barnaby:2011vw}
  N.~Barnaby, R.~Namba and M.~Peloso,
  JCAP {\bf 1104} (2011) 009
  [arXiv:1102.4333 [astro-ph.CO]].



\bibitem{Meerburg:2012id} 
  P.~D.~Meerburg and E.~Pajer,
  JCAP {\bf 1302}, 017 (2013)
  doi:10.1088/1475-7516/2013/02/017
  [arXiv:1203.6076 [astro-ph.CO]].



\bibitem{gauge-GW-discussion}
%
  S.~Eccles, W.~Fischler, D.~Lorshbough and B.~A.~Stephens,
  arXiv:1505.04686 [astro-ph.CO];
%
  K.~Choi, K.~Y.~Choi, H.~Kim and C.~S.~Shin,
  arXiv:1507.04977 [astro-ph.CO];
%
  I.~Ben-Dayan,
  arXiv:1604.07899 [astro-ph.CO];
%


\bibitem{Barnaby:2011qe} 
  N.~Barnaby, E.~Pajer and M.~Peloso,
  Phys.\ Rev.\ D {\bf 85}, 023525 (2012)
  doi:10.1103/PhysRevD.85.023525
  [arXiv:1110.3327 [astro-ph.CO]].


\bibitem{Domcke:2016bkh} 
  V.~Domcke, M.~Pieroni and P.~BinŽtruy,
  arXiv:1603.01287 [astro-ph.CO].


\bibitem{Crowder:2012ik} 
  S.~G.~Crowder, R.~Namba, V.~Mandic, S.~Mukohyama and M.~Peloso,
  Phys.\ Lett.\ B {\bf 726}, 66 (2013)
  doi:10.1016/j.physletb.2013.08.077
  [arXiv:1212.4165 [astro-ph.CO]].


\bibitem{Linde:2012bt} 
  A.~Linde, S.~Mooij and E.~Pajer,
  Phys.\ Rev.\ D {\bf 87}, no. 10, 103506 (2013)
  doi:10.1103/PhysRevD.87.103506
  [arXiv:1212.1693 [hep-th]].

\bibitem{Erfani:2015rqv} 
  E.~Erfani,
  JCAP {\bf 1604}, no. 04, 020 (2016)
  doi:10.1088/1475-7516/2016/04/020
  [arXiv:1511.08470 [astro-ph.CO]].


\bibitem{Cheng:2015oqa} 
  S.~L.~Cheng, W.~Lee and K.~W.~Ng,
  Phys.\ Rev.\ D {\bf 93}, no. 6, 063510 (2016)
  doi:10.1103/PhysRevD.93.063510
  [arXiv:1508.00251 [astro-ph.CO]].

\bibitem{McDonough:2016xvu} 
  E.~McDonough, H.~B.~Moghaddam and R.~H.~Brandenberger,
  JCAP {\bf 1605}, no. 05, 012 (2016)
  doi:10.1088/1475-7516/2016/05/012
  [arXiv:1601.07749 [hep-th]].

\bibitem{Mukohyama:2014gba} 
  S.~Mukohyama, R.~Namba, M.~Peloso and G.~Shiu,
  JCAP {\bf 1408}, 036 (2014)
  [arXiv:1405.0346 [astro-ph.CO]].






\bibitem{odd-TTT} 
  J.~L.~Cook and L.~Sorbo,
  JCAP {\bf 1311}, 047 (2013)
  [arXiv:1307.7077 [astro-ph.CO]]; 
%
\bibitem{Shiraishi:2013kxa} 
  M.~Shiraishi, A.~Ricciardone and S.~Saga,
  JCAP {\bf 1311}, 051 (2013)
  [arXiv:1308.6769 [astro-ph.CO]]; 
%
\bibitem{Shiraishi:2014roa} 
  M.~Shiraishi, M.~Liguori and J.~R.~Fergusson,
  JCAP {\bf 1405}, 008 (2014)
  doi:10.1088/1475-7516/2014/05/008
  [arXiv:1403.4222 [astro-ph.CO]]; 
%
\bibitem{Shiraishi:2014ila} 
  M.~Shiraishi, M.~Liguori and J.~R.~Fergusson,
  JCAP {\bf 1501}, no. 01, 007 (2015)
  doi:10.1088/1475-7516/2015/01/007
  [arXiv:1409.0265 [astro-ph.CO]].



\bibitem{Anber:2009ua} 
  M.~M.~Anber and L.~Sorbo,
  Phys.\ Rev.\ D {\bf 81}, 043534 (2010)
  doi:10.1103/PhysRevD.81.043534
  [arXiv:0908.4089 [hep-th]].





\bibitem{Ade:2015lrj} 
  P.~A.~R.~Ade {\it et al.}  [Planck Collaboration],
  arXiv:1502.02114 [astro-ph.CO].




\bibitem{Kim:2004rp} 
  J.~E.~Kim, H.~P.~Nilles and M.~Peloso,
  JCAP {\bf 0501}, 005 (2005)
  doi:10.1088/1475-7516/2005/01/005
  [hep-ph/0409138].





\bibitem{Peloso:2015dsa} 
  M.~Peloso and C.~Unal,
  JCAP {\bf 1506}, no. 06, 040 (2015)
  doi:10.1088/1475-7516/2015/06/040
  [arXiv:1504.02784 [astro-ph.CO]].



\bibitem{Pajer:2013fsa} 
  E.~Pajer and M.~Peloso,
  Class.\ Quant.\ Grav.\  {\bf 30}, 214002 (2013)
  doi:10.1088/0264-9381/30/21/214002
  [arXiv:1305.3557 [hep-th]].

 

\end{thebibliography}
\end{document}